\documentclass{ws-ijmpe}
\usepackage[super,compress]{cite}
\usepackage[colorlinks=true,linkcolor=red, linktocpage=true]{hyperref}

\usepackage{acro}

\DeclareAcronym{GCM}{
  short=GCM,
  long=Generator Coordinate Method
} 
\DeclareAcronym{PNP}{
  short=PNP,
  long=Particle-Number Projection
} 
\DeclareAcronym{AMP}{
  short=AMP,
  long=Angular-Momentum Projection
} 

\DeclareAcronym{PP}{
  short=PP,
  long=Parity Projection
} 
\DeclareAcronym{SPGCM}{
  short=SPGCM,
  long=Symmetry-Projected Generator Coordinate Method
}  

\DeclareAcronym{CDFT}{
  short=CDFT,
  long=Covariant Density Functional Theory
} 

\DeclareAcronym{RMF}{
  short=RMF,
  long=Relativistic  Mean Field
}

\DeclareAcronym{RHB}{
  short=RHB,
  long=Relativistic  Hartree Bogoliubov
}

\DeclareAcronym{HF}{
  short=HF,
  long=Hartree Fock
}  

\DeclareAcronym{HFB}{
  short=HFB,
  long=Hartree Fock Bogoliubov
}  
\DeclareAcronym{MBPT}{
  short=MBPT,
  long=Many-Body Perturbation Theory
}

\DeclareAcronym{BCS}{
  short=BCS,
  long=Bardeen-Cooper-Schrieffer
}  

\DeclareAcronym{QRPA}{
  short=QRPA,
  long=Quasiparticle Random Phase Approximation
}  
 
\DeclareAcronym{EDF}{
  short=EDF,
  long=Energy Density Functional
}

\DeclareAcronym{ISM}{
  short=ISM,
  long=Interacting Shell Model
}

\DeclareAcronym{CP}{
  short=CP,
  long=Charge-Parity
}  

\DeclareAcronym{PTSM}{
  short=PTSM,
  long=Pair-Truncated Shell Model 
}

\DeclareAcronym{BMF}{
  short=BMF,
  long=Beyond-Mean-Field
}  

\DeclareAcronym{IBM}{
  short=IBM,
  long=Interacting Boson Model
}

\DeclareAcronym{HO}{
  short=HO,
  long=Harmonic Oscillator
}  

\DeclareAcronym{AMD}{
short=AMD,
long=Antisymmetrized Molecular Dynamics
}

\DeclareAcronym{FMD}{
short=FMD,
long=Fermionic Molecular Dynamics
}

\DeclareAcronym{VAP}{
short=VAP,
long=Variation After Projection
} 

\DeclareAcronym{EDM}{
short=EDM,
long=Electric Dipole Moment
}

\DeclareAcronym{NSM}{
short=NSM,
long=Nuclear Schiff Moment
}

\DeclareAcronym{SRG}{
short=SRG,
long=Similarity Renormalization Group 
} 

\DeclareAcronym{EFT}{
short=EFT,
long=Effective Field Theory
} 

 \DeclareAcronym{LEC}{
short=LEC,
long=Low-Energy Constant
} 

 \DeclareAcronym{QOV}{
short=QOV,
long=Quadrupole-Octupole Vibrator
}

\DeclareAcronym{PRM}{
short=PRM,
long=Particle-Rotor Model
}

\DeclareAcronym{WS}{
short=WS,
long=Woods-Saxon
}

\DeclareAcronym{RCC}{
short=RCC,
long=relativistic coupled-cluster
} 
 
\usepackage{graphicx,amsmath,amssymb,bm,multirow} 
\usepackage{amsfonts}
\usepackage{dcolumn,CJK}
\usepackage{bm}
\usepackage{mathrsfs}
\usepackage[usenames,dvipsnames]{color}

\usepackage[T1]{fontenc}
\usepackage[utf8]{inputenc}
\usepackage{graphicx,color,rotating,pifont} 
\usepackage{mathtools}
\usepackage{siunitx}  
\usepackage{ae}

\usepackage{siunitx}
\usepackage{tensor}
\usepackage{braket}
\usepackage{booktabs}
\usepackage{xspace}
\usepackage{mathrsfs} 

\newcommand{\beq}{\begin{equation}}
\newcommand{\eeq}{\end{equation}}
\newcommand{\beqn}{\begin{eqnarray}}
\newcommand{\eeqn}{\end{eqnarray}}
\newcommand{\bsub}{\begin{subequations}}
\newcommand{\esub}{\end{subequations}}
\newcommand{\bpm}{\begin{pmatrix}}
\newcommand{\epm}{\end{pmatrix}}

\begin{document}

\markboth{E. F. Zhou, J. M. Yao}{Generator coordinate method for nuclear octupole excitations: status and perspectives}

\catchline{}{}{}{}{}


\title{Generator coordinate method for nuclear octupole excitations: status and perspectives}

 \author{E. F. Zhou}
 \address{School of Physics and Astronomy, Sun Yat-sen University, Zhuhai 519082, P.R. China.}

 \author{J. M. Yao \footnote{Corresponding author: yaojm8@sysu.edu.cn}}
 \address{School of Physics and Astronomy, Sun Yat-sen University, Zhuhai 519082, P.R. China.}
 
\maketitle


\begin{abstract}
Strong octupole correlations have been observed in the low-lying states of atomic nuclei across various mass regions. In this review, we provide an overview of Beyond Mean-Field (BMF) studies of nuclear octupole collective motions with Generator Coordinate Method (GCM) in combination with quantum-number projections that are implemented to restore the broken symmetries in nuclear mean-field states. We highlight recent developments within this framework and their applications to excitation spectra and electromagnetic transition rates in octupole-shaped nuclei and hypernuclei. We discuss the novel phenomena of nucleon clustering in light nuclei. Additionally, we explore the phase transition from octupole vibrations to rotational motions as spin increases in heavy nuclei. Lastly, we examine the status and future prospects of studies on octupole deformation effects in nuclear Schiff moments. These studies, along with the upper limits of atomic Electric Dipole Moment (EDM), impose stringent constraints on beyond-standard-model time-reversal-violating nucleon-nucleon interactions.
 
\end{abstract}


\ccode{PACS numbers:}

\tableofcontents

\section{Introduction}
 
An atomic nucleus is a self-bound quantum many-body system composed of several to hundreds of nucleons. These nucleons distribute within a range of a few femtometers, approximately given by the formula $R\simeq 1.2A^{1/3}$ fm, where $A$ is the mass number of the nucleus. With a specific combination of neutrons and protons ($N, Z$), an atomic nucleus can exhibit different equilibrium shapes characterized by nonzero values of dimensionless deformation parameters $\beta_{\lambda\mu}$. In mean-field approaches, these parameters are defined by the expectation values of multipole moment operators, represented as:
\begin{equation}
\label{deformation}
\beta_{\lambda\mu} = \dfrac{4\pi}{3AR^\lambda}\bra{\Phi(\mathbf{q})} \hat Q_{\lambda \mu}\ket{\Phi(\mathbf{q})},
\end{equation}
where $\hat Q_{\lambda\mu}=r^\lambda Y_{\lambda \mu}$ with $\mu=-\lambda,\cdots,\lambda$, and $Y_{\lambda \mu}$   stands for the spherical harmonic function of rank $\lambda$. The symbol $\ket{\Phi(\mathbf{q})}$ represents the wave function of a nuclear state in the intrinsic frame.

Most atomic nuclei exhibit either a spherical equilibrium shape, where $\beta_{\lambda\mu}=0$, or intrinsic ellipsoidal shapes in their ground states. These ellipsoidal shapes can be prolate, indicated by $\beta_{20}>0$, or oblate, indicated by $\beta_{20}<0$. In rare instances, some atomic nuclei spontaneously break reflection symmetry by having a nonzero octupole deformation parameter, $\beta_{3\mu}\neq0$. The presence of intrinsic octupole deformation in the nuclear ground state is inferred from the observation of low-lying negative-parity states or parity doublets connected with strong electric octupole ($E3$) transitions~\cite{Ahmad:1993review,Buttler:1996RMP}.

The emergence of deformed equilibrium shapes in atomic nuclei is attributed to the strong residual nucleon-nucleon interaction between valence nucleons. This interaction can be parameterized as separable multipole-multipole interactions.
In nuclei with neutron or proton numbers near 56, 88, 134, and 194, which are only slightly greater than the magic numbers 50, 82, 128, and 184, respectively, a strong octupole-octupole coupling occurs between pairs of orbitals with $\Delta \ell = \Delta j = 3$. These orbitals are located around the Fermi surface and include pairs such as $(d_{5/2}, h_{11/2})$, $(f_{7/2}, i_{13/2})$, $(g_{9/2}, j_{15/2})$, and $(h_{11/2}, k_{17/2})$. Notably,  the isoscalar neutron-proton part~\cite{Chen:2021} plays a significant role in driving atomic nuclei toward reflection-asymmetric octupole shapes in their ground states. Pairing correlation between nucleons also influences the occupation of nucleons on the orbitals around the Fermi surface, and thus, it has a significant impact on the equilibrium shapes. Whether an atomic nucleus exhibits octupole shapes in its ground state depends on the competition between the underlying shell structure and pairing correlation. Both aspects can be effectively addressed in self-consistent mean-field approaches~\cite{Ring:1980,Bender:2003RMP}.  Recent studies, based on \ac{RHB} approaches~\cite{Agbemava:2016PRC}, Skyrme \ac{HF}+\ac{BCS}~\cite{Ebata:2017} and Skyrme \ac{HFB} approaches~\cite{Cao:2020}, have systematically explored atomic nuclei with nonzero octupole deformation $\beta_{30}$ in their ground states. These studies reveal that the majority of octupole-deformed nuclei are indeed found near the intersection of neutron numbers $N\simeq 88, 134,$ and $194$, and proton numbers $Z\simeq 56$ and $88$~\cite{Cao:2020}, consistent with the finding with the macroscopic–microscopic nuclear model~\cite{Moller:2008}.

 Octupole collective motions in atomic nuclei manifest as low-lying octupole vibrational or rotational states. This phenomenon applies even to spherical nuclei with closed-shell configurations for neutrons and/or protons, such as $^{208}$Pb~\cite{Brown:2000, Yao:2016Pb, Robledo:2016EPJA, Isacker:2022}, where the first $3^-$ state is interpreted as an one-octupole-phonon state.
 This octupole-phonon state has been explored systematically in stable even-even nuclei with the  methods of \ac{GCM}+ \ac{PP}~\cite{Robledo:2016EPJA} and \ac{QRPA}~\cite{Bui:2023}. 

\begin{figure}[tb]
\includegraphics[width=\textwidth]{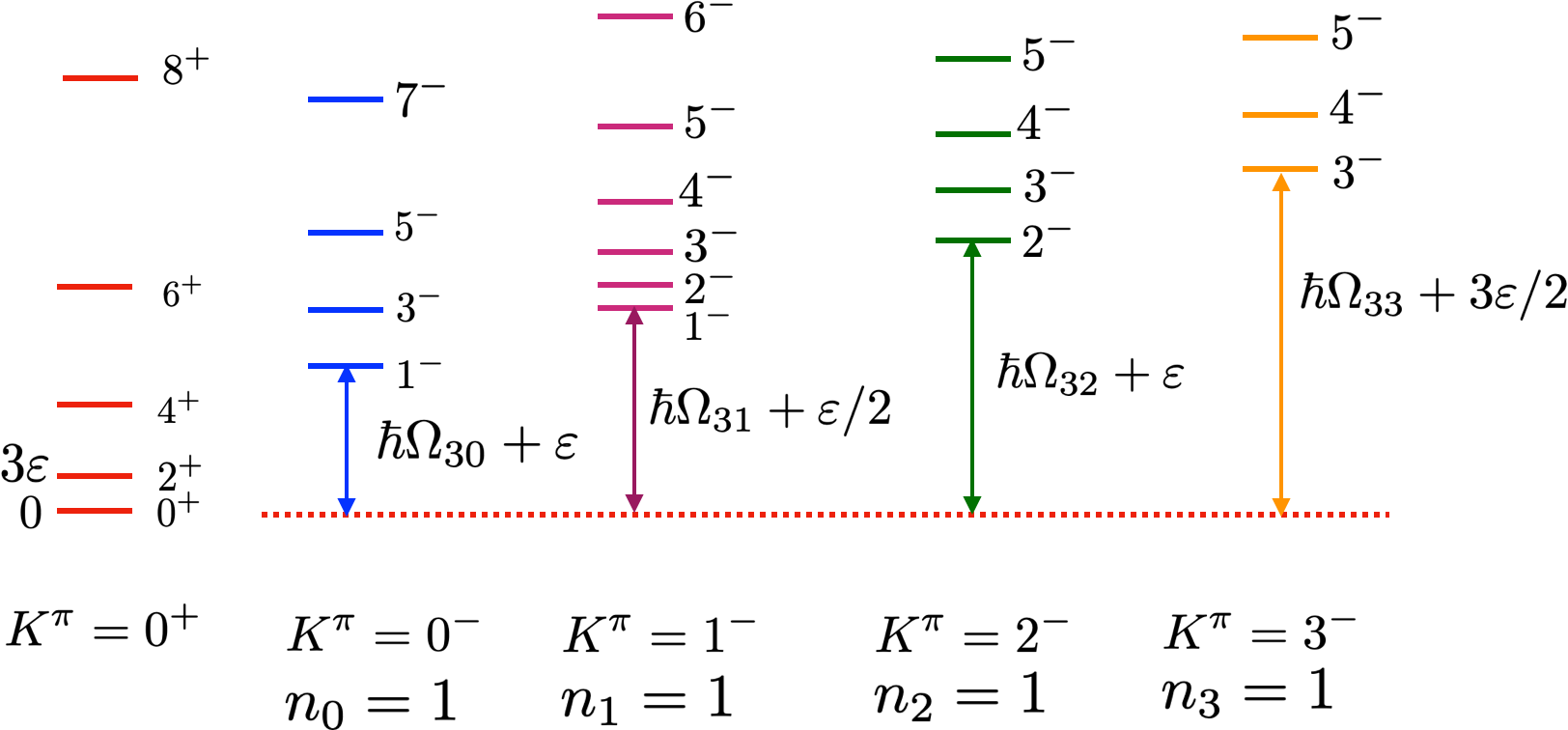}
\caption{Rotational bands in a collective model for an even-even deformed nucleus with different octupole fluctuations. The rotational bands are built on top of one-octupole-phonon states with various values of $K(=0, 1, 2, 3)$.  The labels for the bands include $K^\pi$ (where $K$ represents the angular momentum projection and $\pi$ represents parity) and a nonzero oscillator number $n_\mu$, which represents different types of octupole vibrations. This figure has been adapted from Ref.~\cite{Eisenberg:1970}.}
\label{fig:cartoon4quadrupole-octupole}
\end{figure}

As the nucleon number departs from closed shells, atomic nuclei become deformed, leading to the observation of rotation bands built upon octupole vibrational states. The energy spectrum in this context can be effectively described by a phenomenological octupole collective Hamiltonian. For illustration purpose, we neglect the coupling of rotations and octupole vibrations. Under this approximation, the excitation energies of low-lying states in the collective model are given by~\cite{Eisenberg:1970}
\begin{equation}
E(J, K) = (n_0+1/2)\hbar\Omega_{30} + \sum_{\mu=1}^3 (2n_\mu + 1 + |K_\mu|)\hbar\Omega_{3\mu} + \frac{\varepsilon}{2}[J(J+1)-K^2].
\end{equation}
Here, $n_\mu$ with $\mu=0, 1, 2, 3$ correspond to the number of different types of octupole phonons, $K_\mu=0, \pm1, \pm2, \cdots$, and $K$ is the projection of the total angular momentum $J$ along the intrinsic symmetric axis. The octupole vibrational states of type $\mu$ with the same value of the major quantum number $N_\mu$ defined as $N_\mu\equiv 2n_\mu + 1 + |K_\mu|$ are degenerate. The corresponding energy spectrum with multiple negative-parity bands is schematically depicted in Fig.\ref{fig:cartoon4quadrupole-octupole}. Notably, this energy spectrum has been observed in some isotopes, such as $^{152}$Sm, where octupole bands with $K^\pi=0^-$, $1^-$, and $2^-$ have been observed up to relatively high spin in the $(\alpha, 2n)$ data

\begin{figure}[tb]
\includegraphics[width=\textwidth]{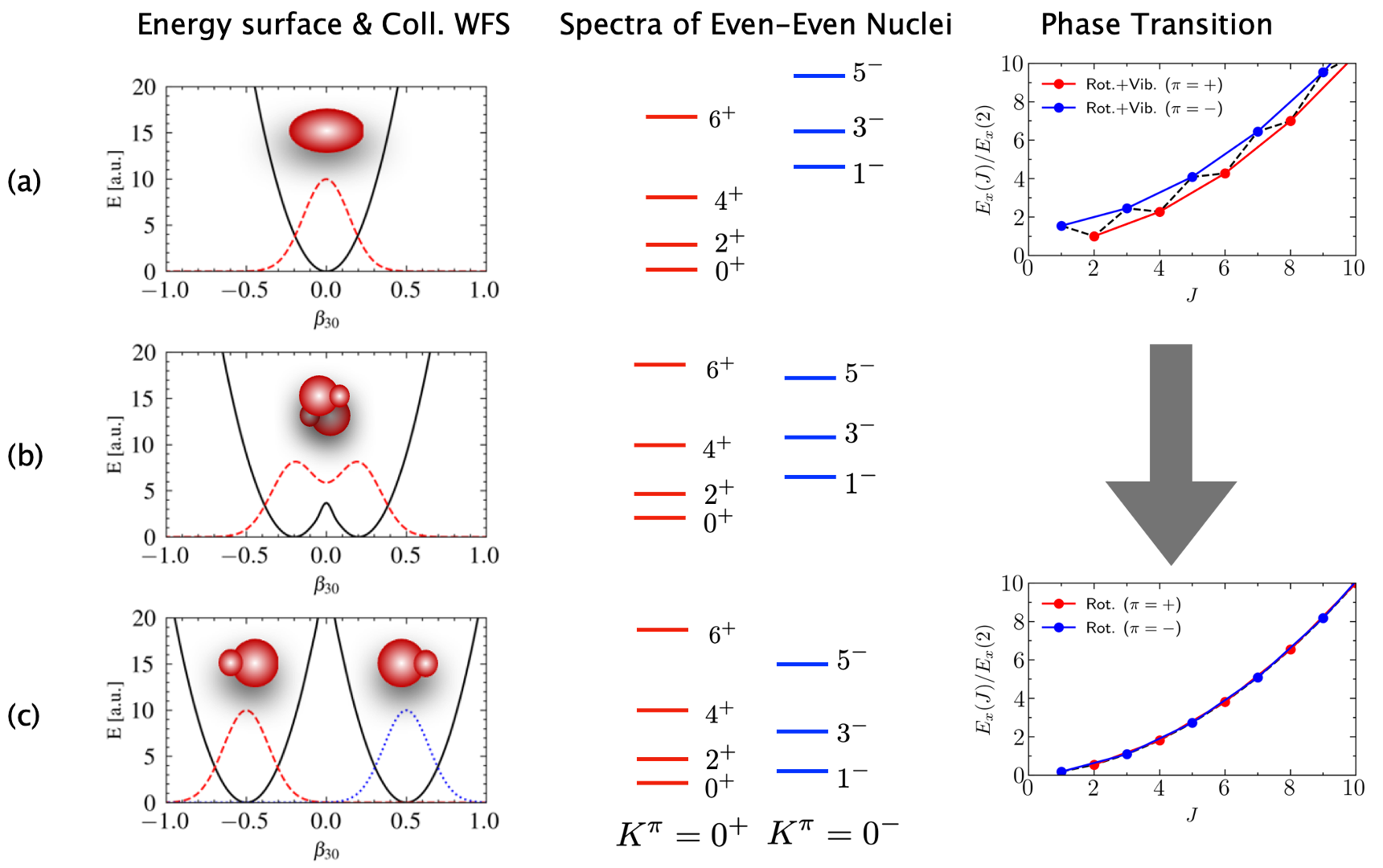}
\caption{The phase transition from an octupole vibrator to an octupole rotor in axially deformed even-even nuclei. (a) The octupole deformation $\beta_{30}$ of the equilibrium shape in the ground state is zero. (b) The nonzero octupole deformation $|\beta_{30}|$ is developed in the ground state, but with a strong quantum tunneling effect. (c) The static octupole deformation $|\beta_{30}|$ is formed without tunneling between the intrinsic states with positive and negative values of $\beta_{30}$, resulting in a rigid octupole rotor. The barrier height separating the degenerate minima is set to infinity, resulting in an energy spectra characteristic of a rigid rotor.  The corresponding energy spectra of $K^\pi=0^+, 0^-$ and the excitation energy ratio $R_{J/2}$ as a function of the angular momentum $J$ ($\hbar$) are also provided. }
\label{fig:cartoon4transition}
\end{figure}

With an increase in the octupole-octupole coupling strength, the excitation energy of the octupole vibrational state decreases and eventually becomes degenerate with the ground state. Beyond a critical value, the nuclear system becomes unstable against octupole vibration, leading to the development of a static octupole-deformed shape in the ground state. As a consequence of this transition, the low-lying states become dominated by the rotational excitations of octupole-deformed shapes. This transition is illustrated in Fig.~\ref{fig:cartoon4transition}, which shows the shift from an octupole vibrator to an octupole rotor. In the rigid rotor with an octupole deformed shape, $\hbar\Omega_{30}=0$. Consequently, the positive and negative-parity states merge into one rotational band. To characterize this phase transition, the ratio of excitation energies, denoted as $R_{J/2}$, provides a signature. It is calculated as the ratio of the excitation energy for a specific angular momentum state $J$ to that of the $2^+$ state, i.e., $R_{J/2}=E_x(J)/E_x(2)$. As the transition progresses from an octupole vibrator to a rigid rotor, the staggering amplitude of the energy ratios $R_{J2}$ decreases, as depicted in Fig.~\ref{fig:cartoon4transition}(c).

In many cases, the barrier height separating the two degenerate minimal states $\ket{\Phi(\beta_{20}, \beta_{30})}$ and $\ket{\Phi(\beta_{20}, -\beta_{30})}=\hat P\ket{\Phi(\beta_{20}, \beta_{30})}$ (as illustrated in Fig.~\ref{fig:cartoon4transition}(c)) may not be high enough to eliminate the tunneling effect between them. In other words, the overlap $\bra{\Phi(\beta_{20}, \beta_{30})}\hat P\ket{\Phi(\beta_{20}, \beta_{30})}$ of the two intrinsic wave functions with positive and negative values of $\beta_{30}$, connected by the space-reversal operator $\hat P$, is nonzero. This quantum tunneling effect leads to an energy shift between the positive and negative-parity states, resulting in staggering in the energy ratio. This phenomenon is observed in the low-spin states of all even-even octupole-deformed nuclei. It implies that the dynamical effect tends to obscure the transition pattern from an octupole vibrator to a rigid octupole rotor in nuclear ground states as a function of nucleon numbers. Strictly speaking, the so-called stable or permanent octupole-deformed shapes with $\bra{\Phi}\hat P\ket{\Phi}=0$ are rarely realized in the ground states of realistic nuclei. Even though strong $E3$ transitions are observed in their low-lying states, such as in $^{144,146}$Ba\cite{Phillips:2020, Bucher:2016PRL, Bucher:2017PRL} and $^{222,224,226}$Ra~\cite{Gaffney:2013nature, Butler:2020PRL}, which are claimed to have permanent octupole deformations. However, there is substantial evidence indicating that octupole shapes in heavy nuclei become stabilized with increasing rotation, leading to the merging of positive and negative-parity states into one rotation band~\cite{Garrote:1997,Yao:2015Ra224,Fu:2018Ba}. This phenomenon highlights the intricate interplay between quantum tunneling effects and rotational motion in nuclear systems. It is worth mentioning that the onset of octupole deformation  in atomic nuclei may also generate other exotic collective motions, such as multiple chiral doublet bands~\cite{Liu:2016_Chiral,Wang:2019_Chiral}.

 Except for the nuclear models based on collective Hamiltonians~\cite{Egido:1989NPA,Li:2013PLB,Xia:2017,Zhao:2018IJMPE,Wang:2019_Chiral} and IBM models~\cite{Nomura:2015,Nomura:2022},  the \ac{GCM}, when combined with symmetry restoration techniques~\cite{Sheikh:2021qv,Yao:2022HandBook}, provides a powerful microscopic tool for studying nuclear collective motions associated with exotic shapes. In the context of a general mean-field wave function $\ket{\Phi(\mathbf{q})}$ with nonzero octupole deformation, both rotational symmetry and reflection symmetry are broken. Consequently, a nuclear state described by the mean-field wave function $\ket{\Phi(\mathbf{q})}$ does not possess a definite spin-parity $J^\pi$. Furthermore, if pairing correlation is considered by the introduction of quasiparticles in the mean-field state $\ket{\Phi(\mathbf{q})}$, nucleon numbers ($N, Z$) are not conserved. The broken symmetries of these mean-field wave functions with nonzero intrinsic deformations are restored using projections and finally mixed in the \ac{GCM}.  This \ac{SPGCM}  has been extensively and successfully applied to describe nuclear low-lying states associated with octupole deformation.   In the early studies, \ac{PP} was implemented into GCM to restore the parities of nuclear states from the mean field calculations with the BKN force~\cite{Marcos:1983NPA},  Skyrme~\cite{Bonche:1991,Skalski:1993,Heenen:1994,Heenen:2001} and Gogny forces~\cite{Egido:1991NPA,Gunzman:2012}. In the recent decade, GCM+\ac{PP}+\ac{PNP}+\ac{AMP} have been implemented into \ac{CDFT}~\cite{Yao:2015Ra224,Zhou:2016,Marevic:2018Ne,Marevi:2019C12} and the HFB with the Gogny force~\cite{Bernard:2016}. Recently, PP+AMP has been implemented into \ac{RHB} for nuclear states with exotic octupole shapes~\cite{Rong:2023PLB}. Within these \ac{SPGCM} frameworks, the picture of rotation-induced phase transition from an octupole vibrator to an octupole-deformed rotor has been illustrated in $^{224}$Ra~\cite{Yao:2015Ra224} and $^{144}$Ba~\cite{Fu:2018Ba}. 

 It is worth mentioning that recent theoretical studies on heavy-ion collisions with a hydrodynamic model have shown that the elliptic flow and triangular flow are sensitive to the nuclear quadrupole and octupole deformations of the colliding nuclei~\cite{Giacalone:2021, Zhang:2022PRL, Bally:2022PRL}. This indicates that flow measurements in heavy-ion collisions, especially when using isobaric systems, provide a complementary method for studying octupole collectivity in nuclear ground states~\cite{Bally:2022}.
 
 The emergence of strong octupole correlations in atomic nuclei may also have a significant impact on the studies of fundamental interactions and symmetries at nuclear level. The \ac{SPGCM} study based on the \ac{CDFT} has demonstrated that the effect of dynamical octupole correlation may quench the nuclear matrix element of neutrinoless double-beta decay in \nuclide[150]{Nd} by about 7\%~\cite{Yao:2016PRC}. See Ref.~\cite{Yao:2022PPNP} for the review on the \ac{BMF} studies of neutrinoless double-beta decay. The nuclei with  octupole deformations  are also of great interest for the search of nonzero permanent atomic \ac{EDM}. The current upper limit provides a valuable constraint on the strengths of time-reversal (T) symmetry-violating interactions and serves as a test for beyond-standard models of \ac{CP} violation. The atomic \ac{EDM} is proportional to the nuclear Schiff moment, which can be enhanced by several orders of magnitude due to octupole deformation~\cite{Engel:2013PPNP,Chupp:2017rkp}. Consequently, the sensitivity of atomic EDM experiments can be significantly increased when the corresponding atomic nucleus exhibits substantial octupole deformation in its ground state.

In this review, we provide an overview of recent GCM studies on nuclear octupole collective motions. We address the role of octupole correlation in the formation of cluster structures in light nuclei, the staggering behavior observed in the low-spin parity-doublet states of medium-mass or heavy nuclei around \nuclide[144]{Ba} and \nuclide[224]{Ra}, and the enhancement of nuclear Schiff moments in odd-mass nuclei. Due to the page limit, lots of exciting developments~\cite{Yao:2020PRL,Frosini:2022EPJA3,Zhang:2023_ML,Matsumoto:2023} in the GCM will not be covered. For a general review of GCM studies on nuclear collective motions, readers can refer to Refs.~\cite{Wong:1975PR,Bender:2003RMP,Robledo:2018,Niksic:2011PPNP}.  
    
The review is arranged as follows: In Sec.\ref{sec:PGCM}, we present the basic formulas for the SPGCM. For the sake of self-containment, we first introduce the general formulas of HFB approaches, which are usually employed to generate reference states for the SPGCM calculations. Afterward, we present the formulas for energy spectrum and electromagnetic multiple transition strengths. In Sec.\ref{sec:collective_motions}, we highlight the SPGCM studies on nuclear collective motions. In Sec.\ref{sec:Schiff_moments}, we introduce the status and perspectives of computing nuclear Schiff moments. A summary of the review is provided in Sec.\ref{sec:summary}.



\section{The \ac{SPGCM}}
\label{sec:PGCM}

 \subsection{The reference states with octupole deformations}

 \subsubsection{The Hartree-Fock-Bogoliubov (HFB) state}

 Both long-range collective correlations associated with deformations and short-range correlations associated with nucleon pairing can be simultaneously described with the HFB approach~\cite{Ring:1980}. The nuclear wave function from the HFB approach is an approximation of the exact nuclear wave function in the intrinsic frame, known as the HFB wave function. This wave function is a general product of wave functions consisting of independently moving quasiparticles
\begin{equation}
\label{eq:HFB_wf}
\ket{\Phi(\bm{q})} = \prod^M_{k=1} \beta_k(\bm{q}) \ket{0},
\end{equation}
where $\ket{0}$ denotes the particle vacuum. It is important to note that the state described by the HFB wave function is a quasiparticle vacuum, i.e., $\beta_k(\bm{q})\ket{\Phi(\bm{q})}=0$. This state is referred to as the HFB state.  The quasiparticle operators $\beta_k(\bm{q}), \beta^\dagger_k(\bm{q})$ are defined as linear combinations of the particle creation and annihilation operators $(c^\dagger_p, c_p)$,
 \beq
 \label{eq:Bogoliubov_Ttransformation}
 \begin{pmatrix}
    \beta(\bm{q})\\
    \beta^{\dagger}(\bm{q})
    \end{pmatrix}
       =
{\cal W}^\dagger(\bm{q})
\begin{pmatrix}
    c\\
    c^\dagger
 \end{pmatrix}
    =
 \begin{pmatrix}
    U^\dagger(\bm{q}) & V^\dagger(\bm{q}) \\
    V^T(\bm{q})       & U^T(\bm{q})
 \end{pmatrix}
\begin{pmatrix}
    c\\
    c^\dagger
 \end{pmatrix},
 \eeq
where $\bm{q}$ labels different HFB states. The unitary transformation (\ref{eq:Bogoliubov_Ttransformation}) is known as the Bogoliubov transformation with the ${\cal W}(\bm{q})$ being a $2M\times2M$ matrix and it preserves the anticommutator relations. The unitarity requires that the  matrices $(U, V)$ satisfy the following relations 
\begin{align}
U^{\dagger} U+V^{\dagger} V &=1, \quad U U^{\dagger}+V^{*} V^{\mathrm{T}}=1, \\
U^{\mathrm{T}} V+V^{\mathrm{T}} U &=0, \quad    U V^{\dagger}+V^{*} U^{\mathrm{T}}=0.
\end{align} 

The HFB wave function in (\ref{eq:HFB_wf}) is defined by the Bogoliubov transformation coefficients ($U, V$), which are determined by the variational principle.  According to {\em Thouless' theorem}~\cite{Thouless:1960NPA}, any general product wave function $\ket{\Phi}$ which is not orthogonal to another product wave function $\ket{\Phi^{(0)}}$ can be expressed as 
\beq 
\label{eq:Thouless}
\ket{\Phi ({\cal Z})} = {\cal N}\exp(\sum_{k<k'}{\cal Z}_{kk'}\beta^\dagger_k\beta^\dagger_{k'})\ket{\Phi^{(0)}}, 
\eeq
where the normalization constant ${\cal N}=\bra{\Phi}\Phi^{(0)}\rangle$, and ${\cal Z}$ is a skew symmetric matrix. The HFB reference state $\ket{\Phi^{(0)}}$ is the vacuum of quasiparticle operator $\beta^{(0)}$, connecting to particle operators $(c^\dagger_p, c_p)$ with the  Bogoliubov transformation coefficients ($U^{(0)}, V^{(0)}$).~\footnote{The proof of this theorem for HFB wave functions is provided in the textbook~\cite{Ring:1980}.} 

With the Thouless theorem (\ref{eq:Thouless}), one can express the energy of HFB wave function $\ket{\Phi}$ in terms of  the unknown matrix ${\cal Z}$,
\beqn
\label{eq:EDF}
E[{\cal Z}] &=&  \frac{\bra{\Phi ({\cal Z})} \hat H \ket{\Phi ({\cal Z})}}{\bra{\Phi ({\cal Z})} \Phi ({\cal Z})\rangle},
\eeqn
where $\hat H$ is the Hamiltonian operator for nuclear systems.  The energy $E[{\cal Z}]$ is a function of ${\cal Z}$ to be determined in variational iteration methods.

Supposing the HFB wave function at $n$-th step is $\ket{\Phi^{(n)}}$, the wave function $\ket{\Phi^{(n+1)}}$ at the $(n+1)$-th step is determined by
\beq 
\ket{\Phi^{(n+1)}} ={\cal N} \exp\Bigg(\sum_{k<k'}{\cal Z}^{(n)}_{kk'}\beta^{(n)\dagger}_k\beta^{(n)\dagger}_{k'}\Bigg)\ket{\Phi^{(n)}}. 
\eeq
The parameter  matrix  $G^{(n)}$ is defined as the gradient of the energy multiplied with a step size parameter $\eta$, 
 \beq   
 \label{eq:variation_step}
G^{(n)}_{kk'}
\equiv -\eta \frac{\partial}{\partial {\cal Z}^\ast_{kk'}} E[{\cal Z}]=-\eta H^{20}_{kk'},\quad G^{(n)}_{kk'}=-G^{(n)}_{k'k}.
\eeq 
 To ensure the commutation relation for the new quasiparticle operator $\beta^{(n+1)}_k$ and guarantee the orthogonality relations for the new HFB wave function ($U^{(n+1)}, V^{(n+1)}$), one finds the following relations~\cite{Ring:1980,Egido:1995NPA}
 \beqn
\label{eq:new_qp_operators}
\left(\begin{array}{l}
\beta^{(n+1)\dagger} \\
\beta^{(n+1)} 
\end{array}\right)
&=&\left(\begin{array}{ll}
{\left[L^{-1}\right]^{*}} & 0 \\
0 & L^{-1}
\end{array}\right)\left(\begin{array}{ll}
I & G^{(n)\dagger} \\
G^{(n)T} & I
\end{array}\right)
\left(\begin{array}{l}
\beta^{(n)\dagger} \\
\beta^{(n)} 
\end{array}\right)\nonumber\\
&=&\left(\begin{array}{ll}
[L^{-1}]^\ast  & [L^{-1}]^\ast G^{(n)\dagger} \\
L^{-1} G^{(n)T} & L^{-1}
\end{array}\right)
\left(\begin{array}{l}
\beta^{(n)\dagger} \\
\beta^{(n)} 
\end{array}\right),
\eeqn
and the coefficients are updated as below,
 \bsub
 \label{eq:HFB_updates}
 \beqn
 U^{(n+1)} &=& \Bigg(U^{(n)} + V^{(n)\ast} G^{(n)\ast}\Bigg)\left[L^{-1}\right]^{\dagger},\\
 V^{(n+1)} &=& \Bigg( V^{(n)} +U^{(n)\ast}G^{(n)\ast}\Bigg)\left[L^{-1}\right]^{\dagger},
 \eeqn
 \esub
 where the $L$ is determined by the Choleski decomposition
 \beq
\label{eq:decomposition_L}
 1 + G^{(n)T}  G^{(n)\ast} = LL^\dagger  
\eeq
and $G^{(n)}$ by the energy gradient (\ref{eq:variation_step}).

With the following two Bogoliubov transformations which relate the quasiparticle operators at the $n$-th and $(n+1)$-th iterations to the single-particle operators,
\bsub
\label{eq:Bogoliubov4both}
\beqn 
\left(\begin{array}{l}
\beta^{(n)\dagger} \\
\beta^{(n)} 
\end{array}\right)
&=&\left(\begin{array}{cc}
U^{(n)T} & V^{(n)T} \\
V^{(n)\dagger} & U^{(n)\dagger}
\end{array}\right)\left(\begin{array}{c}
c^{\dagger} \\
c
\end{array}\right),\\
\left(\begin{array}{l}
\beta^{(n+1)\dagger} \\
\beta^{(n+1)} 
\end{array}\right)
&=&\left(\begin{array}{cc}
U^{(n+1)T} & V^{(n+1)T} \\
V^{(n+1)\dagger} & U^{(n+1)\dagger}
\end{array}\right)\left(\begin{array}{c}
c^{\dagger} \\
c
\end{array}\right)
\eeqn
\esub
 and the unitary transformation between the two sets of quasiparticle operators~\cite{Ring:1980},
 \beqn
\label{eq:general_Bogoliubov}
\left(\begin{array}{l}
\beta^{(n+1)\dagger} \\
\beta^{(n+1)} 
\end{array}\right)
&=&\left(\begin{array}{cc}
{\cal U}^{T} & {\cal V}^{T} \\
{\cal V}^{\dagger} & {\cal U}^{\dagger}
\end{array}\right)
\left(\begin{array}{l}
\beta^{(n)\dagger} \\
\beta^{(n)} 
\end{array}\right),
\eeqn
one finds the expressions for the coefficients of the above unitary transformation,  
\bsub\beqn
{\cal U} &=& U^{(n)\dagger} U^{(n+1)}+V^{(n)\dagger} V^{(n+1)},\\ 
{\cal V} &=& V^{(n)T} U^{(n+1)}+U^{(n)T} V^{(n+1)}.
\eeqn
\esub
Combining (\ref{eq:new_qp_operators}) and (\ref{eq:general_Bogoliubov}), one finds the expression for the inverse of the matrix $L$,
\beq 
\label{eq:Lmatrix}
L^{-1}= {\cal U}^\dagger =U^{(n+1)\dagger}  U^{(n)} +V^{(n+1)\dagger}  V^{(n)}.
\eeq
Therefore, the HFB wave function is updated iteratively based on the Eqs.(\ref{eq:variation_step}), (\ref{eq:HFB_updates}), and (\ref{eq:Lmatrix}). As the matrices $G$ and $H^{20}$ approach zero,  the iteration converges with   $\ket{\Phi^{(n+1)}}\to \ket{\Phi^{(n)}}$.  

It is worth pointing out that the speed of convergence is sensitive to the choice of the step size, $\eta$. If $\eta$ is too large, one can wildly oscillate around energy minima. If $\eta$ is too small, then the minimization procedure becomes extremely slow. The proper size is determined by the inverse of the Hessian matrix, which is defined as the second derivative of the energy expectation value. Since the calculation of the Hessian matrix is usually time-consuming, the step size $\eta$ is thus chosen as an empirical value around $10^{-3}$. Meanwhile, one can improve the iteration by adding a momentum term, which can suppress oscillations in high-curvature directions. This method is also called the heavy-ball algorithm~\cite{Ryssens:2019,Bally:2021EPJA}. The momentum term becomes especially important in situations where the landscape is shallow and flat in some directions and narrow and steep in others. Moreover, one can also implement the conjugate gradient descent (GD) method. This method has been found to reduce the number of iterations by a factor of three to four~\cite{Egido:1995NPA}.

In order to generate a set of mean-field wave functions $\ket{\Phi(\mathbf{q})}$ with correct average nucleon numbers $N_\tau$ for neutrons and protons, and different intrinsic deformations $\beta_{\lambda\mu}$, such as quadrupole and octupole deformations, one needs to add the following constraints to the energy expectation value (\ref{eq:EDF}) in the minimization of energy:
\begin{equation}
\label{eq:HFB_constraints}
\begin{aligned}
\bra{\Phi(\mathbf{q})} \hat N_\tau \ket{\Phi(\mathbf{q})} &= N_\tau, \
\bra{\Phi(\mathbf{q})} \hat Q_{\lambda\mu} \ket{\Phi(\mathbf{q})} &= q_{\lambda\mu},
\end{aligned}
\end{equation}
where $\lambda=1, 2, 3$ and $\mu=0$ (axial symmetry) with the constraint $q_{10}=0$ to ensure the location of the center-of-mass coordinate at the origin. The total Routhian of the nucleus to be minimized becomes,
\begin{equation}
\tilde{E}[\Phi] = \bra{\Phi(\mathbf{q})} \hat H - \sum_{\tau=n,p}\lambda_\tau \hat N_\tau\ket{\Phi(\mathbf{q})} + \frac{1}{2}\sum_{\lambda=1,2,3} C_\lambda \left(\bra{\Phi(\mathbf{q})} \hat Q_{\lambda 0} \ket{\Phi(\mathbf{q})} - q_{\lambda 0}\right)^2,
\end{equation}
 where the Lagrange multipliers $\lambda_\tau$ is called chemical potential introduced in order to fix the average particle number, $C_\lambda$ for deformations.
The variational problem with the constraints (\ref{eq:HFB_constraints}) can be solved similarly with the gradient descent (GD) method, provided that the minimization is carried out within the hypersurface given by the constraints. For more details, see, for instance, Refs.~\cite{Ring:1980,Egido:1980,Bally:2021EPJA}.

Alternatively, the HFB wave function can be solved with the iterative diagonalization method~\cite{Ring:1980}, in which the following eigenvalue problem is solved,
 \beq
 \label{eq:HFB_equation}
 \sum_q \begin{pmatrix}
   \tilde{h}   & \Delta\\
   -\Delta^\ast & -\tilde{h}^\ast 
    \end{pmatrix}_{pq}
    \begin{pmatrix}
      U_{q k}\\
      V_{q k}
 \end{pmatrix}
    =  E_{k} 
    \begin{pmatrix}
     U_{pk}\\
      V_{pk}
 \end{pmatrix},
 \eeq
 where $q$ runs over all single-particle basis, and $E_k$ is the so-called quasiparticle energy. The eigenvectors $(U_{qk}, V_{qk})$ are the coefficients of the Bogoliubov transformation. The matrix element of the Routhian $\tilde{h}$ for single-particle motions reads, 
 \beq 
\tilde{h}^{(\tau)}_{pq}[\rho]=h_{pq}[\rho] - \lambda_\tau \delta_{pq} -  \sum_{\lambda=1,2,3} C_\lambda \Bigg( {\rm Tr}[\rho Q_{\lambda 0}]-q_{\lambda 0}\Bigg) (Q_{\lambda 0})_{pq},
 \eeq
 The matrix element of single-particle hamiltonian reads
 \beq 
 h_{pq}[\rho]=t_{pq}+\sum_{rs} V^{pr}_{qs}\rho^r_s,
 \eeq
 where $V^{pr}_{qs}\equiv\bra{pr}V(1, 2)\ket{qs}$ is the matrix element of two-body interaction. In Eq.(\ref{eq:HFB_equation}), the matrix element of particle-particle (pairing) field is given by $\Delta^{pq} = \dfrac{1}{2}\sum_{rs} V^{pq}_{rs}\kappa_{rs}$. The one-body {\em normal} and {\em abnormal}  density matrices  $\rho$ and $\kappa$ (also known as pairing tensor) of the HFB state $\ket{\Phi(\bm{q})}$ are defined as
\bsub\beqn
\rho^p_q(\bm{q};\bm{q}) &\equiv& \langle\Phi(\bm{q}) \vert c^\dagger_p c_q\vert\Phi(\bm{q})\rangle=(V^\ast V^T)_{qp},\\
\kappa^{pq}(\bm{q};\bm{q}) &\equiv& \langle\Phi(\bm{q}) \vert c^\dagger_p c^\dagger_q\vert\Phi(\bm{q})\rangle^\ast=(U^\ast V^T)_{pq}^\ast,\\
\kappa_{rs}(\bm{q};\bm{q}) &\equiv&  \langle \Phi(\bm{q})\vert c_sc_r\vert\Phi(\bm{q})\rangle=(V^\ast U^T)_{rs}.
\eeqn
\esub 
The solution to (\ref{eq:HFB_equation}) defines the Bogoliubov transformation matrix ${\cal W}$ which diagonalizes the following matrix,
\beqn
 \begin{pmatrix}
    H^{11} & H^{20} \\
    -H^{20\ast} & -H^{11\ast} 
\end{pmatrix} 
= 
 \begin{pmatrix}
   U & V^\ast \\
    V & U^\ast 
\end{pmatrix}^\dagger
 \begin{pmatrix}
   \tilde{h}   & \Delta\\
   -\Delta^\ast & -\tilde{h}^\ast 
    \end{pmatrix}  
 \begin{pmatrix}
   U & V^\ast \\
    V & U^\ast 
\end{pmatrix} 
= \begin{pmatrix}
   E_k & 0\\
   0 & -E_k
\end{pmatrix}. 
\eeqn

The convergence properties of the iterative diagonalization method depend sensitively on the types of constraints.  In many cases one cannot find convergence at all, in particular in regions of level crossings, where the iteration switches back and forth between different occupations. In contrast, the GD method  does not show this kind of numerical instability~\cite{Egido:1980}.

\subsubsection{The HFB state in canonical basis or BCS state}

According to the Bloch-Messiah theorem \cite{Block:1962NP}, the Bogoliubov transformation matrix ${\cal W}(\bm{q})$ in (\ref{eq:Bogoliubov_Ttransformation})  can always be decomposed into the product of three special matrices~\cite{Ring:1980}
\beqn
\label{eq:Bloch-Messiah}
{\cal W}
=\left(\begin{array}{cc}
D & 0 \\
0 & D^{*}
\end{array}\right)\left(\begin{array}{cc}
\bar{U} & \bar{V} \\
\bar{V} & \bar{U}
\end{array}\right)\left(\begin{array}{cc}
C & 0 \\
0 & C^{*}
\end{array}\right).
\eeqn
In comparison with Eq.(\ref{eq:Bogoliubov_Ttransformation}), one finds the relations $U=D\bar U C$ and $V=D^\ast\bar V C$, where $D$ and $C$ are $M\times M$ unitary matrices that transform particle operators and quasiparticle operators, respectively.
The matrix $D_{pl}$ transforms from the original basis $p$ to the so-called canonical basis consisting of pairs ($l,\bar{l}$) of canonically conjugate states. The second transformation given by the real matrices $\bar U$ and $\bar V$
     \beqn
     \label{eq:uv-coefficients}
      \bar U_{ll'}=\left(\begin{array}{cc}
      u_l & 0 \\
      0 & u_l  \\
     \end{array}\right)\delta_{ll'},\quad
      \bar V_{ll'}=\left(\begin{array}{cc}
      0 & v_l \\
      -v_l & 0  \\
     \end{array}\right)\delta_{l\bar l'},
     \eeqn
 is the Bogoliubov-Valatin transformation connecting the quasiparticle operators ($\alpha^\dagger_l, \alpha^\dagger_{\bar l}$) in canonical basis with the single-particle operators $(a^\dagger_l, a_l)$ in HF basis,
 \bsub\begin{align}
 \alpha^\dag_l&= u_l a^\dag_l - v_l a_{\bar{l}}, \\
 \alpha^\dag_{\bar{l}}&=u_l a^\dag_{\bar{l}} + v_l a_l,
 \label{eq:Bogoliubov-Valatin}
 \end{align}
 \esub
 where the index $l=1, 2, \ldots, M/2$ and $u_{l}^{2}+v_{l}^{2}=1$. The $v^2_l \in[0,1]$ is interpreted as the occupation probability of the $l$-th single-particle state in canonical basis, in which the one-body density $\rho$ is diagonal. The third matrix $C$ in Eq. (\ref{eq:Bloch-Messiah}) describes a transformation in quasiparticle space from the operators $\alpha$ to the operators $\beta$ in Eq. (\ref{eq:Bogoliubov_Ttransformation}).

 In the canonical basis or BCS wave function, the pairing field matrix element $\Delta_{k\bar l}=\Delta_k\delta_{k l}$ becomes diagonal, and the occupation probabilities $v^2_k$ of neutron $(\tau=n)$ or proton $(\tau=p)$  states are determined by the  
 \beq
 \label{eq:occupation_v2}
  v^2_k = \frac{1}{2}\Bigg(1-\frac{\epsilon_k-\lambda_\tau}{\sqrt{(\epsilon_k-\lambda_\tau)^2+\Delta^2_k}}\Bigg),
 \eeq
 where $\epsilon_k$ is the single-particle energy from the HF solution, the Fermi energy $\lambda_\tau$ determined by the condition $2\sum^\tau_{k>0} v^2_k=N_\tau$, and the pairing gap $\Delta_l$ by the employed pairing force~\cite{Ring:1980},
 \beq
 \Delta_k 
 = -\frac{1}{2}\sum_{l}V^{k\bar k}_{l \bar l}\frac{\Delta_{l}}{\sqrt{(\epsilon_l-\lambda_\tau)^2+\Delta^2_l}}.
 \eeq

\subsubsection{The HFB state with odd-number parity}

In the HFB approaches, the lowest-energy state of an odd-mass nucleus is described as a one-quasiparticle state, 
\beq
\ket{\Phi_{k_0}(\mathbf{q})} = \beta^\dagger_{k_0} \ket{\Phi(\mathbf{q})}.
\eeq
One can prove that this state is the vacuum of another type of quasiparticle operators $\gamma_{j=1,2,\cdots, M}$ defined as,  
\beqn
\gamma_j 
=\left\{\begin{array}{cc}
    \beta^\dagger_j, & \quad j=k_0\\
    \beta_j, & \quad~{\rm others}.
\end{array}
\right.
\eeqn 
and
\beq
\gamma_j \ket{\Phi_{k_0}(\mathbf{q})}=0.
\eeq
Since the quasiparticle operator $\beta^\dagger_{k_0}$ is a linear combination of single-particle creation and annihilation operators, it changes the number parity $(-1)^{N_\tau}$ of the state. Thus the number parities of the wave functions $\ket{\Phi(\mathbf{q})}$ and $\ket{\Phi_{k_0}(\mathbf{q})}$ are different.

In the canonical basis, the wave function of an odd-mass nucleus can be constructed as
\begin{eqnarray} \label{odd-A function}
 \ket{\Phi_{k_0}(\mathbf{q})} = \alpha^\dagger_{k_0} \prod_{k>0}(u_k+v_ka^\dagger_ka^\dagger_{\bar k})\ket{0},   
\end{eqnarray} 
where $u_k$ and $v_k$ are determined in the BCS calculation (\ref{eq:occupation_v2}) and $\alpha^\dagger_{k_0}$ the quasiparticle creation operator (\ref{eq:Bogoliubov-Valatin}) defined in canonical basis. If the same Fermi energy $\lambda_\tau$ is used for the odd-mass nucleus, the mean value of the particle-number operator, i.e., $\bra{\Phi_{k_0}(\mathbf{q})}\hat N\ket{\Phi_{k_0}(\mathbf{q})}=N_\tau+u^2_{k_0}-v^2_{k_0}\neq N_\tau\pm1$. Thus, one needs to readjust the Fermi energy. Here, there are different ways to construct mean-field reference states with odd-number parity corresponding to different levels of approximations. See Ref.~\cite{Zhou:2023_GCM4OA} for details.  

\subsubsection{Effective nuclear Hamiltonians}

The Hamiltonian operator of nuclear systems is usually composed of kinetic energy, two-body nuclear forces, and even three-body nuclear forces. In most applications, the effect of the three-body nuclear force is considered using a density-dependent two-body force. Particularly, phenomenological Hamiltonians or EDFs with parameters fitted to the properties of nuclear many-body systems are employed. These include non-relativistic zero-range Skyrme forces, finite-range Gogny forces, and relativistic effective Lagrangian densities~\cite{Bender:2003RMP}. 

In the relativistic framework of \ac{CDFT}, also known as the \ac{RMF} approach, nucleons interact via the exchange of different effective meson fields or through different types of coupling vertices~\cite{Vretenar:2005PR, Meng:2005PPNP, Meng:2017}. These approaches are commonly referred to as the meson-exchange model and point-coupling model, respectively. In both cases, an effective nuclear Hamiltonian is derived from the Lagrangian density. When applying the mean-field approximation, one obtains an Energy Density Functional (EDF) that depends on various four-component currents and their derivatives.
Within this theoretical framework, a reflection-asymmetric CDFT was developed by expanding nucleons and mesons on the eigenfunctions of the two-center harmonic-oscillator potential~\cite{Geng:2007CPL, Li:2023}. This approach has been applied to investigate the shape evolution from spherical to octupole-deformed shapes in isotopes of Sm~\cite{Zhang:2010PRC}, Th~\cite{Guo:2010PRC}, and Ra~\cite{Yu:2012}. Simultaneously, CDFT was developed using the standard (single-center) harmonic-oscillator basis for superheavy nuclei, incorporating axial octupole shape degrees of freedom~\cite{Abusara:2012}, as well as accommodating general shapes~\cite{Lu:2012PRC}.

It is worth mentioning that one may encounter singularity and self-interaction problems in the \ac{EDF}-based \ac{BMF} approaches~\cite{Bender:2009, Duguet:2009}. Many efforts have been devoted to finding an energy functional free of these problems. These issues do not exist in Hamiltonian-based approaches. Remarkable progress has been achieved in the development of nuclear {\em ab initio} methods in the past decade, starting from Hamiltonians derived from chiral (\ac{EFT})\cite{Weinberg:1991}. The obtained \ac{HFB} state serves as a reference state for {\em ab initio} many-body approaches\cite{Hergert:2020}. With the help of the (\ac{SRG})\cite{Bogner:2010} and in-medium SRG\cite{Hergert:2016jk}, one is able to derive effective nuclear interactions suitable for \ac{GCM} calculations~\cite{Yao:2020PRL, Frosini:2022EPJA3}. The implementation of symmetry-restoration methods into {\em ab initio} methods, starting from a realistic nuclear force, has become very attractive. Notably, the aforementioned singularity problems are not present in these {\em ab initio} frameworks.

 \subsection{Symmetry restoration with projections}

 Since symmetries are allowed to be violated during the variational process in the \ac{HFB} approach, particle number, translational, rotational, and parity symmetries are all broken in the octupole-deformed HFB states $\ket{\Phi(\mathbf{q})}$. The wave function of a state with quantum-number projections is constructed as
\begin{equation}
\label{eq:basis_func}
\ket{NZ JK \pi; \mathbf{q}} = \hat P^J_{MK} \hat P^{N}\hat P^Z \hat P^\pi \ket{\Phi(\mathbf{q})}.
\end{equation}
Here, the generator coordinate $\mathbf{q}$ stands for the discretized deformation parameters (${\beta_{2\mu}, \beta_{3\mu}}$) of the reference states $\ket{\Phi(\mathbf{q})}$ obtained from deformation-constrained self-consistent mean-field calculations using HF+BCS, HFB, relativistic \ac{RHB} or \ac{CDFT}. The $\hat P^{J}_{MK}$, $\hat{P}^{N, Z}$, and $\hat P^\pi$ are projection operators that extract components with the right angular momentum $J$, neutron number $N$, proton number $Z$, and parity $\pi=\pm$~\cite{Ring:1980}. The projection operators are defined as follows
\begin{subequations}
\begin{align}
\label{eq:Euler_angles}
\hat P^{J}_{MK} &= \frac{2J+1}{8\pi^2}\int d\Omega D^{J\ast}_{MK}(\Omega) \hat R(\Omega),\\
\hat P^{N_\tau} &= \frac{1}{2\pi}\int^{2\pi}_0 d\varphi_{\tau} e^{i\phi_{\tau}(\hat N_\tau-N_\tau)},\\
\hat P^\pi &= \frac{1}{2}(1+\pi\hat P),
\end{align}
\end{subequations}
where the operator $\hat P^J_{MK}$ extracts the component of angular momentum along the intrinsic axis $z$ given by $K$. The Wigner-D function is defined as $D^{J}_{MK}(\Omega)\equiv\bra{JM}e^{i\varphi\hat J_z}e^{i\theta\hat J_y}e^{i\psi\hat J_z}\ket{JK}$ with $\Omega=(\varphi, \theta, \psi)$ representing the three Euler angles.

The energy of the projected state $\ket{\Phi(\mathbf{q})}$ is given by the ratio of the diagonal element of the Hamiltonian to  that of norm kernels,
\beq 
\label{eq:projected_energy}
E^{NZJK \pi}(\mathbf{q})
=\frac{\mathscr{H}^{J\pi}_{\kappa, \kappa}}{\mathscr{N}^{J\pi}_{\kappa, \kappa}},
\eeq
where the hamiltonian kernel $\mathscr{H}^{J\pi}_{\kappa_a, \kappa_b}$ and norm kernel
$\mathscr{N}^{J\pi}_{\kappa_a, \kappa_b}$ are given by,
\begin{equation}
\label{eq:kernel}
 \mathscr{O}^{J\pi}_{\kappa_a, \kappa_b}
 =\bra{NZ JK_a \pi; \mathbf{q}_a} \hat O \ket{NZ JK_b \pi; \mathbf{q}_b}
 =\bra{\Phi(\mathbf{q}_a)} \hat O \hat P^J_{K_aK_b} \hat P^N\hat P^Z  \hat P^\pi \ket{\Phi(\mathbf{q}_b)},
\end{equation}
with the operator $\hat O$ representing $\hat H$ and $1$, respectively. The matrix elements of a Hamiltonian operator $\hat H$ can be evaluated using the generalized Wick’s theorem~\cite{Balian:1969} 
\begin{eqnarray}
\label{eq:Hkernel}
 \mathscr{H}^{J\pi}_{\kappa_a, \kappa_b} 
 &=&\frac{2J+1}{8\pi^2}\int d\Omega D^J_{K_aK_b}(\Omega) \frac{1}{2\pi}\int^{2\pi}_0 e^{i\varphi_nN_0} \frac{1}{2\pi}\int^{2\pi}_0  e^{i\varphi_pZ_0}\nonumber\\
 &&\times\frac{1}{2}\Bigg[ \bra{\Phi(\mathbf{q}_a)} \hat H \hat R(\Omega) e^{-i\varphi_n\hat N} e^{-i\varphi_n\hat Z}    \ket{\Phi(\mathbf{q}_b)}\nonumber\\
 &&
 + \pi\bra{\Phi(\mathbf{q}_a)} \hat H \hat R(\Omega) e^{-i\varphi_n\hat N} e^{-i\varphi_n\hat Z} \hat P  \ket{\Phi(\mathbf{q}_b)}\Bigg].
\end{eqnarray}
In contrast to the Hamiltonian-based GCM, the  hamiltonian kernel $\mathscr{H}^{J\pi}_{\kappa_a, \kappa_b}$ in the EDF-based GCM calculations is usually evaluated with the {\em mixed density} prescription, i.e.,  the hamiltonian overlap takes the same functional form of mean-field EDF, provided that all the densities $\rho$ are replaced with the mixed densities $\tilde\rho$ and $\tilde\rho^p$, 
\bsub\begin{eqnarray}
  \frac{\bra{\Phi(\mathbf{q}_a)} \hat H \hat R(\Omega) e^{-i\varphi_n\hat N} e^{-i\varphi_p\hat Z}    \ket{\Phi(\mathbf{q}_b)}}{ \bra{\Phi(\mathbf{q}_a)}   \hat R(\Omega) e^{-i\varphi_n\hat N} e^{-i\varphi_p\hat Z}    \ket{\Phi(\mathbf{q}_b)}}
&=&E[\tilde \rho, \nabla^2\tilde\rho, \cdots],\\
  \frac{\bra{\Phi(\mathbf{q}_a)} \hat H \hat R(\Omega) e^{-i\varphi_n\hat N} e^{-i\varphi_p\hat Z}     \hat P\ket{\Phi(\mathbf{q}_b)}}{ \bra{\Phi(\mathbf{q}_a)}   \hat R(\Omega) e^{-i\varphi_n\hat N} e^{-i\varphi_p\hat Z}  \hat P  \ket{\Phi(\mathbf{q}_b)}}
&=&E[\tilde \rho^p, \nabla^2\tilde\rho^p, \cdots].
\end{eqnarray}
\esub
The matrix elements of mixed (transition) densities are defined as
\bsub\beqn
\tilde \rho_{ij}(g)&=&
\frac{\bra{\Phi(\mathbf{q}_a)}  c^\dagger_j c_i \hat R(\Omega) e^{-i\varphi_n\hat N} e^{-i\varphi_p\hat Z}    \ket{\Phi(\mathbf{q}_b)}}{\bra{\Phi(\mathbf{q}_a)}   \hat R(\Omega) e^{-i\varphi_n\hat N} e^{-i\varphi_p\hat Z}    \ket{\Phi(\mathbf{q}_b)}}=[V^\ast (\mathbf{q}_b, g)\mathbb{U}^{-1}V^T(\mathbf{q}_a)]_{ij},\\
\tilde \rho^p_{ij}(g)&=&
\frac{\bra{\Phi(\mathbf{q}_a)}  c^\dagger_j c_i \hat R(\Omega) e^{-i\varphi_n\hat N} e^{-i\varphi_p\hat Z}    \hat P\ket{\Phi(\mathbf{q}_b)}}{\bra{\Phi(\mathbf{q}_a)}   \hat R(\Omega) e^{-i\varphi_n\hat N} e^{-i\varphi_p\hat Z}  \hat P  \ket{\Phi(\mathbf{q}_b)}}=[V^{(p)\ast} (\mathbf{q}_b, g)\mathbb{U}^{(p)-1}V^T(\mathbf{q}_a)]_{ij}, \nonumber\\
\eeqn
\esub
with the symbol $g$ standing for $(\Omega, \varphi_n, \varphi_p)$, and $V^{(p)}_{qk}=(-1)^{\ell_q}V_{qk}$, where $q$ is for the spherical \ac{HO} basis $\ket{q}=\ket{n\ell j m_j}_q$. The matrix $\mathbb{U}$ is defined as~\cite{Ring:1980,Yao:2022PPNP}
\bsub\beqn 
\mathbb{U} &=& U^\dagger(\mathbf{q}_b, g) U(\mathbf{q}_a)
+V^\dagger(\mathbf{q}_b, g) V(\mathbf{q}_a),\\
\mathbb{U}^{(p)} &=& U^{(p)\dagger}(\mathbf{q}_b, g) U(\mathbf{q}_a)
+V^{(p)\dagger}(\mathbf{q}_b, g) V(\mathbf{q}_a),
\eeqn
\esub
with $U(\boldsymbol{q}, g)$  and  $V(\boldsymbol{q}, g)$ defined by,
\beq
U(\boldsymbol{q}, g)=D(g) U(\boldsymbol{q}), \quad V(\boldsymbol{q}, g)=D^{\ast}(g) V(\boldsymbol{q}),
\eeq
where $D(g)$ is the product of the representation matrices of the rotation operators $\hat R(\Omega) e^{-i\varphi_n\hat N} e^{-i\varphi_p\hat Z}$ in the \ac{HO} basis.

\subsection{Configuration mixing with \ac{GCM}}
In the \ac{SPGCM}, the wave function $\vert \Psi^{J\pi}_\alpha\rangle$ of the $\alpha$-th state with spin-parity $J^\pi$ is constructed as a linear combination of the projected states (\ref{eq:basis_func})~\cite{Yao:2022HandBook} 
\begin{equation}
\label{eq:gcmwf}
\vert \Psi^{J\pi}_\alpha\rangle
=\sum_{\kappa\in\{\mathbf{q}, K\}} f^{J\pi\alpha}_\kappa  \ket{NZ JK \pi; \mathbf{q}},
\end{equation} 
where the weight function $f^{J\pi\alpha}_\kappa$ is determined by the Hill-Wheeler-Griffin (HWG) equation~\cite{Hill:1953, Ring:1980},
\begin{equation}
\label{eq:HWG}
\sum_{\kappa_b} \left[  \mathscr{H}^{J\pi}_{\kappa_a, \kappa_b}-E^{J\pi}_{\alpha} \mathscr{N}^{J\pi}_{\kappa_a, \kappa_b}\right]f^{J\pi\alpha}_{\kappa_b}=0
\end{equation}
where the hamiltonian kernel $\mathscr{H}^{J\pi}_{\kappa_a, \kappa_b}$ and norm kernel
$\mathscr{N}^{J\pi}_{\kappa_a, \kappa_b}$ have been defined in (\ref{eq:kernel}). 

 The solution of the HWG equation (\ref{eq:HWG}) provides energy $E^{J\pi}_{\alpha}$ and weight function $f^{J\pi \alpha}_{\kappa}$ of the nuclear state.  For even-even octupole-deformed nuclei built on the eigenstates of the simplex operator $\hat S=\hat P\hat R_y(\pi)$, the projected states with angular momentum $J$ must necessarily have the natural parity $\pi=(-1)^J$. Therefore, for the states of the $K=0$ band, only quantum numbers of angular momentum and parity such as $0^+, 1^-, 2^+, 3^-$, etc., are possible.

 Since the basis functions $\ket{NZ JK \pi; \mathbf{q}}$ are nonorthogonal to each other, one usually defines the collective wave function $g^{J\pi}_\alpha(K, \mathbf{q})$ as below
\begin{equation}
\label{eq:coll_wf}
g^{J\pi}_\alpha(K, \mathbf{q})=\sum_{\kappa'} (\mathscr{N}^{1/2})^{J\pi}_{\kappa,\kappa'} f^{J\pi\alpha}_{\kappa'},
 \end{equation}
 which fulfills the normalization condition 
 \begin{equation}
\sum_{K, \mathbf{q}} |g^{J\pi}_\alpha(K, \mathbf{q})|^2
=\sum_{\kappa}\sum_{\kappa'} f^{J\pi\alpha\ast}_{\kappa} \mathscr{N}^{J\pi}_{\kappa, \kappa'} f^{J\pi\alpha}_{\kappa'}
=\bra{\Psi^{J\pi}_\alpha}\Psi^{J\pi}_\alpha\rangle = 1.
 \end{equation}
The distribution of $g^{J\pi}_\alpha(K, \mathbf{q})$ over $K$ and $\mathbf{q}$ reflects the contribution of each basis function (\ref{eq:basis_func}) to the nuclear state $\vert \Psi^{J\pi}_\alpha\rangle$.

\subsection{Electromagnetic transition rates}

The probabilities of $\gamma$-ray transitions between nuclear states serve as an excellent probe  to the details  of nuclear structure. By combining experimentally measured half-lives, transition energies, and partial gamma-ray and conversion line intensities (including branching ratios, mixing ratios, and conversion coefficients), it becomes possible to determine the absolute $\gamma$-ray transition probabilities between nuclear states. These data offer a rigorous test of nuclear models
 
With the wave functions of initial state $\ket{\Psi^{J_i\pi_i}_i}$ and final state $\ket{\Psi^{J_f\pi_f}_f}$, one can compute the $\gamma$-ray transition rate of the tensor operator $\hat T_{\lambda\mu}$ between these two states~\cite{Ring:1980} 
\begin{eqnarray}
 T_{i\to f}\left(T\lambda;  J^{\pi_i}_{i} \rightarrow  J^{\pi_f}_{f}\right)
 &=&8 \pi c \alpha\frac{(\lambda+1)}{\lambda [(2 \lambda+1)!!]^2}\left(\frac{E_\gamma}{\hbar c}\right)^{2\lambda+1} B\left(T\lambda;  J^{\pi_i}_{i} \rightarrow  J^{\pi_f}_{f}\right),
\end{eqnarray}
where $E_\gamma$ (in MeV) is the energy difference of the two states. Substituting the following constants,
\begin{equation}
  c=3.0\times 10^{23}~{\rm fm/s},\quad \hbar c= 197.327~{\rm fm\cdot MeV},\quad  \alpha=\frac{e^2}{4\pi \hbar c}=1/137.035,
\end{equation}
one finds the transition rate (in $s^{-1}$)
\begin{eqnarray}
T_{i\to f}\left(E\lambda;  J^{\pi_i}_{i} \rightarrow  J^{\pi_f}_{f}\right)
=\alpha_\lambda E^{2\lambda+1}_\gamma B\left(E\lambda;  J^{\pi_i}_{i} \rightarrow  J^{\pi_f}_{f}\right)
\end{eqnarray}
where the constant $\alpha_\lambda$ for $\lambda=1, 2, 3$ reads
\begin{eqnarray}
\alpha_1 \simeq 1.59 \times 10^{15},\quad \alpha_2 \simeq 1.23 \times 10^{9},\quad\alpha_3 \simeq 5.71 \times 10^{2},
\end{eqnarray}
and the reduced transition  strength $B(E\lambda)$ (in $e^2 {\rm fm}^{2\lambda}$) reads
 \begin{eqnarray}
 B\left(E\lambda;  J^{\pi_i}_{i} \rightarrow  J^{\pi_f}_{f}\right)
 =\frac{1}{2J_i+1} \Bigg|M(E\lambda; J^{\pi_i}_{i} \rightarrow  J^{\pi_f}_{f})\Bigg|^2.
 \end{eqnarray}
The reduced transition matrix element $M(E\lambda)$ is given by
  \begin{eqnarray}
 M(E\lambda; J^{\pi_i}_{\alpha_i} \rightarrow  J^{\pi_f}_{\alpha_f}) 
 &=& \sum_{\mathbf{q}_i\mathbf{q}_f}  f^{J_f\pi_f\alpha_f}_{\kappa_f}  f^{J_i\pi_i\alpha_i}_{\kappa_i}\nonumber\\
 &&\times \bra{NZ J_fK_f \pi_f; \mathbf{q}_f}\vert \hat T_{\lambda}\vert\ket{NZ J_iK_i \pi_i; \mathbf{q}_i}, 
 \end{eqnarray}
 where the deformation-dependent reduced matrix element is determined by
 \begin{eqnarray}
 \label{eq:RME}
&& \bra{NZ J_fK_f \pi_f; \mathbf{q}_f}\vert \hat T_{\lambda}\vert\ket{NZ J_iK_i \pi_i; \mathbf{q}_i}\nonumber\\
&=&\delta_{1,\pi_{f} \pi_\lambda\pi_{i}} (2J_f+1) (-1)^{J_{f}-K_{f}} \sum_{M^\prime M^{\prime \prime}}
\left(\begin{array}{c c c}
J_{f} & \lambda & J_{i} \\
-K_{f} & M^{\prime} & M^{\prime \prime}
\end{array}\right)  \nonumber\\
&&\times\bra{\Phi(\mathbf{q}_f)}\hat{T}_{\lambda M^{\prime}} \hat{P}_{M^{\prime\prime} K_{i}}^{J_{i}}\hat{P}^{Z}\hat{P}^{N} \hat{P}^{\pi}\ket{\Phi(\mathbf{q}_i)}.
 \end{eqnarray}
 The nonzero value of the transition strength requires that the spin-parity quantum numbers of the initial and final states fulfill the following conditions, 
 \begin{equation}
    \pi_{f} \pi_\lambda\pi_{i} = 1,\quad |J_f - J_i|\leq \lambda \leq J_f + J_i,
 \end{equation}
 where the parity $\pi_\lambda=(-1)^\lambda$ for electric multipole transitions $\hat{T}_{\lambda}=r^\lambda Y_{\lambda}$, and $\pi_\lambda=(-1)^{\lambda+1}$ for magnetic multipole  transitions $\hat{T}_{\lambda}=(g_s\mathbf{S} + \frac{2}{\lambda+1}g_\ell \mathbf{L})\cdot (\nabla r^\lambda Y_{\lambda})$ in units of nuclear magneton ($\mu_N$).  We note that free-space values of electric charges and $g_{s/\ell}$ factors are usually used in the calculations of full single-particle space. If a valence-space Hamiltonian is employed, effective values are  used.

 In the case that the intrinsic states $\ket{\Phi(\mathbf{q})}$ are restricted to have axial symmetry, the integration  in (\ref{eq:Euler_angles}) over the two Euler angles $(\varphi, \psi)$ can be carried out analytically. The reduced matrix element between the states with different deformations and $K_f=K_i=K$ is simplified as follows,
 \begin{eqnarray}
&&\bra{NZ J_fK \pi_f; \mathbf{q}_f}\vert \hat T_{\lambda}\vert\ket{NZ J_iK \pi_i; \mathbf{q}_i} \nonumber\\
&=& (-1)^{J_{f}-K} \delta_{1,\pi_{f} \pi_\lambda\pi_{i}}  
\frac{(2J_i+1)(2J_f+1)}
{2 } \sum_{\nu M}
\left(\begin{array}{c c c}
J_{f} & \lambda & J_{i} \\
-K & \nu & M
\end{array}\right)\nonumber\\
&&\times\int^{\pi}_0 \sin\theta d\theta
d^{J_i}_{MK}(\theta)\bra{\Phi(\mathbf{q}_f)}\hat{T}_{\lambda \nu} e^{i\theta \hat{J_y}}
\hat{P}^{Z}\hat{P}^{N}\hat{P}^{\pi_i}\ket{\Phi(\mathbf{q}_i)}.
 \end{eqnarray}

\section{Nuclear collective motions with octupole shapes}
\label{sec:collective_motions}


\subsection{Reflection-asymmetric cluster structures in light nuclei} 

Cluster structure appears in the excited states of many light nuclei, particularly in those of self-conjugate $4n$ nuclei close to the particle emission threshold, as illustrated by the Ikeda diagram~\cite{Ikeda:1968}. Advances in nuclear theory have made it possible to study nucleon clustering phenomena in atomic nuclei, not only from the degrees of freedom of different clusters but also from the individual nucleons, with a full account of the Pauli exclusion principle. Cluster structures emerge automatically from the $A$-body wave function starting from realistic or effective nucleon-nucleon interactions~\cite{Freer:2018}. 

Self-consistent mean-field approaches, with all nucleons treated on the same footing, provide a good tool for understanding these phenomena. Previous studies in this context have shown that nucleons tend to form cluster structures in nuclear states at high spin~\cite{Ichikawa:2011}, isospin~\cite{Zhao:2015PRL}, large deformation~\cite{Arumugam:2005,Ebran:2013,Yao:2014_O16,Matsumoto:2022}, deep confining nuclear potential~\cite{Ebran:2012}, and with low density~\cite{Girod:2013}. The GCM provides a \ac{BMF} study of cluster structures in excited states of atomic nuclei by mixing mean-field reference states with both shell-model-like structure and cluster structure, with their weights determined by variational principles. As it will be illustrated below,  the inclusion of reflection asymmetric shapes in the reference states is necessary for the reproduction of clustering structure in the light nuclei of concerned, including the Hoyle state of $^{12}$C and molecular-like structure in \nuclide[20]{Ne}.  However, this breaks rotational symmetry and space-reversal symmetry of the nuclear Hamiltonian. These symmetries need to be restored to describe cluster structures in nuclear excited states characterized by the quantum numbers of spin and parity.   

$^{12}$C is a well-known example of a nucleus with nucleon clustering structures. Its excited states are predominantly characterized by configurations involving three $\alpha$ particles, which can arrange themselves into various geometric shapes, such as triangular, linear, or others. Of particular importance is the second $0^+$ state, commonly referred to as the Hoyle state, which plays a crucial role in nucleosynthesis. Researchers have extensively studied its cluster structure using various nuclear models, including ab initio methods~\cite{Epelbaum:2011, Epelbaum:2012, Otsuka:2022, Shen:2023}. In this context, our primary focus will be on investigations conducted within the SPGCM framework.

The nucleon clustering structure in $^{12}$C has been extensively studied with the models of \ac{AMD}\cite{Kanada:1998, Kanada:2007} and \ac{FMD}\cite{Feldmeier:2000}. In both of these models, wave functions of configurations are constructed as Slater determinants of Gaussian wave packets. In the AMD model, parameters such as the center of the Gaussian wave packet, the width parameter, and the orientation of intrinsic spin for each nucleon are treated as variational parameters determined by the principles of variation. A study using this model with an effective nuclear interaction, including a zero-range three-body force and two-body interaction, has revealed that the $^{12}$C ground state contains both sub-shell closure and $3\alpha$ cluster components. Excited states in $^{12}$C exhibit various $3\alpha$ cluster structures. Additionally, the study highlights the importance of the $\alpha$-cluster breaking component in the excited states~\cite{Kanada:2007}. In contrast, the FMD model treats the width parameter as a complex variational parameter, allowing it to vary for each single-particle state~\cite{Neff:2004NPA}. An effective interaction in the FMD model is derived from realistic Bonn or Argonne nucleon-nucleon interactions using the unitary correlation operator method. To simulate the effects of three-body correlations and genuine three-body forces, an additional two-body force is introduced, which includes momentum-dependence and spin-orbit forces with parameters adjusted to match binding energies and radii of finite nuclei. The configurations in the FMD model are generated through VAP calculations with constraints on radii and quadrupole deformation. Notably, the FMD model predicts that the Hoyle state in $^{12}$C is a dilute self-bound gas of bosonic $\alpha$ particles, akin to a Bose-Einstein condensate~\cite{Chernykh:2007PRL}. It is also found that the spin-orbit force tends to break the $\alpha$ clusters, leading to a significant shell-model component in the FMD ground state. However, in the Hoyle state, $\alpha$ clustering remains dominant, albeit with a sizable component of shell-model nature.

In the past decade, researchers have also explored cluster structures in the low-lying states of $^{12}$C using EDF-based GCM frameworks. For instance, Fukuoka et al. conducted configuration-mixing calculations of mean-field states with projections onto different parities and angular momenta~\cite{Fukuoka:2013}. In their approach, mean-field states were generated from Skyrme \ac{HF} calculations, represented in three-dimensional coordinate-space allowing for various shapes. Initially, single-particle wave functions were prepared as Gaussian wave packets with random centers, evolving using the imaginary-time method. As the configurations evolved, various cluster structures emerged. To perform GCM calculations, configurations were selected based on a specific criterion: if the overlap between a given wave function and all selected configurations fell below a predefined cutoff value, that configuration was included in the GCM subspace. This approach allowed for a comprehensive exploration of cluster structures in $^{12}$C. A similar concept has been applied in recent GCM studies, such as those related to neutrinoless double-beta decay~\cite{Romero:2021PRC,Zhang:2023_ML}.

\begin{figure}[tb]
\centerline{\includegraphics[width=\textwidth]{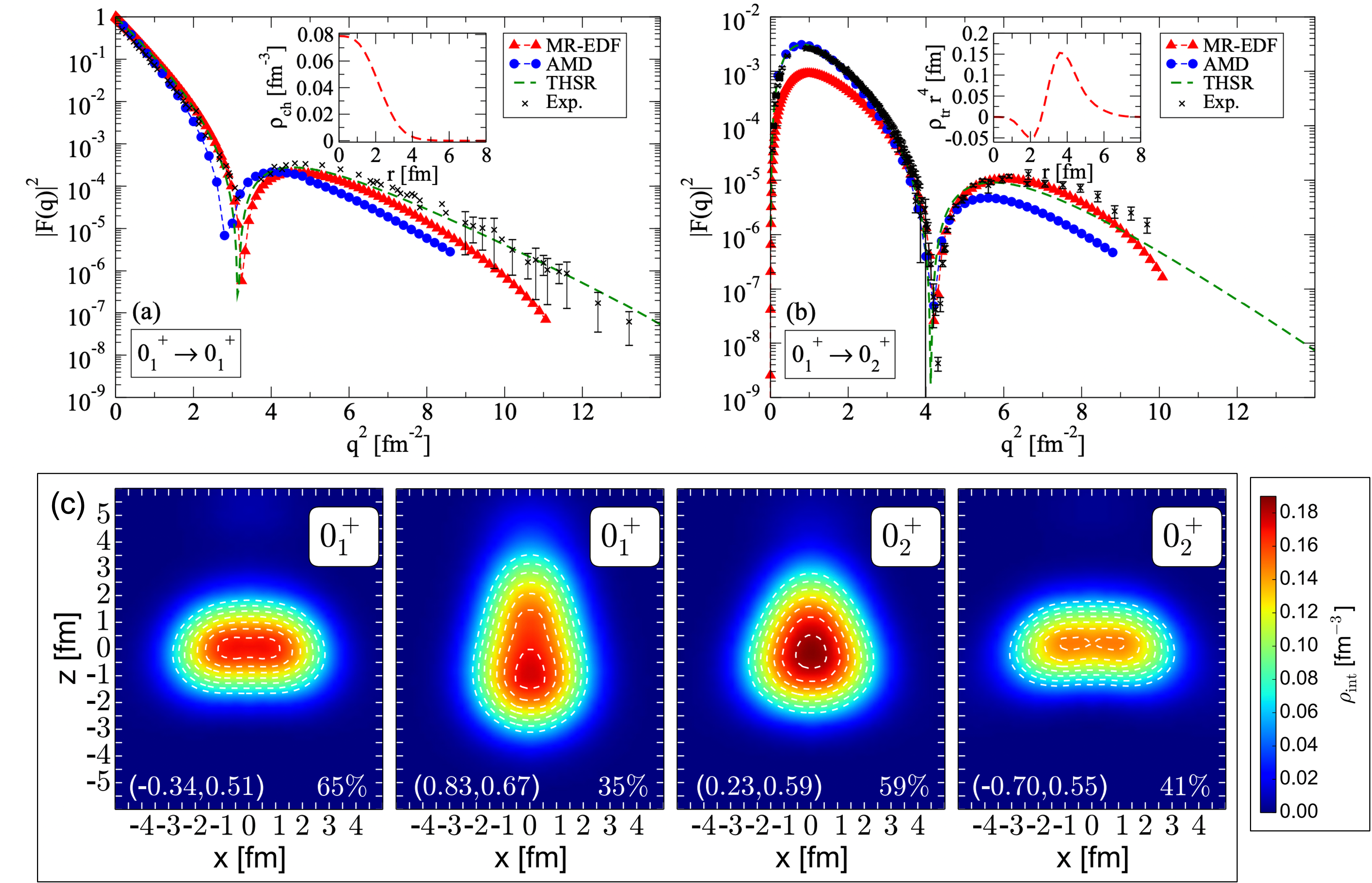}}
\caption{The form factors of electron scatterings on $^{12}$C for the (a) elastic $0^+_1\to 0^+_1$ and (b) inelastic $0^+_1\to 0^+_2$ processes from different calculations, in comparison with data. (c) The nucleon densities of the predominant configurations in the first two $0^+$ states. Figure adapted from Ref.~\cite{Marevi:2019C12}. Figure reprinted with permission from the American Physical Society.}
\label{fig:Marevi2019_density}
\end{figure}

In recent years, extensive investigations into the low-lying states of $^{12}$C have been carried out using the  \ac{SPGCM}. These studies involved mixing axial and reflection-asymmetric reference states generated from the \ac{RHB} calculation~\cite{Marevi:2019C12}. The techniques of \ac{AMP},  \ac{PNP}, and \ac{PP} were implemented simultaneously to restore the broken symmetries in the reference states.

One common approach to revealing cluster structures in these states is by computing nucleon density profiles and electron-scattering form factors. For instance, the longitudinal Coulomb form factor $F_{L}(q)$, which describes transitions between initial $\ket{\Psi_{\alpha_{i}}^{J_{i}\pi_{i}}}$ and final $\ket{\Psi_{\alpha_{f}}^{J_{f} \pi_{f}}}$ states can be determined using the plane-wave Born approximation
\beq
F_{L}(q)=\frac{\sqrt{4 \pi}}{Z} \int_{0}^{\infty} d r r^{2} \rho_{J_{i} \alpha_{i}, L}^{J_{f} \alpha_{f}}(r) j_{L}(q r)
\eeq
Here, $q$ represents the momentum transfer for angular momentum $L$, $j_{L}(q r)$ is the spherical Bessel function of the first kind, and $\rho_{J_{i} \alpha_{i}, L}^{J_{f} \alpha_{f}}(r)$ are the reduced transition densities of protons. These reduced transition densities can be expressed as follows~\cite{Yao:2015PRC_scattering}
\beqn
\rho_{J_{i} \alpha_{i}, L}^{J_{f} \alpha_{f}}(r)
&= & (-1)^{J_{f}-J_{i}} \frac{2 J_{f}+1}{2 J_{i}+1} \sum_{K}\left\langle J_{f} 0 L K \mid J_{i} K\right\rangle \nonumber\\
& &\times \int d \hat{\mathbf{r}} \rho_{\alpha_{f} \alpha_{i}}^{J_{f} J_{i} K 0}(\mathbf{r}) Y_{L K}^{*}(\hat{\mathbf{r}}),
\eeqn
where  $\rho_{\alpha_{f} \alpha_{i}}^{J_{f} J_{i} K 0}(\mathbf{r})$  stands for the pseudo-GCM density, which in the axial case is given by~\cite{Yao:2015PRC_scattering}
\beqn
\rho_{\alpha_{f} \alpha_{i}}^{J_{f} J_{i} K 0}(\mathbf{r})
&=&\sum_{\mathbf{q},\mathbf{q}^\prime}
f^{J_f\pi_f\alpha_f}_{\mathbf{q}^\prime}f^{J_i\pi_i\alpha_i}_{\mathbf{q}} \frac{2J_i+1}{2} \int_{0}^{\pi} d \theta \sin (\theta) d_{K 0}^{J_{i}}(\theta)  \nonumber\\
&&\times \bra{\Phi(\mathbf{q}^{\prime})} \sum^A_{i=1} \delta(\mathbf{r}_i-\mathbf{r}) \hat{P}^{N} \hat{P}^{Z} e^{i\theta\hat J_y}\ket{\Phi(\mathbf{q})}.
\eeqn

The quality of the theoretical description for nucleon density and transition density can be examined by looking at the elastic and inelastic form factors.  To illustrate this, Fig.\ref{fig:Marevi2019_density} displays the form factors of electron elastic and inelastic scattering, as well as nucleon densities of the predominant configurations in the first two $0^+$ states of $^{12}$C. The results of \ac{SPGCM}+RHB calculation \cite{Marevi:2019C12}, labeled as MR-EDF, are compared with those obtained from the calculations of \ac{SPGCM}+AMD ~\cite{Kanada:2007} and 
Tohsaki-Horiuchi-Schuck-Ropke (THSR) wave function
\cite{Funaki:2015_C12}. It was found that the form factors are well reproduced by the MR-EDF calculation. In this calculation, the  wave function of the $0^+_2$ state is predominated by two configurations: a weakly prolate deformed configuration and a strongly oblate deformed one, as shown in Fig.\ref{fig:Marevi2019_density}(c). Both two configurations have very large octupole deformations.
 
 \begin{figure}[bt]
\centerline{\includegraphics[width=\textwidth]{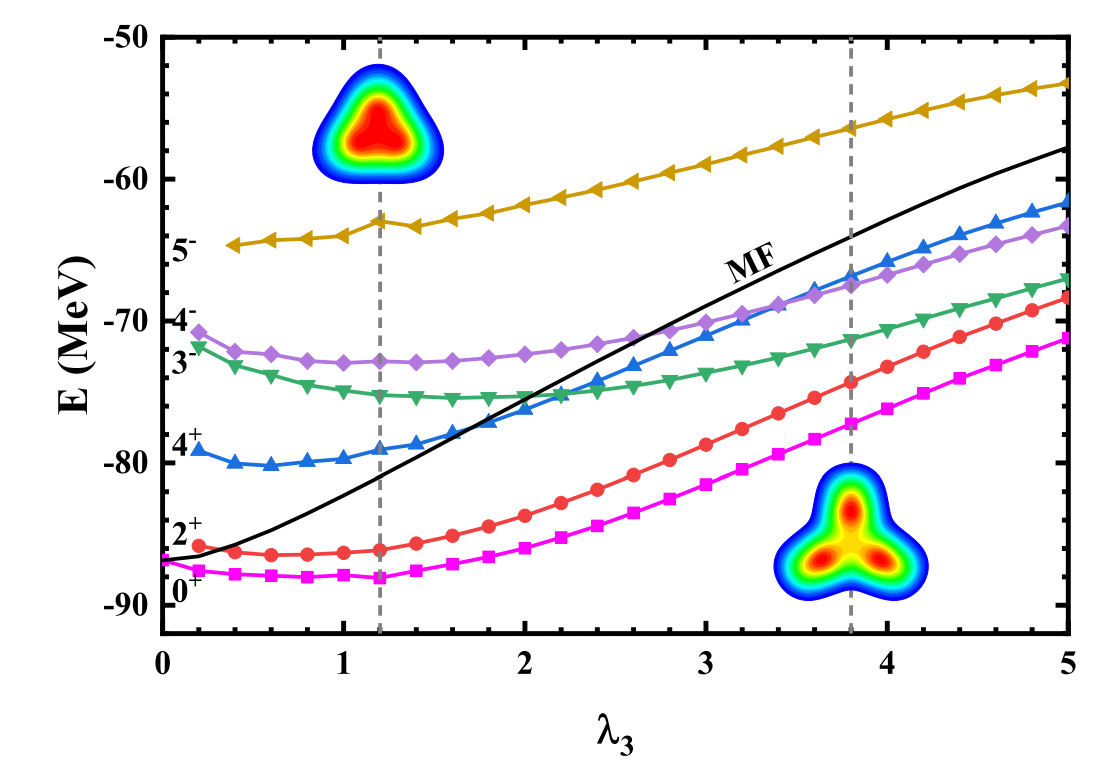}}
 \caption{The energies of mean-field (black line) and spin-parity  projected states with different $J^\pi$ (scatter plot points) for $^{12}$C  as a function of the intrinsic triangular deformation parameter $\lambda_3$ defined in (\ref{eq:lambda3}) from the RHB calculation.  The insets display the density profiles of the intrinsic states at $\lambda_3 = 1.2$ and $3.8$, indicated by two gray dashed  lines. 
 Taken from Ref.~\cite{Wang:2022CTP}.}
 \label{fig:Wu_GCM4C12_PES}
\end{figure}

It is worth noting that the study conducted with the \ac{SPGCM}+RHB framework~\cite{Marevi:2019C12} did not take into account triaxial deformation. To achieve the expected triangular distribution of the Hoyle state, axial symmetry needs to be broken. Actually,  the constraint on multipole moments may not be an effective way for the study of cluster structure in atomic nuclei. Recent progress was made in the development of multi-dimensional RHB approach~\cite{Lu:2012PRC} in which atomic nuclei are allowed to have arbitrary geometric shapes. The techniques of \ac{PP} combined with 3D Angular Momentum Projection (3DAMP) were employed to restore rotation symmetry in order to investigate the $3\alpha$ cluster structure in $^{12}$C~\cite{Wang:2022CTP}.  In this study, a constraint on the triangular moment 
\beq
\label{eq:lambda3}
\lambda_3= \frac{4\pi}{3NR^3} \bra{\Phi(\mathbf{q})}\hat S_3\ket{\Phi(\mathbf{q})},
\eeq
was introduced to control the distance among the three $\alpha$ particles, arranged in an equilateral triangle, in the reference states. The corresponding operator $\hat S_3$ is defined as
\beq
\hat S_{3}=\left(z^{2}+\rho^{2} \cos ^{2} \varphi\right)^{3 / 2} \cos (3 \theta),
\eeq
where $\rho$, $\varphi$, and $z$ represent the cylindrical coordinates, while $\theta$ is the polar angle on the $x$-$z$ plane. Furthermore, to account for pairing correlations between nucleons, a separable pairing force of finite range was employed~\cite{TMR:2009PLB}. This approach allowed for a comprehensive exploration of the $3\alpha$ cluster structure in $^{12}$C with more degrees of freedom.

As seen from Fig.~\ref{fig:Wu_GCM4C12_PES}, spin-parity projection favors configurations with nonzero values of $\lambda_3$, suggesting the formation of $3\alpha$ cluster structures. It is worth noting that the projected energy surface exhibits a significant softness. Thus, the extension of this study with the mixing of particle-number projected states with different values of the intrinsic triangular moment would provide a deeper insight into the $3\alpha$ cluster structure in $^{12}$C.

\begin{figure}[th]
\centerline{\includegraphics[width=\textwidth]{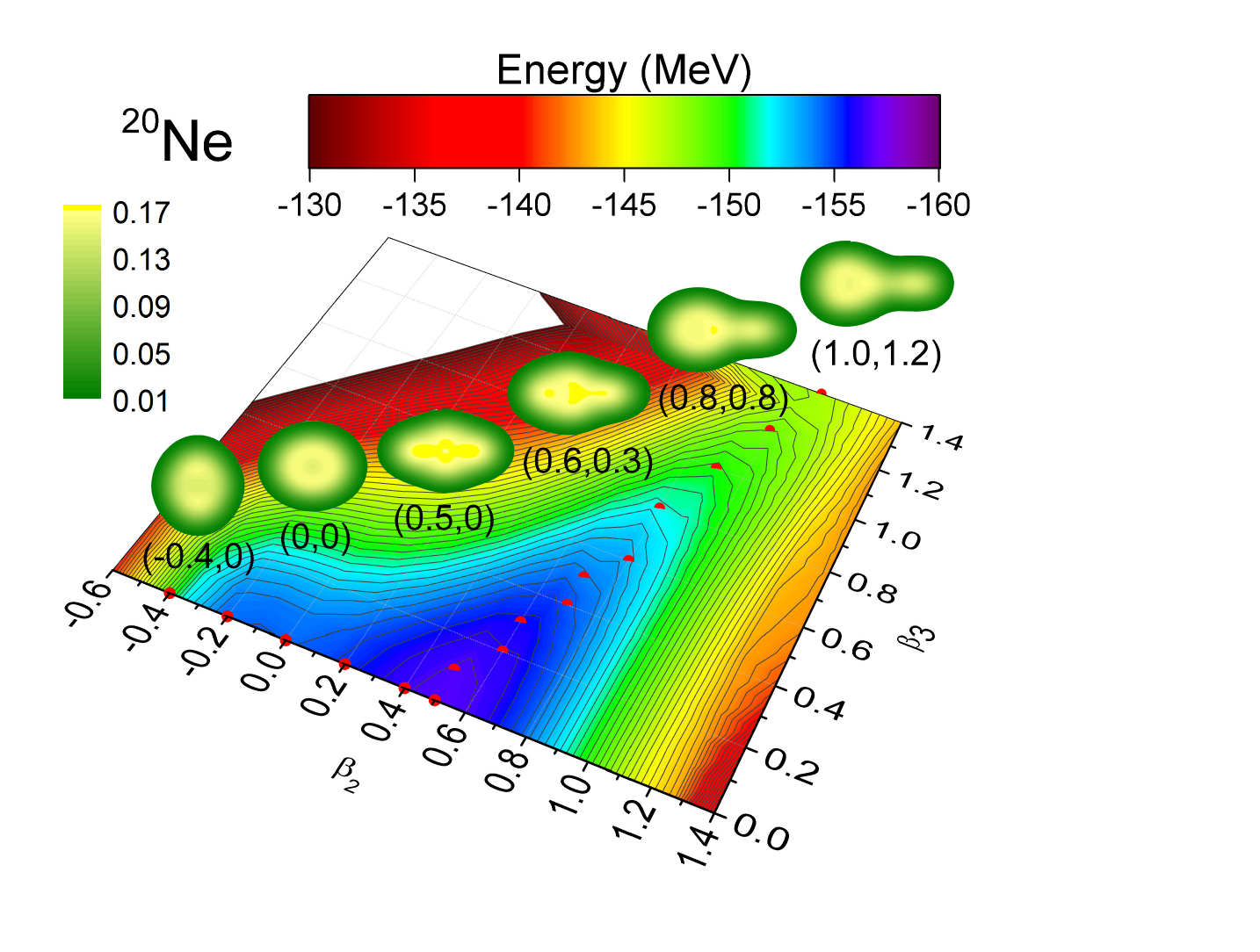}}
\caption{ The mean-field energy surface of $^{20}$Ne in two-dimensional $(\beta_{20}, \beta_{30})$ deformation plane, where the optimal configurations (along the valley) are indicated with red dots. The density profiles of several selected configurations are plotted (in fm$^{-3}$). Taken from Ref.~\cite{Zhou:2016}.}
\label{fig:PES4Ne20_Zhou2016}
\end{figure}

\begin{figure}[th]
\centerline{\includegraphics[width=\textwidth]{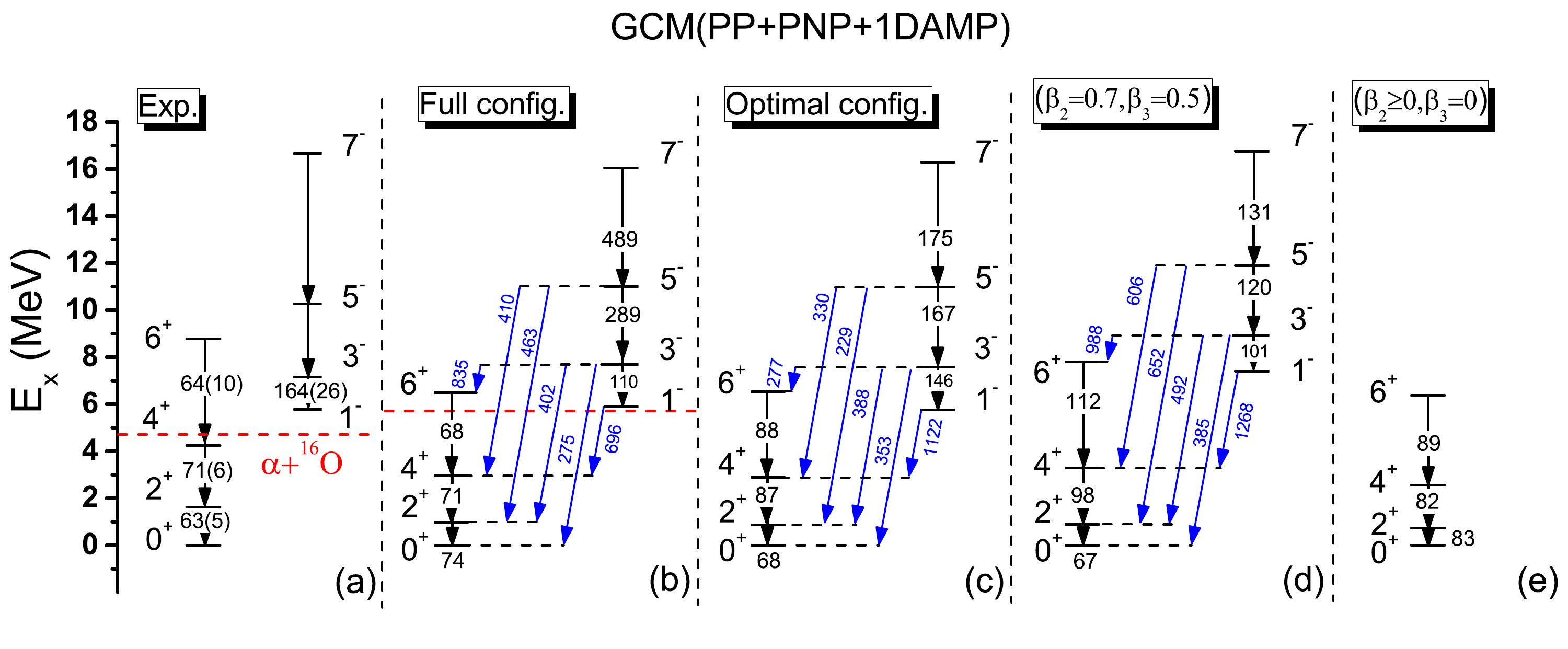}}
\caption{Energy spectra of $^{20}$Ne from different calculations, in comparison with data.  
The optimal configurations in (c) are those indicated by red dots in Fig.~\ref{fig:PES4Ne20_Zhou2016}. The single configuration with deformation parameters $(\beta_2=0.7,\beta_3=0.5)$ in (d) corresponds to the energy minimum of the energy surface with \ac{AMP}.
Taken from Ref.~\cite{Zhou:2016}.}
\label{fig:Spectra4Ne20_Zhou2016}
\end{figure}

\begin{figure}[tb]
\centerline{\includegraphics[width=\textwidth]{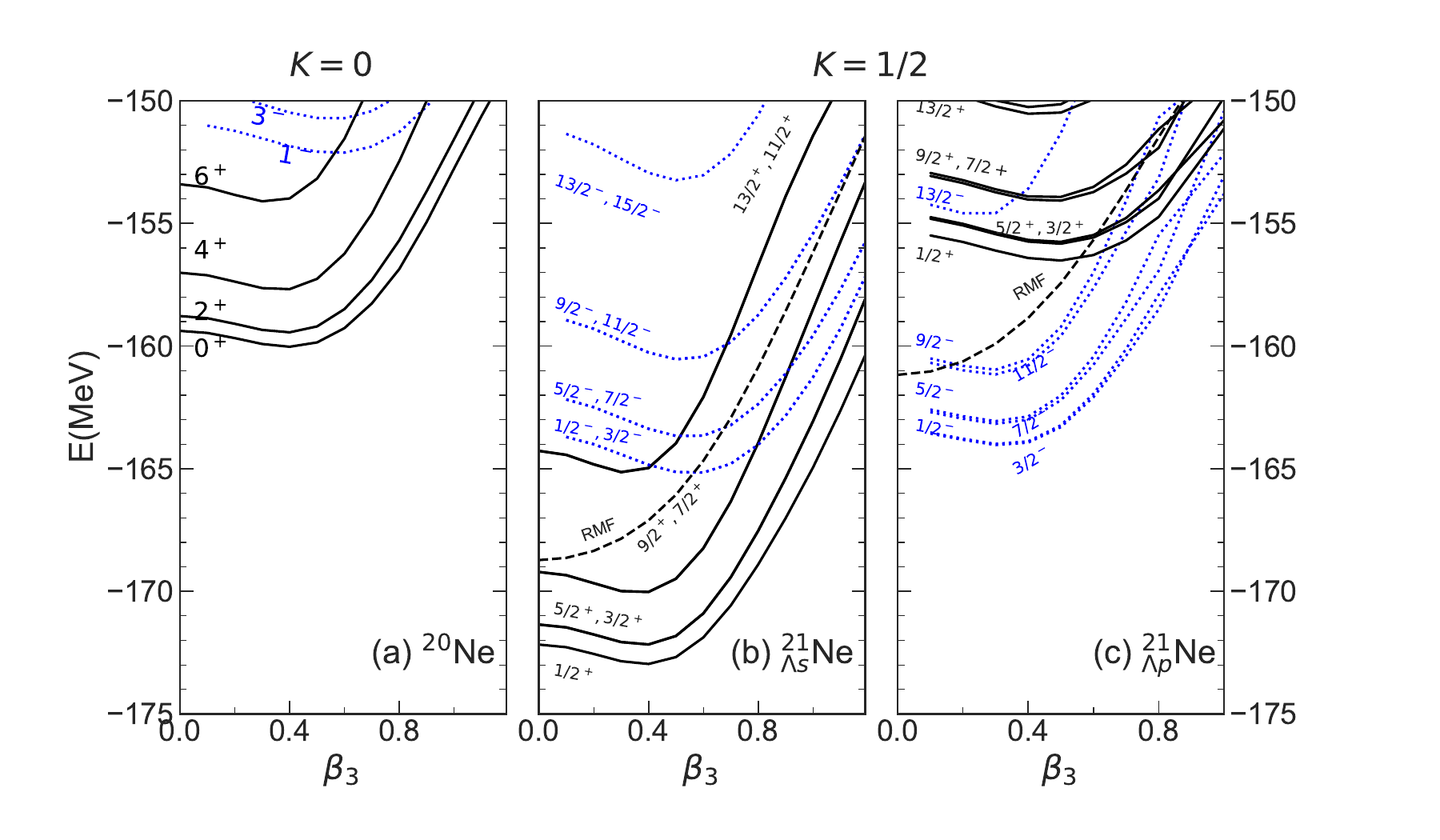}}\vspace{-0.5cm}
\caption{The projected energy curves as a function of the octupole deformation $\beta_3$ of the intrinsic state ($\beta_2=0.60$ is fixed) in $^{20}$Ne, $^{21}_{\Lambda_s}$Ne and $^{21}_{\Lambda_p}$Ne, respectively. The symbols $\Lambda_s$ and $\Lambda_p$ label the configurations where the $\Lambda$ hyperon is put on the first and second lowest-energy orbits, respectively. The levels with positive (negative) parity are plotted with solid (dotted) curves. Taken from Ref.~\cite{Xia:2019}.}
\label{fig:PES_21_L_Ne}
\end{figure}

\begin{figure}[tb]
\centerline{\includegraphics[width=\textwidth]{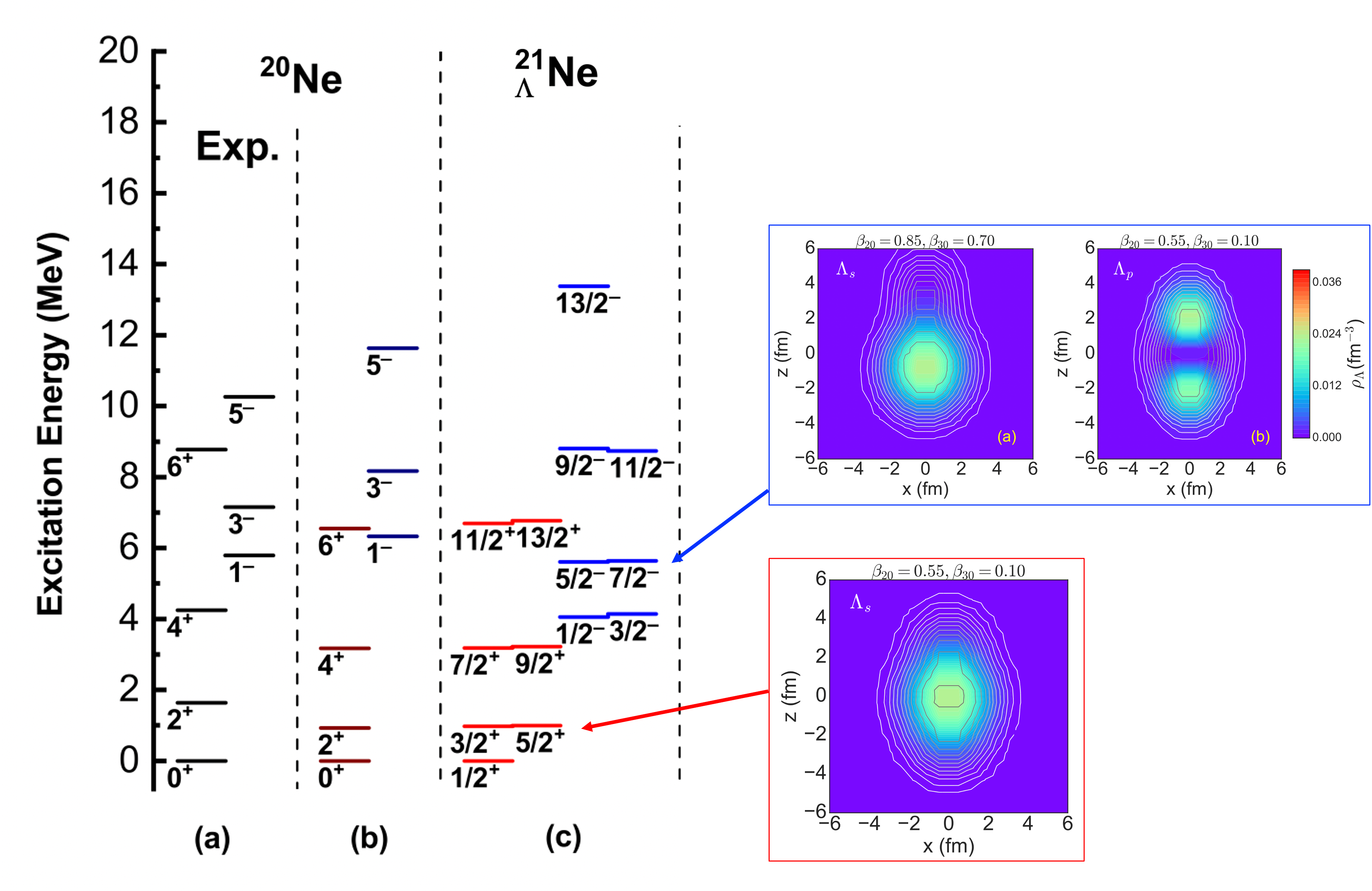}}
\caption{The low-lying energy spectra of \nuclide[20]{Ne} and $^{21}_\Lambda$Ne from the HyperGCM calculation with the mixing of quadrupole-octupole deformed configurations. The contour plots of the nucleon density and hyperon of the predominated configurations of positive and negative-parity states are also displayed. Results are taken from Ref.~\cite{Xia:2023}. }
\label{fig:Spectra_21_L_Ne}
\end{figure}

Another example nucleus with different cluster structures is $\nuclide[20]{Ne}$. It has been investigated  using AMD models, which were based either on spherical Gaussian basis functions (SphAMD)~\cite{Kanada-Enyo:1995PTP} or triaxially deformed (TriAMD) Gaussian basis functions~\cite{Taniguchi:2004,Kimura:2004}. These studies employed phenomenological Hamiltonians or the Gogny D1S force, respectively.  It was found in the SphAMD study   that  both positive and negative parity low-spin states are dominated by the diatomic molecular structures $\alpha+^{16}$O. With the consideration of triaxial deformation in the basis, the ground band with $K^\pi=0^+$ turns out to have a mixed character of the deformed mean-field structure and the cluster structure. In contrast, the negative-parity state with $K^\pi=0^-$ has an almost pure $\alpha+^{16}$O cluster structure. In particular, it was shown that the cluster structure moderates and the average distance between $\alpha$ and \nuclide[16]{O} 
becomes smaller as the angular momentum increases. 

In recent years, the cluster structures in $\nuclide[20]{Ne}$ were examined with the CDFT~\cite{Zhou:2016,Marevic:2018Ne}. Figure~\ref{fig:PES4Ne20_Zhou2016} illustrates the mean-field energy surface of $\nuclide[20]{Ne}$ in the $(\beta_{20}, \beta_{30})$ deformation plane from the calculation using the relativistic PC-PK1~\cite{Zhao:2010PRC}. Although a global energy minimum is found at the reflection-symmetric prolate shape ($\beta_2=0.5,\beta_3=0$), this equilibrium shape is shown to be unstable against quantum fluctuations. A schematic representation of nuclear shapes corresponding to selected configurations along the optimal path on the energy surface highlights the progressive development of the reflection-asymmetric diatomic molecular structure ($\alpha+^{16}$O) as the values of $\beta_3$ increase along the optimal path. This is consistent with the finding in the TriAMD study based on the Gogny D1S force~\cite{Kimura:2004}. 
It is worth pointing out that only the configurations with nonzero value of $\beta_3$ contribute to the negative-parity states.  Thus, the finding that  the negative-parity states are predominated by the configuration with $\alpha+^{16}\mathrm{O}$ structure does not necessary mean that the nucleus $^{20}$Ne has a permanent octupole deformation. Actually, it has been shown in Ref.~\cite{Zhou:2016,Marevic:2018Ne} that the collective wave functions of the negative-parity states are broadly distributed in the $\beta_2$-$\beta_3$ plane, confirming the picture of octupole vibration rather than the rotation with a permanent octupole deformation for the low-lying states of $\nuclide[20]{Ne}$.

Figure~\ref{fig:Spectra4Ne20_Zhou2016} presents the low-energy spectra for $\nuclide[20]{Ne}$ obtained from different calculations. These calculations consider the dynamical octupole vibration effects and successfully reproduce both energy spectra and electric multipole transition strengths of low-lying parity-doublet bands. Compared to the results of calculations with the mixing of only prolate deformed configurations in Fig.~\ref{fig:Spectra4Ne20_Zhou2016}(e), the energy spectra become slightly stretched and  $E2$ transition strengths are reduced in the calculations with octupole-deformed configurations. Moreover, it is shown in Fig.~\ref{fig:Spectra4Ne20_Zhou2016} that the energy spectra from the calculations using  full configurations and optimal configurations are close to each other, and are more compressed than those of  single-configuration calculation and the corresponding data. Among all the calculations, the $E2$ transition strengths in the ground-state band from the full configuration-mixing calculations are closest to the data. An in-depth analysis of the distributions of collective wave functions for the low-lying states of $\nuclide[20]{Ne}$ reveals  that the positive-parity states are dominated by the $\alpha+^{12}$C+$\alpha$ structure, and the negative-parity states by the $\alpha+^{16}$O structure.

The low-lying states of hypernuclei offer a wealth of information about hyperon-nucleon interactions in nuclear medium and the impurity effect of hyperon in atomic nuclei. Extracting this valuable information from experimental data necessitates the use of accurate hypernuclear models. The SPGCM has been extended for hypernuclear spectroscopy based on different nuclear EDFs, including the Gogny force~\cite{Isaka:2011,Isaka:2012}, Skyrme force~\cite{Cui:2017}, or relativistic Lagrangian density~\cite{Mei:2016PRC}. Within this framework, the inclusion of dynamical octupole shapes may change significantly the interpretation of the spectroscopy of some hypernuclei, like $^{21}_\Lambda$Ne.  Figure~\ref{fig:PES_21_L_Ne} display the mean-field and projected energy curves as a function of the octupole deformation $\beta_3$ for the intrinsic states in $^{20}$Ne and $^{21}_\Lambda$Ne. The quadrupole deformation parameter $\beta_2$ is fixed to the value around the energy minimum. Notably, the energies of the negative-parity states  with the configuration $\ket{^{20}{\rm Ne}(K^\pi=0^-)}\otimes \ket{\Lambda_s}$  (blue dashed curves in Fig.~\ref{fig:PES_21_L_Ne}(b)) are close to those with the configuration $\ket{^{20}{\rm Ne}(K^\pi=0^+)}\otimes \ket{\Lambda_p}$  (blue dashed curves in Fig.~\ref{fig:PES_21_L_Ne}(c))~\cite{Xia:2019}, where in these two types of configurations  the $\Lambda$ hyperon occupies either the first or second lowest-energy states. It turns out in the study~\cite{Xia:2023} that these two configurations are strongly mixed due to the octupole correlations, producing the low-lying negative-parity states in Fig.~\ref{fig:Spectra_21_L_Ne}. Moreover, the electric multipole transition strengths of the negative-parity states are significantly quenched with this configuration mixing effect. The density distribution of these two dominated configurations are also shown in Fig.~\ref{fig:Spectra_21_L_Ne}. One can see that the negative-parity states in $^{21}_\Lambda$Ne are dominated by the mixing of the cluster structures $\alpha+^{16}$O+$\Lambda_s$ and $\alpha+^{12}$C+$\alpha+\Lambda_p$.  In contrast, the positive-parity states are dominated by the $\alpha+^{12}$C+$\alpha+\Lambda_s$ structure, which are only slightly changed by the dynamical octupole  effect.

From all the above studies, one may conclude that octupole shape is one of the relevant degrees of freedom for describing the clustering structure in $^{12}$C and  \nuclide[20]{Ne}. However, the use of a constraint on the quadrupole and octupole moments might not be sufficient  for the description of nucleon clustering structures. It is thus of importance to explore alternative ways to generate configurations with all possible cluster structures for the nuclei of interest within the SPGCM.   

\subsection{Octupole vibrational states in near-spherical nuclei }

Nuclei with proton and/or neutron numbers near magic numbers often exhibit octupole correlations through vibrational motions. One of the most notable observations is the low-lying $3^-$ state, which is often interpreted as a one-octupole-phonon state. The higher-lying states are dominated by  the weak coupling between the octupole $3^-$ phonon and other single-particle excited states, or multi-phonon states with anharmonic characteristics. This picture has been illustrated through SPGCM studies that incorporate octupole shape fluctuation effects~\cite{Yao:2016Pb,Robledo:2016,Rong:2023PLB}.

\begin{figure}[tb]
\centerline{\includegraphics[width=0.8\textwidth]{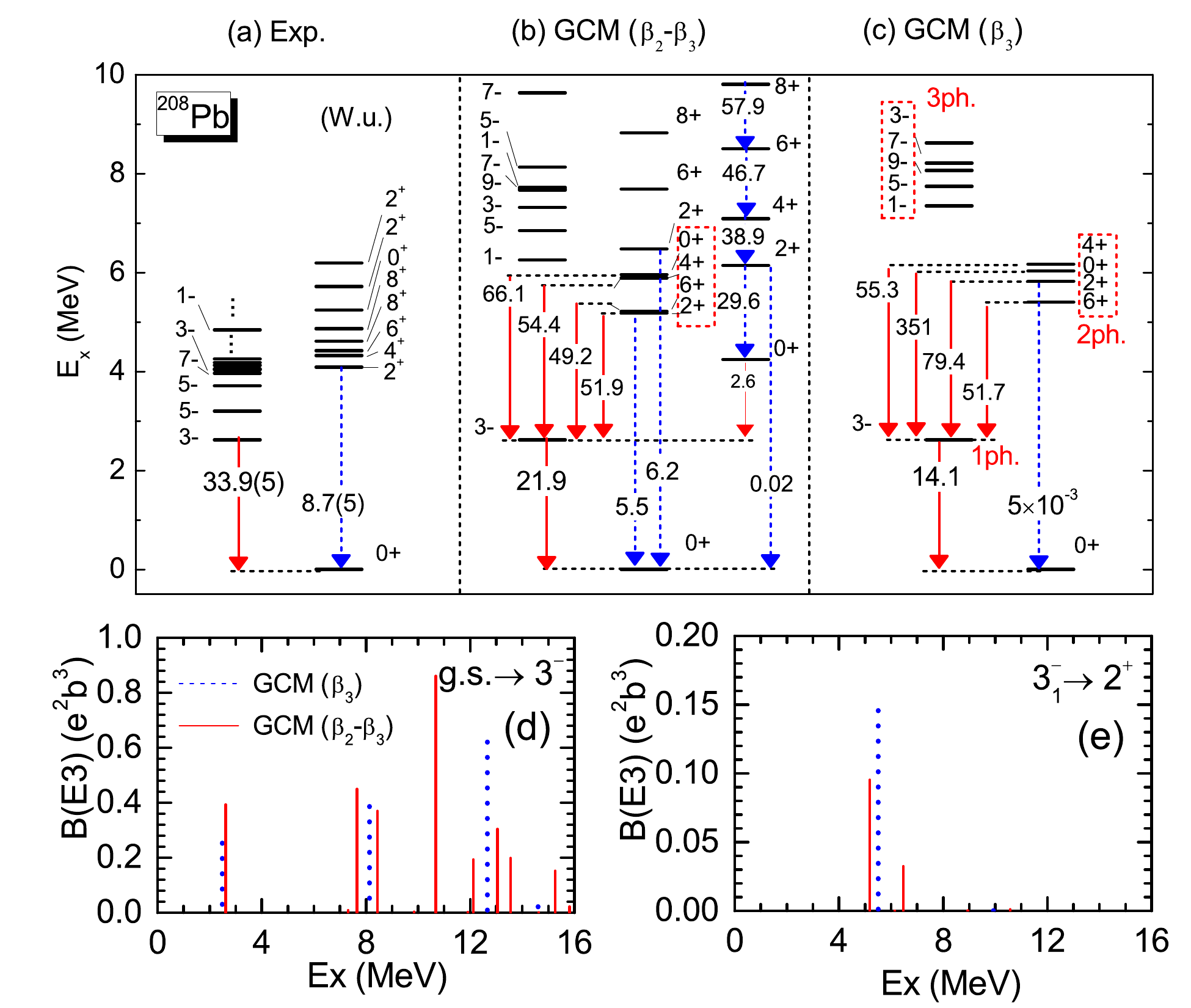}}
\caption{ The low-lying energy spectra of $^{208}$Pb obtained from the GCM calculation with the mixing of quadrupole-octupole ($\beta_2$-$\beta_3$) deformed configurations (b) or the mixing of only octupole ($\beta_3$) deformed configurations with $\beta_2=0$ (c), in comparison with the experimental data taken from Ref.~\cite{NNDC}. The panel (d) and (e) show the  $E3$ transition strengths from the the ground state to $3^-$ states and from the first $3^-$ state to $2^+$ states as a function of the excitation energy of the final states. In (a)-(c), the red solid and the blue dashed lines indicate the $E3$ and $E2$ transition strengths (in W.u.), respectively. All the calculated excitation energies are scaled to the empirical excitation energy of the lowest $3^-$ state by dividing them by a factor of 1.78 and 1.64 in the panels (b) and (c), respectively. Figure adapted from Ref.~\cite{Yao:2016Pb}.  Figure reprinted with permission from the American Physical Society.}
\label{fig:Spectra4Pb208}
\end{figure}

 The double-magic nucleus \nuclide[208]{Pb} serves as a prime example of a near-spherical nucleus with intriguing low-lying octupole phonon states, which have been the focus of extensive research utilizing various nuclear models~\cite{Brown:2000}. Numerous experiments have been conducted to search for multi-octupole-phonon states in \nuclide[208]{Pb}~\cite{Yeh:1996,Vetter:1998}. A study in Ref.\cite{Yao:2016Pb} extended the SPGCM to investigate the low-lying states of \nuclide[208]{Pb}. This study employed a CDFT, with the generator coordinates chosen as the parameters of quadrupole deformation ($\beta_2$) and octupole deformation ($\beta_3$). The results, displayed in Fig.~\ref{fig:Spectra4Pb208}, showcase the low-lying energy spectra of \nuclide[208]{Pb}.  One can see from Fig.~\ref{fig:Spectra4Pb208}(c) that the multiples of two- and three-“phonon” states (labeled as 2ph and 3ph, respectively) appear at similar excitation energies to each other. The predicted spin-average of the excitation energies for the 2ph and 3ph states is 9.6 MeV and 13.1 MeV, respectively, which is about twice and three times that of the one-octupole phonon state, 4.3 MeV. We note that the energy spectra in Fig.~\ref{fig:Spectra4Pb208} are normalized to the data for the excitation energy of the $3^+_1$ state. However, there are some energy displacements, indicating the existence of anharmonicity. The anharmonicity is shown to be stronger in the transition strengths. The $B(E3)$ value for the transition from the 2ph multiplets to the $3^+_1$ state is  much larger than twice the $B(E3)$ value from the $3^+_1$ state to the ground state. Fig.~\ref{fig:Spectra4Pb208}(b) shows that the inclusion of the quadrupole shape fluctuation slightly alters the excitation energies, and the transition strengths for the $3^+_1\to 0^+_1$ and the  $2^+_1\to 0^+_1$ transitions are significantly improved.
 Fig.~\ref{fig:Spectra4Pb208}(d) and (e) show that the E3 transitions become more fragmented after taking into account the interplay of the quadrupole and octupole shape fluctuations. 
 
\begin{figure}[tb]
\centerline{\includegraphics[width=0.6\textwidth]{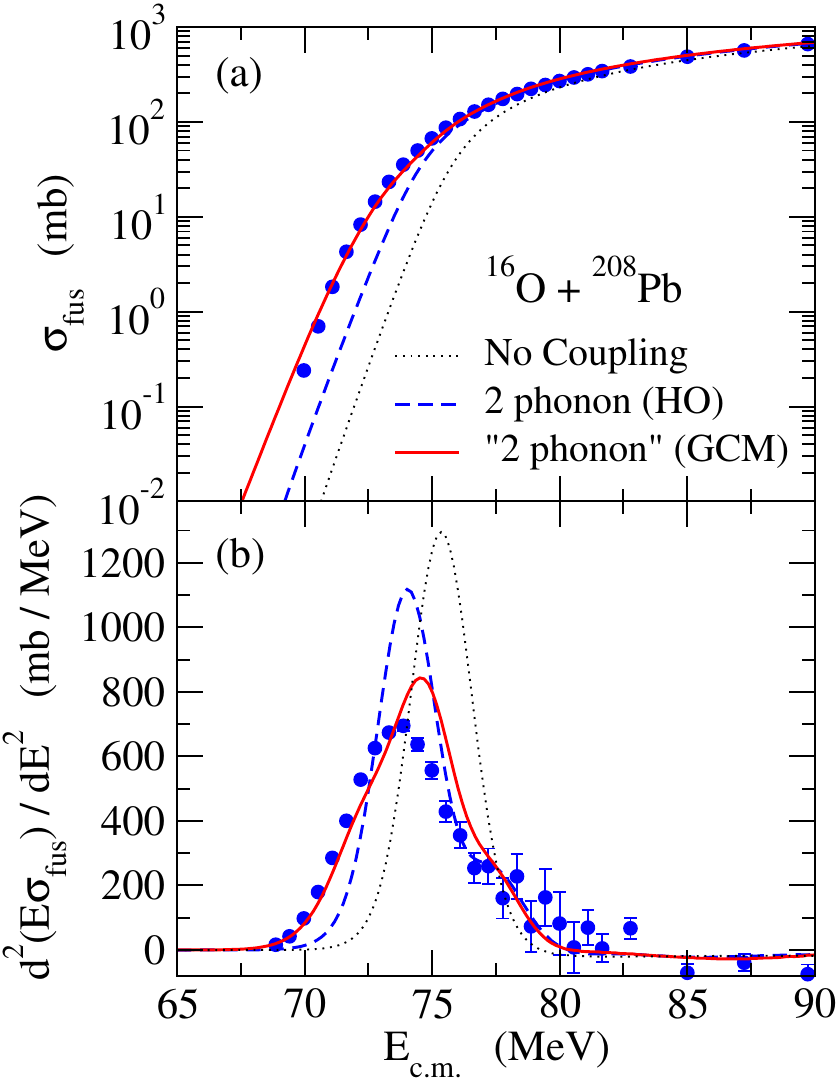}}
\caption{ The fusion cross sections (upper panel) and the fusion
  barrier distributions (lower panel) for the $^{16}$O+$^{208}$Pb
  system obtained with the semi-microscopic coupled-channels calculation
with the coupling strengths from the SPGCM calculations (the solid line). The dashed and the dotted lines show the results of the two-phonon coupling in the harmonic oscillator limit and of the no-coupling limit, respectively. Figure adapted from Ref.~\cite{Yao:2016Pb}.  Figure reprinted with permission from the American Physical Society.}
\label{fig:FusionPb208}
\end{figure}

 The low-lying $3^-$ states in \nuclide[208]{Pb} turns out to have a significant impact on the cross section of sub-barrier fusion reaction of $^{16}$O+$^{208}$Pb, as shown in Fig.~\ref{fig:FusionPb208}. The solid line is the fusion cross sections (the upper panel) and the fusion barrier distribution (the lower panel) calculated with the  {\tt CCFULL} code~\cite{Hagino:1999CCFULL} using the excitation energies and transition strengths from the SPGCM calculation as inputs. The fusion barrier distribution is defined as the second energy derivative of the product of the energy $E$ and fusion cross section $\sigma_{\rm fus}$, that is, $d^2(E\sigma_{\rm fus})/dE^2$. The results are compared to the results of calculation based on the two-phonon harmonic oscillator approximation and to the single-channel calculation. For the former, the 3$_1^-$, 2$_1^+$, $(3_1^-)^2$, $(2_1^+)^2$, and $3_1^-\otimes 2_1^+$ states are included within the harmonic oscillator coupling scheme~\cite{Hagino:2012PTP}. It has been a long-standing problem that for this particular system the coupled-channels calculation with the harmonic oscillator couplings overestimates the height of the main peak in the barrier distribution \cite{Morton:1999PRC} and underestimates the fusion cross sections at energies below the Coulomb barrier. In contrast,  the calculation using nuclear structure properties from the GCM calculation (see Fig.~\ref{fig:Spectra4Pb208}) yields a much lower peak in the fusion barrier distribution, leading to a better agreement with the experimental data both for the fusion cross sections and for the barrier distribution.  It turns out that the couplings between the $3^-_1$ and the $2^+_1$ states, the two-phonon states, and the excited negative-parity states, play an important role.  

 Nuclear octupole shapes may also play a significant role in heavy-ion collisions. The research by Zhang {\em et al.}\cite{Zhang:2022PRL} demonstrated that the triangular flow and anisotropy $\epsilon_3$ in \nuclide[96]{Zr}+\nuclide[96]{Zr} collisions are more pronounced than those in \nuclide[96]{Ru}+\nuclide[96]{Ru} collisions. This discrepancy was attributed to the stronger octupole correlation in \nuclide[96]{Zr}, a conclusion supported by recent RHB studies that include projections onto parity and angular momentum~\cite{Rong:2023PLB}.

 \subsection{Transition from octupole vibrations to rotations}

The transition from octupole vibrational to rotational motions with the change of isospin or spin, is a fascinating phenomenon in atomic nuclei. This transition can be deduced from the systematic evolution of energy spectra and electric multipole transition strengths, see Fig.~\ref{fig:cartoon4transition}.

\begin{figure}[tb]
\centerline{\includegraphics[width=0.8\textwidth]{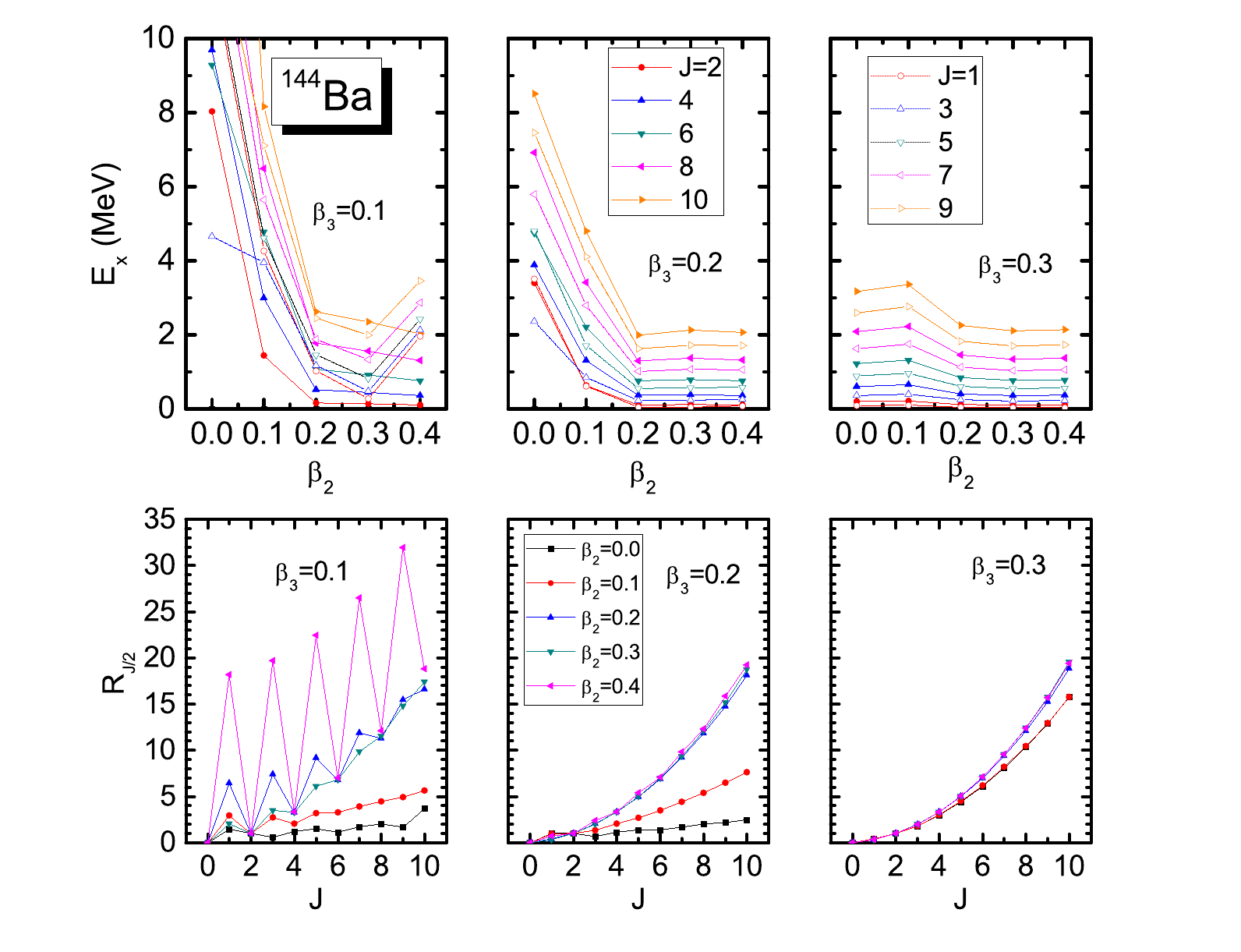}}
\caption{The excitation energy of  states projected onto particle number, angular momentum and parity based on an individual configuration for $^{144}$Ba. (Lower panels) The ratio $R_{J/2}$ as a function of angular momentum $J (\hbar)$.  Figure adapted from Ref.~\cite{Fu:2018Ba}. Figure reprinted with permission from the American Physical Society.}
\label{fig:Ba144_projection}
\end{figure}

Figure~\ref{fig:Ba144_projection} provides a clear illustration of how the energy spectra and the energy ratio $R_{J/2}$ of low-spin states in $^{144}$Ba change as the quadrupole $\beta_2$ and octupole $\beta_3$ deformation parameters increase. These results are obtained through symmetry-projection calculations based on different intrinsically deformed states. For cases where $\beta_3=0.1$, the amplitude of staggering in $R_{J/2}$ increases with the quadrupole deformation parameter $\beta_2$. Octupole-deformed nuclei typically have octupole deformation parameters ranging from 0.1 to 0.2, resulting in energy spectra that exhibit staggering in low-spin states, as discussed in the introduction.

It is worth noting that the $3^-$ state is predicted to be the first excited state for the weakly deformed state with $\beta_2=0$ and $\beta_3=0.1$ or $0.2$. This observation aligns with findings in $^{208}$Pb, as shown in Figure~\ref{fig:Spectra4Pb208}. However, for states with larger octupole deformations, such as $\beta_3=0.2$ or $0.3$, the odd-even parity states become interleaved, irrespective of the value of the quadrupole deformation $\beta_2$, and the characteristic staggering behavior disappears.

\begin{figure}[tb]
\centerline{\includegraphics[width=0.8\textwidth]{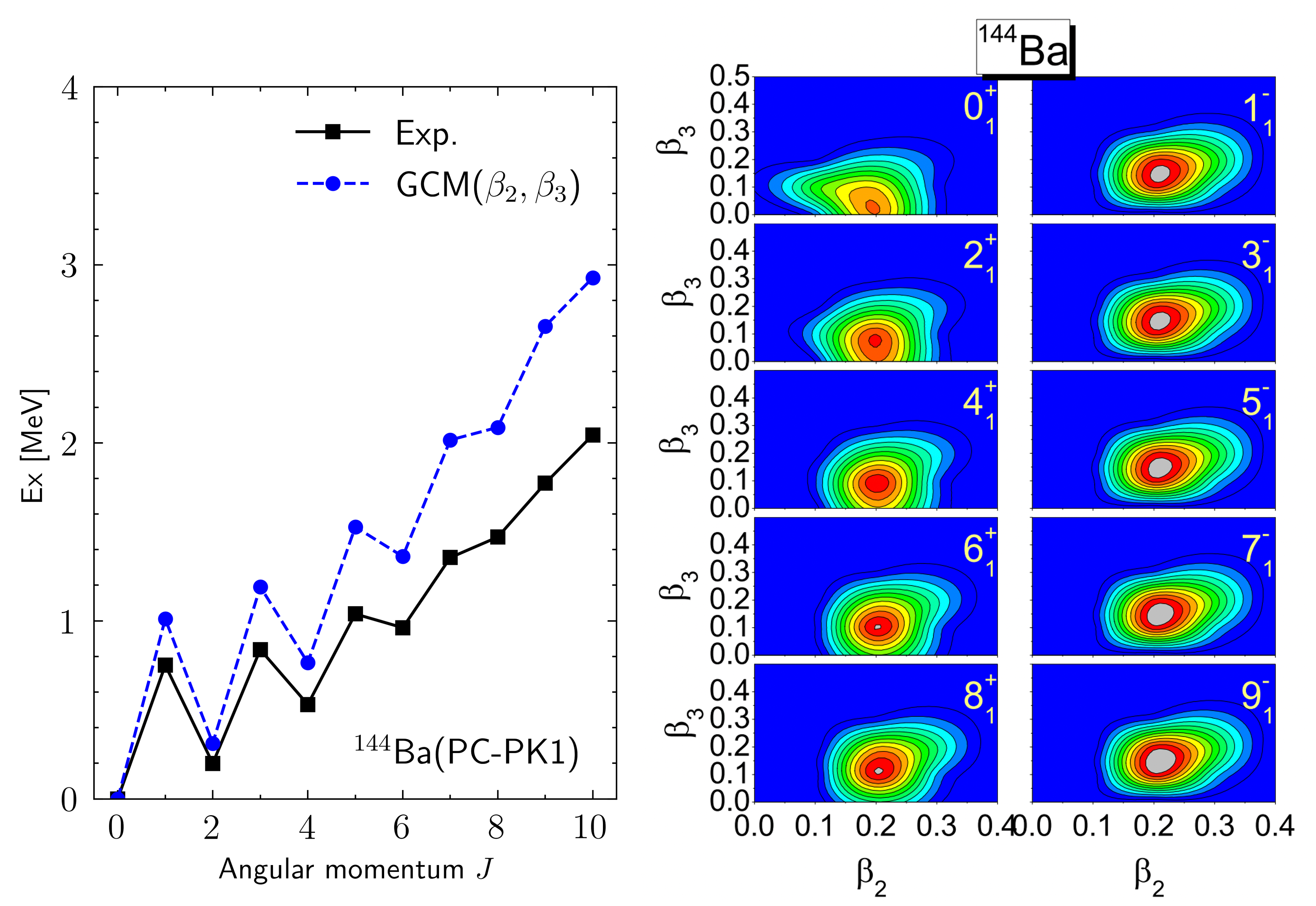}}
\caption{The excitation energies (left panel)  and   the distributions of collective wave functions $\vert g^{J\pi}_\alpha(K=0, \mathbf{q})\vert^2$ (\ref{eq:coll_wf}) in the deformation parameters $\mathbf{q}(\beta_2, \beta_3)$ plane (right panel) for the parity-doublet states of $^{144}$Ba.  }
\label{fig:Ba144_wfs}
\end{figure}

\begin{figure}[t]
\centerline{\includegraphics[width=\textwidth]{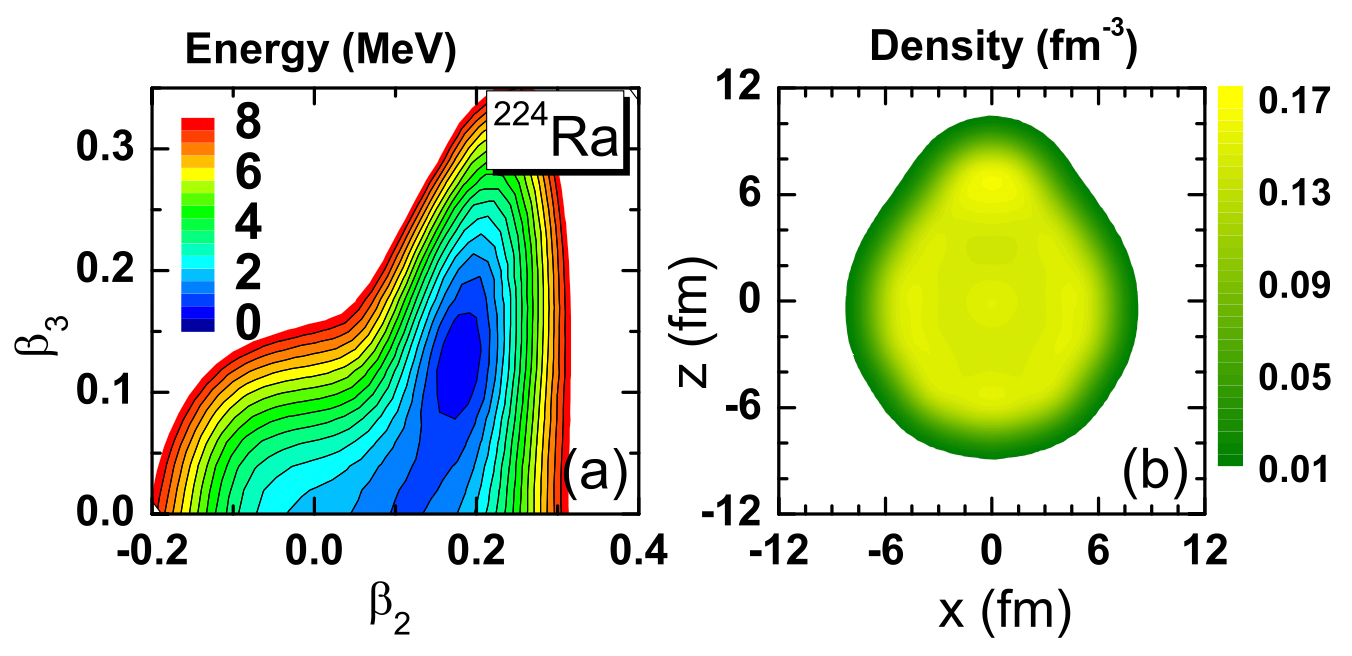}}
\caption{(a) Mean-field energy surface of \nuclide[224]{Ra} in the $(\beta_{20}, \beta_{30})$ plane and (b) distribution of nucleons for the energy-minimal state $\ket{\Phi(\beta_2, \beta_3)}$ in $x$-$z$ plane from the RMF+BCS calculation using the PC-PK1 force~\cite{Zhao:2010PRC}. Figure adapted from Ref.~\cite{Yao:2015Ra224}. Figure reprinted with permission from the American Physical Society.}
\label{fig:PES_density_Ra224}
\end{figure}

\begin{figure}[tb]
\centerline{\includegraphics[width=0.9\textwidth]{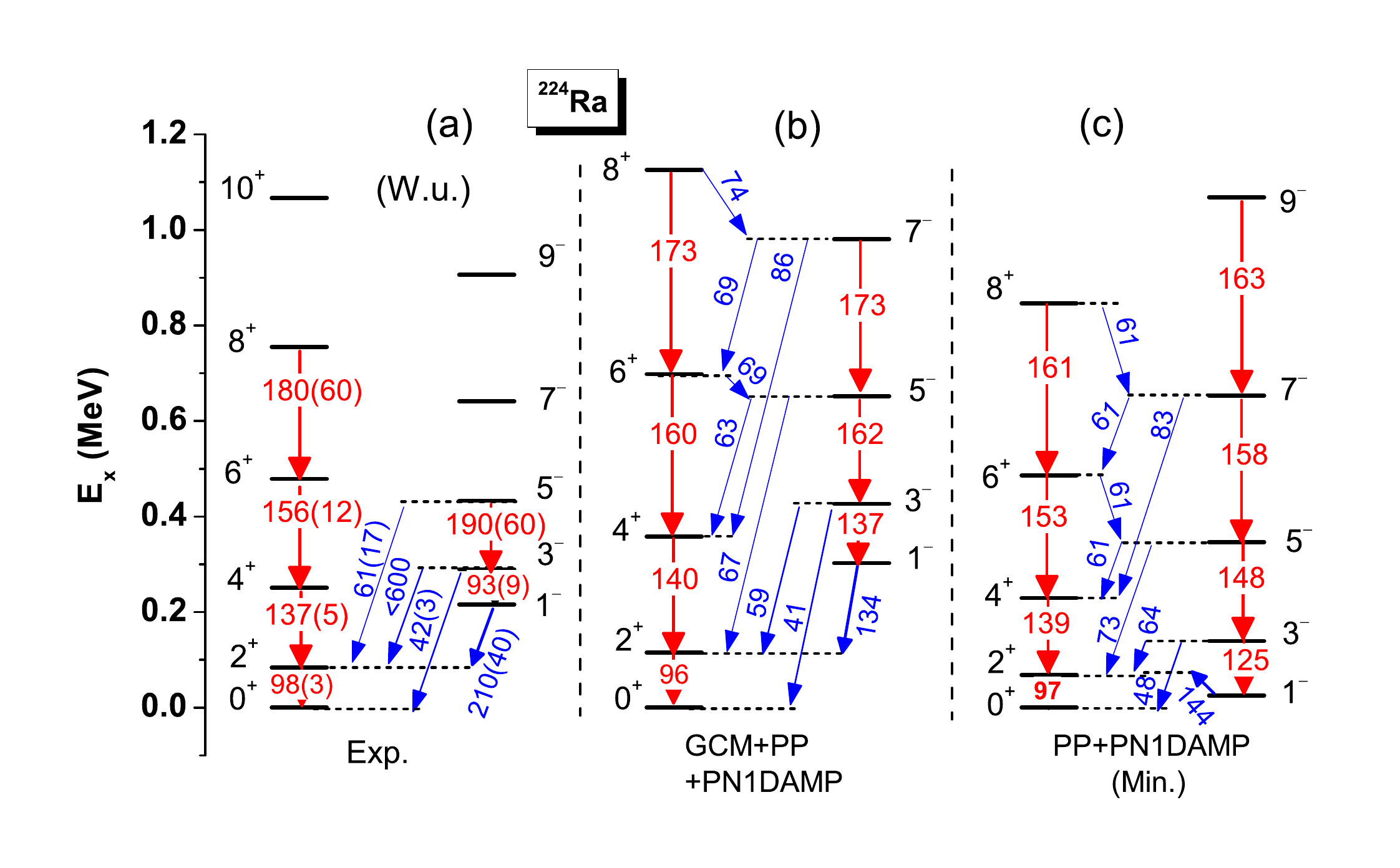}}
\vspace{-0.5cm}\caption{Low-lying energy spectra for $^{224}$Ra from two different calculations, in comparison with data. The numbers on arrows are $E2$ (red color) and $E3$ (blue color) transition strengths (Weisskopf units). Figure adapted from Ref.~\cite{Yao:2015Ra224}. Figure reprinted with permission from the American Physical Society.}
\label{fig:Ra224_spectra}
\end{figure}

\begin{figure}[tb]
\centerline{\includegraphics[width=0.7\textwidth]{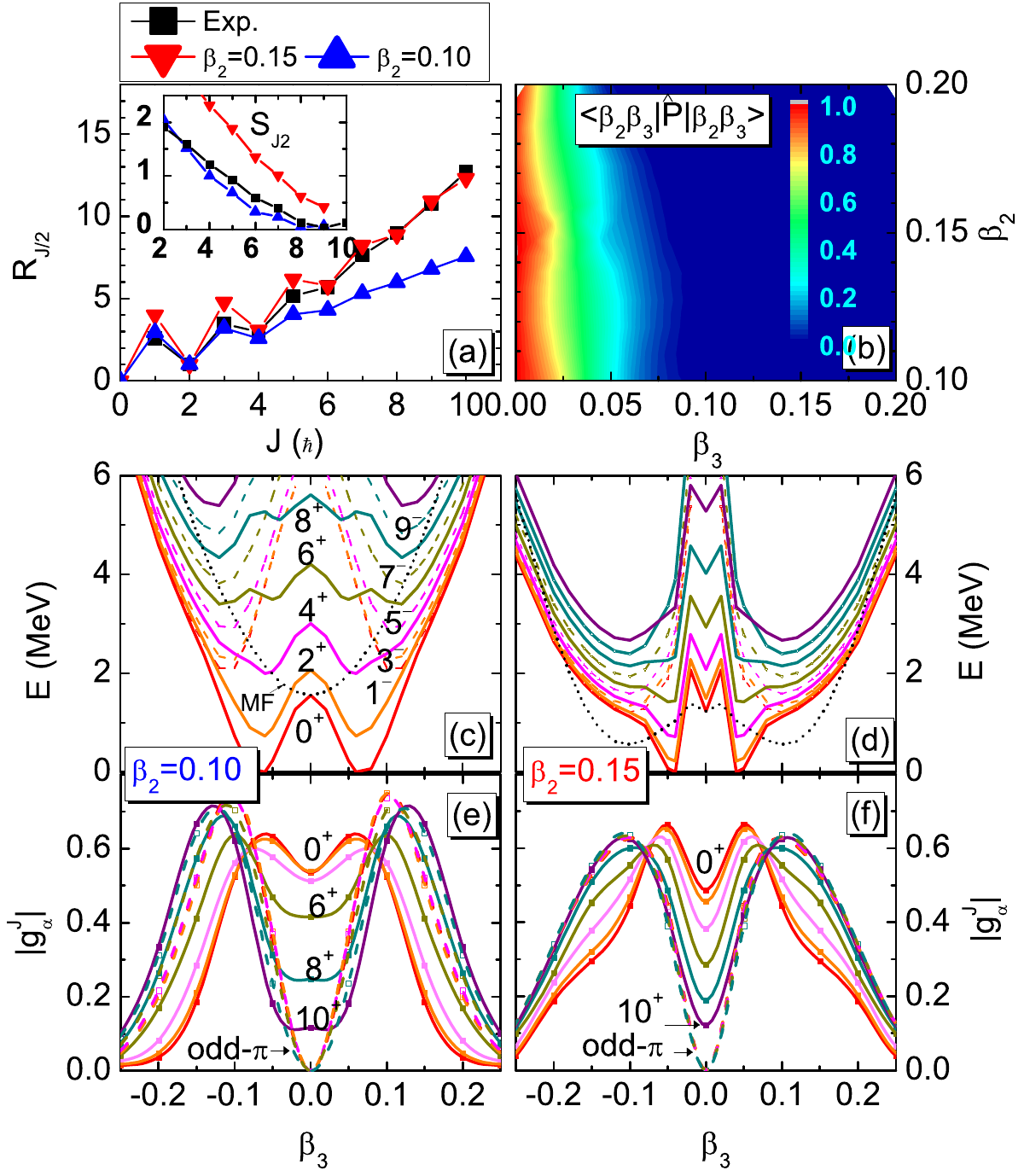}}
\caption{(a) The energy ratio $R_{J/2}=E_x(J^\pi)/E_x(2^+)$ and
 (inset) staggering amplitude $S_{J/2}$ of (\ref{eq:SJ2}) from the SPGCM calculation with the mixing of the configurations with different values of $\beta_3$, keeping the $\beta_2$ being fixed at 0.15 and 0.10, respectively. (b) norm overlap $\bra{\Phi (\beta_2,\beta_3}\hat P \ket{\Phi(\beta_2,\beta_3)}$ of the parity operator; (c) and (d) mean-field and projected energy curves (solid lines for even $J$, dashed lines for odd $J$) in \nuclide[224]{Ra}.   The mean-field energy surfaces in (c) and (d) are shifted down by 3.43 MeV and 4.15 MeV, respectively. (e) and (f) absolute value of collective wave functions for parity-doublet states. The peak at $\beta_3=0.02$ in (d) is due to the collapse of pairing correlation between protons. Figure adapted from Ref.~\cite{Yao:2015Ra224}. Figure reprinted with permission from the American Physical Society.}
\label{fig:Ra224_wfs}
\end{figure}

 Figure~\ref{fig:Ba144_wfs} displays the excitation energies of parity-doublets in \nuclide[144]{Ba} from the SPGCM calculation using the PC-PK1, in comparison with data. It is shown that the staggering amplitude decreases gradually with the increase of angular momentum of the state. The distribution of the collective wave function shows that the predominant configuration of the positive-parity state is drifting to that of negative-parity state with the increase of the angular momentum. At the angular momentum $J\ge8$, the distributions of the wave functions for the positive and negative states are close to each other, demonstrating clearly that the two bands of parity-doublet states merge into one band. 

 A similar rotation-induced shape transition picture is also observed in \nuclide[224]{Ra}. As illustrated in Fig.\ref{fig:PES_density_Ra224}(a), the energy surface and density distribution of the energy-minimal state of \nuclide[224]{Ra} from the CDFT calculation\cite{Yao:2015Ra224} are shown. The global energy minimum is located at the state with deformation parameters $\beta_2=0.179, \beta_3=0.125, \beta_4=0.146$, which are slightly larger than the values $\beta_2=0.154, \beta_3=0.097, \beta_4=0.15$ reported in Ref.\cite{Gaffney:2013nature}. The total nucleon distribution of this energy-minimum state is depicted in Fig.\ref{fig:PES_density_Ra224}(b), revealing a well-developed pear shape. However, the energy difference between the energy minimum and the lowest reflection-symmetric configuration is not large enough to prevent shape fluctuations
   
Figure~\ref{fig:Ra224_spectra} displays the low-lying parity doublets of \nuclide[224]{Ra}. The spectra and the $E2$ and $E3$ transitions can be reproduced reasonably well using only the energy-minimum configuration. However, the predicted alternating parity rotation band as expected for a stable octupole shaped nucleus is not supported by the data. The energy displacement between the parity doublets can only be reproduced in the GCM calculation by mixing the configurations around the equilibrium shape, which does not affect appreciably the good description for the E$\lambda$ strengths.  To understand the connection between spin-dependent parity splitting and shape evolution in a simple and intuitive way, the energy ratio $R_{J/2}$ and normalized staggering amplitude $S_{J/2}$ from the GCM calculations by mixing the octupole configurations at either $\beta_2 = 0.10$ or 0.15 are shown in Fig.~\ref{fig:Ra224_wfs}(a), where the normalized  staggering amplitude is defined as
 \beq 
 \label{eq:SJ2}
 S_{J/2}\equiv \Big\vert E_x(J)- \frac{J+1}{2J + 1} E_x(J-1)-\frac{J}{2J + 1} E_x(J+1)\Big\vert/E_x(2^+).
 \eeq 
 The gradually reduced energy staggering is shown in both cases. Figure~\ref{fig:Ra224_wfs}(c) and (d) shows that the projected energy surfaces with positive parity evolve evidently with increasing spin, presenting a transition from weakly deformed octupole shape to a large deformed one, which is consistent with the behavior of the spin-dependent collective potentials introduced phenomenologically in Ref.~\cite{Jolos:1994}. This spin-controlled shape stabilization picture is demonstrated more clearly in the distributions of collective wave functions in Fig.~\ref{fig:Ra224_wfs}(e) and (f). With the increase of spin, the dominated configuration of positive-parity state drifts gradually from the weak octupole configurations ($\beta_3\simeq0.05$) to those with large octupole shapes ($\beta_3\in[0.10, 0.15]$), for which the overlap  $\bra{\Phi}\hat P\ket{\Phi}\to 0$. The origin of spin-dependent parity splitting in low-spin states is related to the octupole shape stabilization of positive-parity states, the dominated shapes of which drift gradually to that of negative-parity ones.

\section{Schiff moments in odd-mass nuclei and octupole deformation}
\label{sec:Schiff_moments}

The observation of nonzero permanent \ac{EDM}s in elementary or composite particles, such as neutrons, nuclei, atoms, or molecules, would signal the violation of T-symmetry. This violation also implies the violation of \ac{CP} symmetry, as predicted by the CPT theorem~\cite{Luders:1954_CPT}. CP violation is one of Sakharov's conditions for explaining the matter-antimatter asymmetry of the Universe~\cite{Sakharov:1967}. However, CP violation within the standard model of particle physics cannot account for the observed asymmetry~\cite{David:2012,Michael:2003}, necessitating the existence of new CP-violating sources beyond the standard model. The EDM signals arising from PT-violating interactions allow for the differentiation of these much weaker CP-violating interactions from the dominant strong and electromagnetic interactions. Therefore, the study of \ac{EDM}s in various systems has been at the forefront of particle physics, nuclear physics, and atomic physics~\cite{Chupp:2017rkp}, since the first attempt to measure the EDM of neutrons~\cite{Smith:1957_EDM_neutron}.

\begin{figure}[tb]
\centerline{\includegraphics[width=0.8\textwidth]{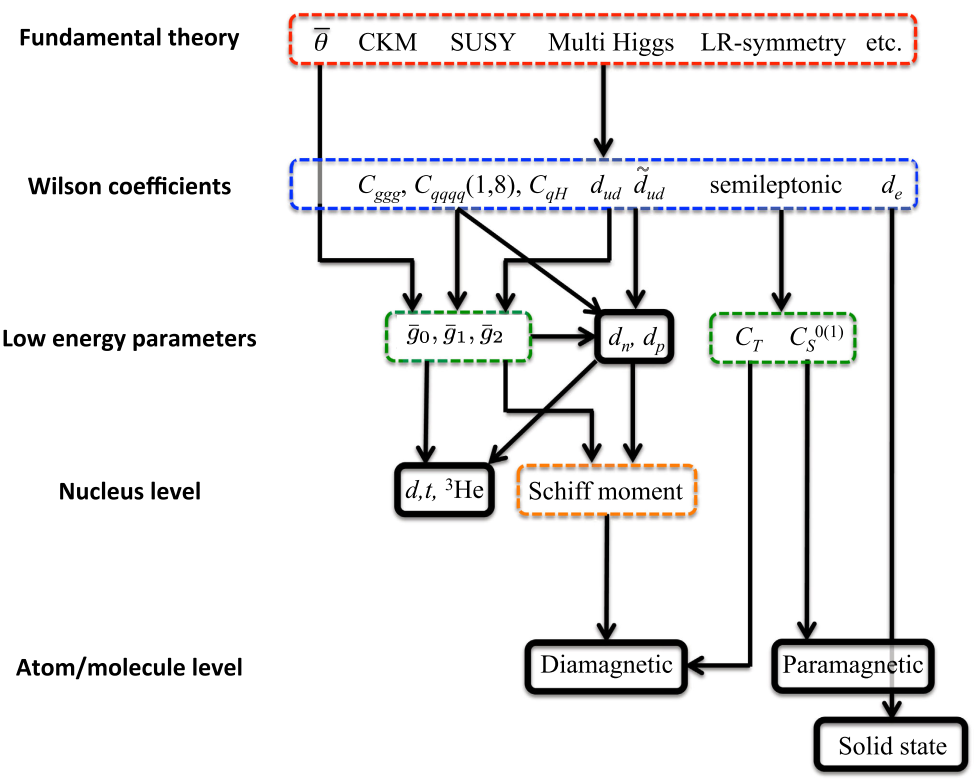}}
\caption{The hierarchical diagram depicting the connections between a fundamental theory at a high-energy scale and an EDM in a measurable low-energy system. The dashed boxes represent levels that are dominated by theory, while the solid boxes identify systems that are the object of current and future experiments. Figure adapted from Ref.~\cite{Chupp:2017rkp}. Figure reprinted with permission from the American Physical Society.}
\label{fig:EDM}
\end{figure}

The odd-mass atomic nuclei with static intrinsic octupole deformation or a soft octupole vibrational mode have an enhanced Schiff moment, which is important for searching for permanent \ac{EDM} in atoms~\cite{Engel:2013PPNP}. Over the past several decades, extensive research has been conducted on EDM at various levels. Fig.\ref{fig:EDM} illustrates the connections from fundamental theory, including the standard model and beyond-standard-model physics, through a series of theoretical levels at different energy scales to experimentally accessible P-odd and T-odd observables in a variety of systems. More details can be found in the review papers\cite{Engel:2013PPNP,Chupp:2017rkp}. Here, we provide a brief overview of the studies on nuclear Schiff moments, with a focus on the discrepancies among the predictions of different nuclear models and the impact of nuclear octupole deformation. In particular, we demonstrate the great potential of extending the \ac{SPGCM} for calculating nuclear Schiff moments.

\subsection{\ac{NSM}}

It was pointed out by Schiff~\cite{Schiff:1963zz} that for an atom composed of a point-like nucleus and non-relativistic electrons, the effect of any nonzero nuclear \ac{EDM} is completely screened by the atomic electrons, regardless of the magnitude of the external potential. As a result, the net atomic \ac{EDM} is zero. However, this hindrance is lifted when the finite size of the nucleus is taken into account.
 
Let us consider the charge distribution $\rho(\mathbf{r})$ of an atomic nucleus inside an atom. The corresponding electrostatic potential at the position $\mathbf{R}$ (much larger than the nuclear size) is given by,
\begin{eqnarray}
\label{eq:epotential}
\varphi(R)=\int \cfrac{e\rho(\mathbf{r})}{|\mathbf{R}-\mathbf{r}|}d^3r+\cfrac{1}{Z}(\mathbf{d}\cdot\mathbf{\nabla})\int \frac{\rho(\mathbf{r})}{|\mathbf{R}-\mathbf{r}|}d^3r\label{2}
\end{eqnarray}
where the second term is for the screen effect induced from the electrons of the atom. Only the dipole (PT-odd) term is kept, where $\mathbf{d}$ is the electric dipole moment of the atomic nucleus, and $Z$ is the proton number,
\begin{eqnarray}
\mathbf{d}=\int d^3r \mathbf{r}\rho(\mathbf{r}),\quad Z=\int d^3r \rho(\mathbf{r}).
\end{eqnarray}
With the help of multipole expansion
\begin{eqnarray} 
 \cfrac{1}{|\mathbf{R}-\mathbf{r}|} 
 &=&\sum_{\ell=0}\frac{r^\ell}{R^{\ell+1}}P_\ell(\cos\theta)\nonumber\\
 &=&\frac{1}{R} -\mathbf{r}\cdot \mathbf{\nabla}\cfrac{1}{R}+\frac{1}{2}(\mathbf{r}\cdot\mathbf{\nabla})^2 \cfrac{1}{R}
-\frac{1}{6}(\vec{r}\cdot\mathbf{\nabla})^3 \cfrac{1}{R}+\cdots
\end{eqnarray}
the electrostatic potential $\varphi(R)$ is simplified as
 \begin{eqnarray}
\varphi(R)
&=&\frac{Ze}{R} - e\int d^3 r \rho(\mathbf{r})
\mathbf{r}\cdot \mathbf{\nabla}\cfrac{1}{R}
+\frac{e}{2}\int d^3 r \rho(\mathbf{r})(\mathbf{r}\cdot\mathbf{\nabla})^2 \cfrac{1}{R}
-\frac{e}{6}\int d^3 r e\rho(\mathbf{r})(\mathbf{r}\cdot\mathbf{\nabla})^3 \cfrac{1}{R}\nonumber\\
&&+\cfrac{1}{Z}(\mathbf{d}\cdot\mathbf{\nabla})\int  d^3r \cfrac{\rho({r})}{R}
-\cfrac{1}{Z}(\mathbf{d}\cdot\mathbf{\nabla})\int  d^3r
\mathbf{r}\cdot \mathbf{\nabla}\cfrac{1}{R}
+\cfrac{1}{Z}(\mathbf{d}\cdot\mathbf{\nabla})\frac{1}{2}\int  d^3r
(\mathbf{r}\cdot\mathbf{\nabla})^2 \cfrac{1}{R}\nonumber\\
&&
-\cfrac{1}{Z}(\mathbf{d}\cdot\mathbf{\nabla})\frac{1}{6}\int  d^3r
(\mathbf{r}\cdot\mathbf{\nabla})^3 \cfrac{1}{R}+\cdots\nonumber\\
&\equiv&  \varphi^{\rm (even)}(R) +  \varphi^{\rm (odd)}(R)
\end{eqnarray}
where the terms in the second and third lines are from the screen effect. One can see that the sub-leading term in the first line cancels the leading term in the second line, as shown in Refs.~\cite{Schiff:1963zz,Flambaum:1984fb}. The potential is divided into P-even and P-odd terms, where the P-odd terms are
 \begin{eqnarray}
 \varphi^{\rm (odd)}(R)
 &=&-\frac{e}{6}\int d^3 r \rho(\mathbf{r})(\mathbf{r}\cdot\mathbf{\nabla})^3 \cfrac{1}{R}
+\cfrac{1}{Z}(\mathbf{d}\cdot\mathbf{\nabla})\frac{1}{2}\int  d^3r
(\mathbf{r}\cdot\mathbf{\nabla})^2 \cfrac{1}{R}+\cdots\nonumber\\
&=&-\mathbf{S}\cdot \mathbf{\nabla} \nabla^2\cfrac{1}{R}
-\cfrac{1}{6}\sum_{i,j,k} Q_{ijk}\nabla_i\nabla_j\nabla_k\cfrac{1}{R}
+ \cdots
\end{eqnarray}
In the last equivalence, the following relation is used,
\begin{eqnarray}
\nabla_i\nabla_j\nabla_k\cfrac{1}{R}
&=&\left[\nabla_i\nabla_j\nabla_k-\cfrac{1}{5}(\delta_{ij}\nabla_k+\delta_{ik}\nabla_j+\delta_{kj}\nabla_i   )\nabla^2\right]\cfrac{1}{R}\nonumber\\
&&+\cfrac{1}{5}(\delta_{ij}\nabla_k+\delta_{ik}\nabla_j+\delta_{kj}\nabla_i)\nabla^2\cfrac{1}{R},
\end{eqnarray} 
and the  Schiff moment  $\mathbf{S}$ is defined as 
\begin{eqnarray}
\mathbf{S}=\cfrac{1}{10}\left(\int e\rho(\mathbf{r})~r^2~\mathbf{r}~d^3r-\cfrac{5}{3Z} ~\mathbf{d} \int \rho(\mathbf{r}) ~r^2~d^3r\right).
\end{eqnarray}
In the first quantization, the operator of Schiff moment is written as~\cite{Flambaum:1984fb}
\begin{eqnarray}
\hat{\mathbf{S}}
= \hat{\mathbf{S}}^{\rm (c)} +  \hat{\mathbf{S}}^{\rm (EDM)}, 
\end{eqnarray}
where the first term is due to the charge distribution of the nucleus  \footnote{Here, $\hat{\mathbf{r}}_p$ stands for the coordinate operator for protons, instead of unit vector. }
\begin{eqnarray}
\hat{\mathbf{S}}^{\rm (c)}
=\cfrac{1}{10}~\sum_{p=1}^Ze\left(\hat r_p^2-\cfrac{5}{3}R^2_{c}
\right)\hat{\mathbf{r}}_p 
\end{eqnarray}
and the second term accounts for the contribution from the EDMs of neutron and proton inside the atomic nucleus,
 \begin{eqnarray}
 \label{eq:Schiff_EDM_nucleon}
\hat{\mathbf{S}}^{\rm (EDM)}
=\cfrac{1}{6}\sum_{N=1}^A {\mathbf{d}}_N(\hat r_N^2-R^2_{ch})
+\cfrac{1}{5}\Bigg[(\hat{\mathbf{r}}_N\cdot {\mathbf{d}}_N)\hat{\mathbf{r}}_N
-\cfrac{1}{3} {\mathbf{d}}_N\hat{r}_N^2\Bigg].
\end{eqnarray}
where $R^2_{c}$ is the mean squared radius of the nuclear charge distribution,
\beq
R^2_{c}
=\frac{1}{Z}\int d^3r \mathbf{r}^2 \rho(\mathbf{r}),
\eeq
and ${\mathbf{d}}_N$ is the EDM of a neutron $(N=n)$ or proton ($N=p$),
\beq
{\mathbf{d}}_N
=\int d^3r \mathbf{r} \rho^{(N)}_c(\mathbf{r}),
\eeq
where $\rho^{(N)}_c(\mathbf{r})$ is the charge distribution inside the neutron or proton, $\mathbf{r}_N$ is the coordinate of nucleon.
The sum $\sum_N$ runs over all nucleons, while  $\sum_p$ is restricted to protons. It is shown in Eq.(\ref{eq:Schiff_EDM_nucleon}) that the $\mathbf{S}^{\rm (EDM)}$ is zero if  $\mathbf{d}_N=0$. In the below discussion, we exclude this contribution to the Schiff moment.

The PT-violating electrostatic interaction $\hat{H}^{(eN)}_{PT} $ between electrons and  nuclear Schiff moment ${\mathbf{S}}$ in the atom  is given by~\cite{Dzuba:2002kg}
 \begin{eqnarray}
 \hat{H}^{(eN)}_{PT}   = -4\pi e\sum_i   \mathbf{S}\cdot \boldsymbol{\nabla}\delta^3(R_i)
\end{eqnarray}
which mixes atomic states of opposite parity and induces a T-violating EDM in the atomic  ground state~\cite{Dzuba:2002kg},
\begin{equation}
d_{A}
\label{eq:EDM_atom}
\equiv\bra{\Psi^{(at)}_0} \hat D_z\ket{\Psi^{(at)}_0}
\simeq 2 \sum_{m\neq0} \frac{\bra{\Phi^{(at)}_0} \hat{H}^{(eN)}_{PT}\ket{\Phi^{(at)}_m}\bra{\Phi^{(at)}_m} \hat D_{z}\ket{\Phi^{(at)}_0}}{{\cal E}^A_{0}-{\cal E}^A_{m}},
\end{equation}
where $\ket{\Psi^{(at)}_0}$ is the perturbed atomic ground state with the mixture of parities due to $\hat{H}^{(eN)}_{PT}$. The sum $\sum_m$ runs over a complete set of  unperturbed atomic states $\ket{\Phi^{(at)}_m}$.  The $\ket{\Phi^{(at)}_0}$ and $\ket{\Phi^{(at)}_m}$ denote the atomic ground state and excited states with
${\cal E}^A_0, {\cal E}^A_m$ being their energies, and $\hat D_z=-e\hat r_z$ is the atomic electric dipole operator.  The atomic states have been calculated with different atomic many-body approaches, including Dirac-Fock method~\cite{Dzuba:2002kg,Dzuba:2009kn}, the multiconfiguration Dirac-Hartree-Fock method~\cite{Radziute:2013sba,Radziute:2015apa}, RPA~\cite{Dzuba:2002kg,Dzuba:2009kn}, and \ac{RCC} theory~\cite{Latha:2009nq,Singh:2013PRA_Xe129,Singh:2015_Ra225,Sahoo:2016zvr,Sahoo:2018ile}. Here we quote the results by the Dirac-Fock method and RCC theory.
\begin{itemize}
\item  In the Dirac-Fock calculation, atoms are treated as systems of closed-shell plus some valence electrons. The wave functions of electrons are determined by the single-particle relativistic HF equation. Correlations between the valence electrons were considered using the configuration-interacting method and the valence-core correlations were included using \ac{MBPT}. The results  for the three popular atoms $^{129}$Xe, $^{199}$Hg,  and $^{225}$Ra are  as  follows~\cite{Dzuba:2002kg,Dzuba:2009kn} 
\begin{subequations}
\label{eq:Dirac-Fock-relations}
\begin{align}
    d_{A}(^{129}\mathrm{Xe})&=0.38\times 10^{-17} S (e~ \mathrm{fm}^3)^{-1} (e~\mathrm{cm}),\\
    d_{A}(^{199}\mathrm{Hg})&=-2.8\times 10^{-17} S (e~ \mathrm{fm}^3)^{-1} (e~\mathrm{cm}),\\
        d_{A}(^{225}\mathrm{Ra})&=-8.5\times 10^{-17} S (e~ \mathrm{fm}^3)^{-1} (e~\mathrm{cm}).
        \end{align}
\end{subequations}
\item In the RCC  calculation, the wave functions of atomic states are constructed as superposition of configurations truncated to singlets and doublets excitations on top of a reference state. The results of the RCC calculations for $^{129}$Xe\cite{Singh:2013PRA_Xe129}, $^{199}$Hg\cite{Latha:2009nq},  and $^{225}$Ra\cite{Singh:2015_Ra225} are as follows
\begin{subequations}
\label{eq:RCC-relations}
\begin{align}
    d_{A}(^{129}\mathrm{Xe})&=  0.336\times 10^{-17} S (e~ \mathrm{fm}^3)^{-1} (e~\mathrm{cm}),\\
    d_{A}(^{199}\mathrm{Hg})&=-2.46\times 10^{-17} S (e~ \mathrm{fm}^3)^{-1} (e~\mathrm{cm}),\\
        d_{A}(^{225}\mathrm{Ra})&= -6.79\times 10^{-17} S (e~ \mathrm{fm}^3)^{-1} (e~\mathrm{cm}).
 \end{align}
\end{subequations} 
\end{itemize}

It is shown above that the atomic EDM $d_A$ also depends on the Schiff moment $S$ of atomic nucleus,  
\begin{eqnarray}
S\equiv \langle \Psi_0|\hat S_z\ket{\Psi_0},
\end{eqnarray}
which is zero if parity is conserved in the nuclear state $\ket{\Psi_0}$. However, if there is a PT-violating nucleon-nucleon interaction $\hat V_{PT}$, parity will be violated in the nuclear state. Considering the fact that the $\hat V_{PT}$ is much weaker (if not zero) compared to the standard nuclear force,  the wave function of parity-mixed nuclear ground state $\ket{\Psi_0}$ can be obtained with the first-order perturbation theory, i.e., 
\beq 
\ket{\Psi_0}
\simeq\ket{\Phi_0}+\sum_{i\neq0}
\cfrac{\bra{\Phi_i}\hat{V}_{PT}\ket{\Phi_0}}{E_0-E_i}\ket{\Phi_i}.
\eeq
Here, $\ket{\Phi_0}$ and $\ket{\Phi_{i\neq0}}$ represent  the wave functions of nuclear ground-state and excited states of the Hamiltonian $\hat H_0$  with PT-even nuclear strong interaction.  
With this approximation, the Schiff moment $S$ of the ground state $\ket{\Psi_0}$ becomes
\begin{eqnarray}
\label{eq:NSM}
    S\simeq
    \sum_{i\neq 0}\cfrac{\langle \Phi_0|\hat{S}_z|\Phi_i\rangle\langle \Phi_i|\hat{V}_{PT}|\Phi_0\rangle}{E_0-E_i}+c.c..
\end{eqnarray}
Since the PT-violating nuclear interaction $\hat V_{PT}$ is a scalar and changes parity, and $\hat{S}_z$ is a tensor operator of rank one, the intermediate excited states $\ket{\Phi_i}$ are required to have the same angular momentum as the ground state but with opposite parity, and the angular momentum should be nonzero. 
With these considerations, the atoms of odd-mass nuclei are chosen to search for permanent EDMs.

\begin{figure}[tb]
\centerline{\includegraphics[width=0.3\textwidth]{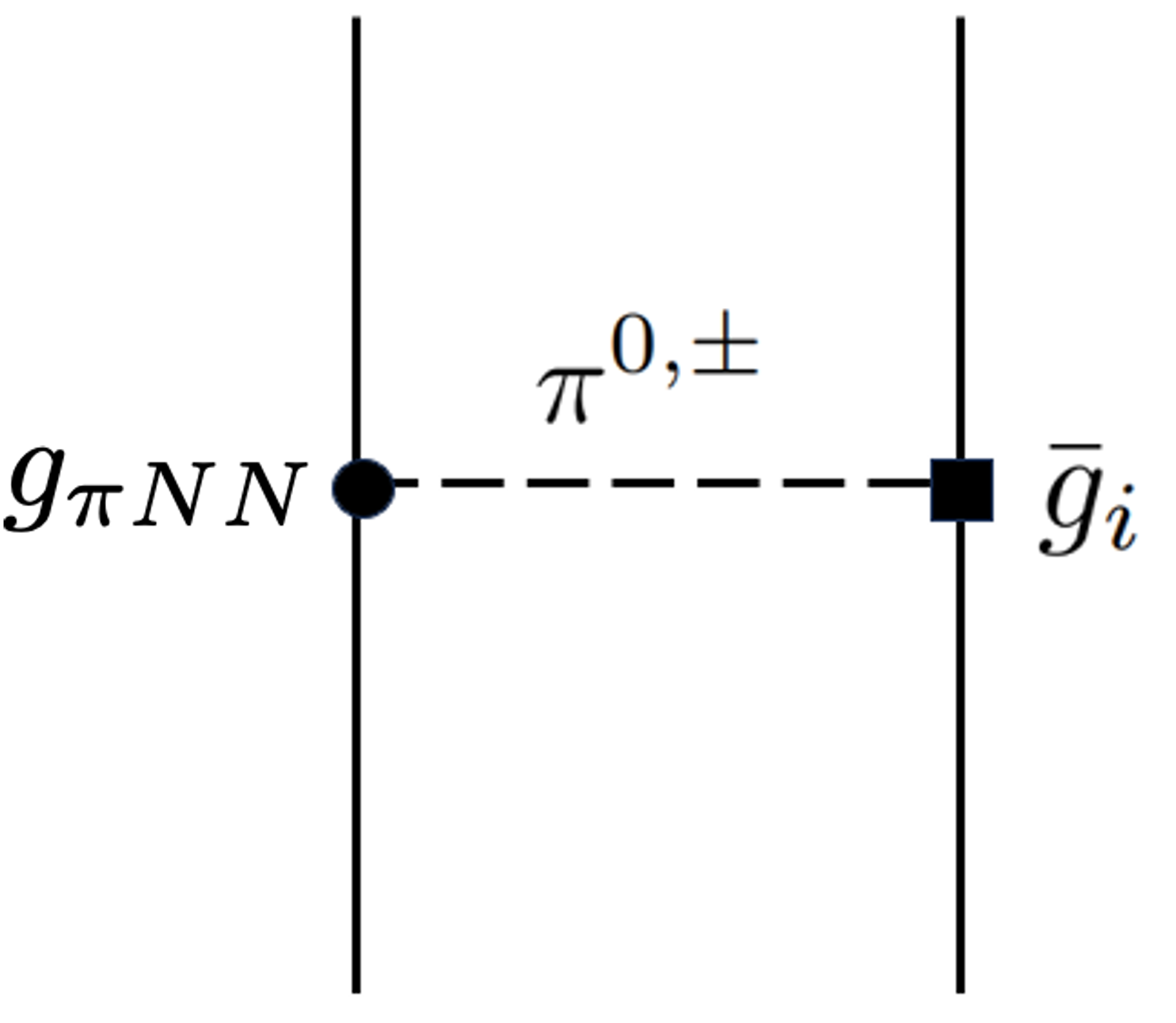}}
\caption{The long-range $1\pi$-exchange diagram contributing to the leading-order PT-odd nucleon-nucleon interactions. The solid and dashed lines represent nucleon and pion, respectively. The filled circle represents the standard PT-even $\pi N$ interaction with the LEC $g_{\pi NN}$, while the square stands for the PT-violating $\pi N$ interactions of isoscalar, isovector and isotensor types with the LECs $\bar g_{0,1,2}$, respectively. }
\label{fig:TCTV}
\end{figure}

No matter which fundamental theory is considered for the CP-violating source at high-energy scale, as listed in Fig.\ref{fig:EDM}, the PT-violating interaction $\hat V_{PT}$ at the nuclear energy scale contains the contribution from the long-range one-pion exchange, as schematically depicted in Fig.\ref{fig:TCTV}. The standard PT-even $\pi N$ coupling is described by the following Lagrangian density in the nonrelativistic framework, 
\begin{eqnarray}
\mathcal{L}^{\mathrm{(PT-even)}}_{\pi N} 
= -\frac{g_{\pi NN}}{2m_N} \bar{N} (\boldsymbol{\sigma}\cdot \boldsymbol{\nabla}) \vec{\tau}\cdot \vec{\pi} N,
\end{eqnarray}
where $N$ stands for nucleon field, $g_{\pi NN}=m_Ng_A/f_\pi$ for the coupling constant  with the pion decay constant $f_\pi\simeq 92.4$ MeV, and axial-vector coupling $g_A\simeq1.27$. The nonrelativistic Lagrangian for the  PT-odd $\pi N$ coupling is~\cite{Engel:2013PPNP}
\beq
\label{eq:Lagrangian_PVTV}
\mathcal{L}_{\pi N}^{\mathrm{(PT-odd)}}=  \bar{N}\left[\bar{g}_0 \vec{\tau} \cdot \vec{\pi}+\bar{g}_1\pi^{0}+\bar{g}_2\left(3 \tau_{3} \pi^{0}-\vec{\tau} \cdot \vec{\pi}\right)\right] N,
\eeq
where $\bar g_{i=0,1,2}$ are the unknown \ac{LEC}s for the isoscalar, isovector, and isotensor types of PT-violating $\pi N$ interactions. With the above two Lagrangian densities, one can derive the  PT-violating two-body interaction $\hat V_{PT}(\mathbf{r}_{12})$ in the coordinate-space representation~\cite{Haxton:1983dq,Herczeg:2001,Engel:2003rz,Maekawa:2011vs}
\begin{eqnarray}
\label{eq:vptp}
    \hat V_{PT}(\mathbf{r}_{12}) 
&=&g_{\pi NN}\Bigg(\bar g_0 {\cal V}^{\rm (PT-odd)}_0
+\bar g_1 {\cal V}^{\rm (PT-odd)}_1
+\bar g_2 {\cal V}^{\rm (PT-odd)}_2\Bigg),
\end{eqnarray}
where the PT-odd potentials ${\cal V}^{\rm (PT-odd)}_i(\mathbf{r}_{12})$ are given by
\begin{subequations} 
\begin{eqnarray} 
  {\cal V}^{\rm (PT-odd)}_0(\mathbf{r}_{12}) &=&-\cfrac{m_\pi^2}{8\pi m_N} (\vec{\tau}_1 \cdot \vec{\tau}_2)(\boldsymbol{\sigma}^-_{12}\cdot{\hat{\mathbf{r}}}_{12})\frac{e^{-x}}{x}(1+\frac{1}{x}),\\
  {\cal V}^{\rm (PT-odd)}_1(\mathbf{r}_{12}) &=&+\cfrac{m_\pi^2}{16\pi m_N}\Bigg[ (\tau_{1z}+\tau_{2z})(\boldsymbol{\sigma}^-_{12}\cdot{\mathbf{r}}_{12}) \nonumber\\
  &&+(\tau_{1z}-\tau_{2z})(\boldsymbol{\sigma}^+_{12}\cdot\hat{{\mathbf{r}}}_{12})
    \Bigg] \frac{e^{-x}}{x}(1+\frac{1}{x}),\\
  {\cal V}^{\rm (PT-odd)}_2(\mathbf{r}_{12}) &=&-\cfrac{m_\pi^2}{8\pi m_N}(3\tau_{1z}\tau_{2z}-\vec{\tau}_1\cdot\vec{\tau}_2)(\boldsymbol{\sigma}^-_{12}\cdot\hat{{\mathbf{r}}}_{12})\frac{e^{-x}}{x}(1+\frac{1}{x}).
\end{eqnarray}
\end{subequations} 
Here, $\mathbf{r}_{12}\equiv\mathbf{r}_1-\mathbf{r}_2$ is for the relative coordinate of two interacting nucleons, and $\boldsymbol{\sigma}^\pm_{12}\equiv\boldsymbol{\sigma}_{1}\pm\boldsymbol{\sigma}_{2}$. Two dimensionless quantities $x\equiv m_\pi r_{12}$, and $\hat{{\mathbf{r}}}_{12}\equiv\mathbf{r}_{12}/r_{12}$ are introduced. The third-component of isospin $\tau_z$ is $+1$
for neutrons and $-1$ for protons, $m_\pi$, and $m_N$ are the masses of pion and nucleon, respectively. 

Substituting the PT-violating nuclear interaction (\ref{eq:vptp}) into (\ref{eq:NSM}), one can express the Schiff moment in terms of the LECs $\bar g_i$,
\begin{eqnarray}
    S=g_{\pi NN}(a_0 \bar g_0+a_1 \bar g_1+a_2 \bar g_2),
\end{eqnarray}
where the coefficients $a_{i=0, 1, 2}$ contain all the nuclear-structure information from theoretical calculations and thus vary with nuclear models.

\subsection{Calculations of \ac{NSM}}

\begin{figure}[tb]
\centerline{\includegraphics[width=0.8\textwidth]{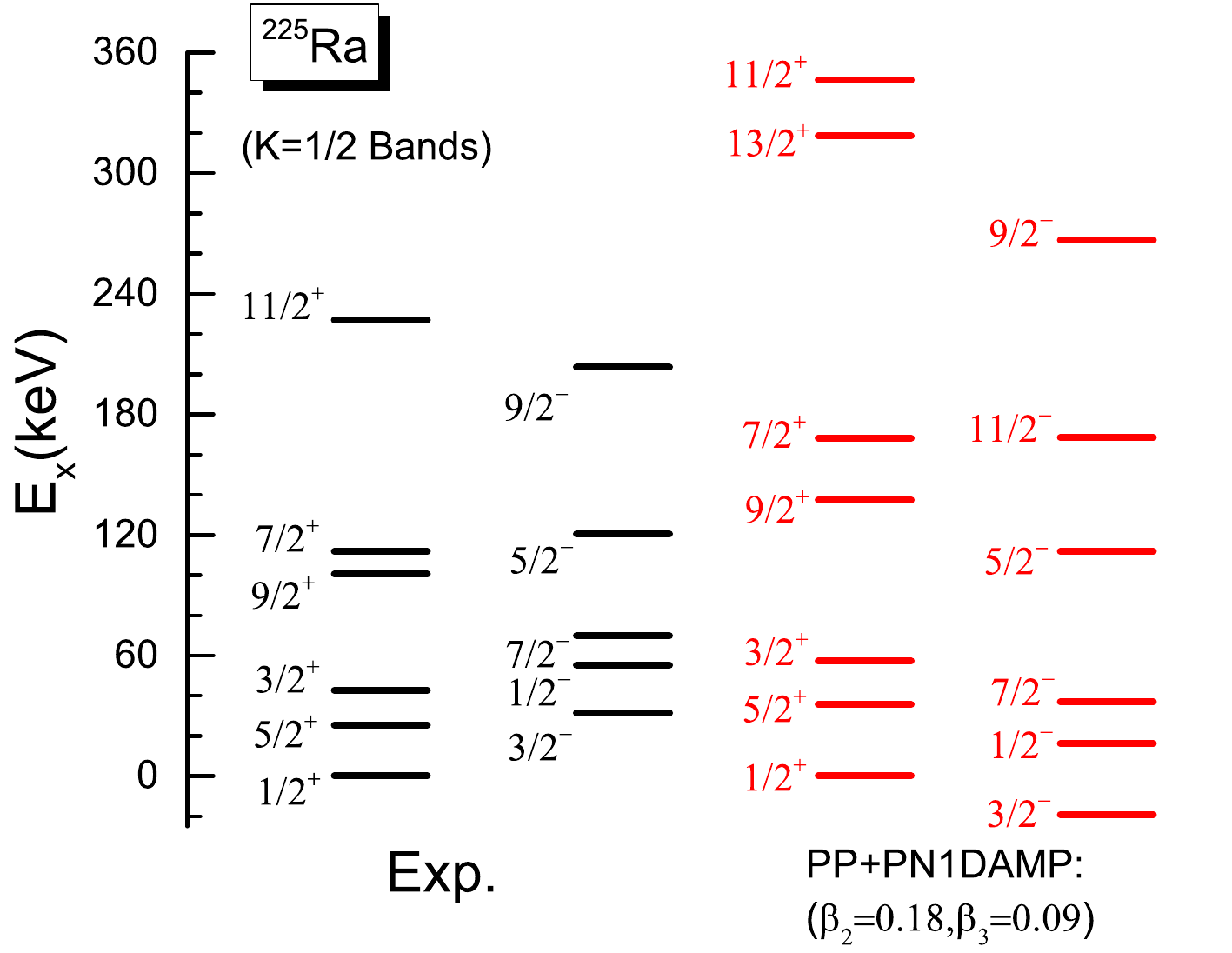}}
\caption{The energy spectra of low-lying parity-doublet states with $K=1/2$ in \nuclide[225]{Ra} from the PP+PN1DAMP calculation based on the quadrupole-octupole deformed  state with deformation parameters $(\beta_2=0.18, \beta_3= 0.09)$ by the CDFT, in comparison with corresponding data~\cite{NNDC}. The results are taken from Ref.~\cite{Zhou:2023_GCM4OA}.}
\label{fig:Ra225 spectra}
\end{figure}

According to Eq. (\ref{eq:NSM}), the nuclear Schiff moment $S$ may be enhanced in nuclei with a low-lying excited state having a similar energy and the same angular momentum as the ground state but with opposite parity. This type of low-energy structure is expected to exist in nuclei with intrinsic octupole deformation or a soft octupole vibration mode, which could lead to strongly enhanced nuclear Schiff moments by a factor of up to two to three orders of magnitude~\cite{Engel:2000,Engel:2003rz,Flambaum:2020tym}. Therefore, the Schiff moments of odd-mass nuclei with strong octupole correlations have garnered significant attention in the nuclear theory community. However, determining Schiff moments presents a significant challenge because they are sensitive to the excitation energies and wave functions of parity-doublet states. The nucleus $^{225}$Ra serves as a notable example of octupole-deformed nuclei that are of great interest to experimentalists~\cite{Chupp:2017rkp}. The energy splitting between the parity doublets (ground state with spin-parity $1/2^+$ and excited state $1/2^-$) is approximately 55 keV\cite{Helmer:1987fim}, as shown in Fig.\ref{fig:Ra225 spectra}.

A variety of nuclear models based on different levels of approximation have been employed to investigate the low-lying states and Schiff moments (\ref{eq:NSM}) of nuclei that are relevant for the measurements of atomic EDMs. 
These models include the simple individual-particle model (IPM) with a phenomenological \ac{WS} potential~\cite{Flambaum:1985gv}, and the IPM+RPA  with core-polarization corrections determined from RPA method~\cite{Dmitriev:2004fk}, the self-consistent mean-field approach of Skyrme \ac{HFB}/HF+BCS~\cite{Dobaczewski:2005hz,Ban:2010ea,Dobaczewski:2018nim} and the SHFB+QRPA with core-polarization corrections by QRPA~\cite{deJesus:2005nb}, the collective models of \ac{PRM}~\cite{Spevak:1996tu} and \ac{QOV}~\cite{Auerbach:2006pp}, and the valence-space \ac{ISM} with different effective interactions and model truncation schemes~\cite{Yoshinaga:2013,Teruya:2017don,Yanase:2020agg,Yanase:2022atk}.
In the following, we sketch the basic ideas of the three types of popularly used nuclear models for the Schiff moments.  

\begin{figure}[t]
\centerline{\includegraphics[width=1.0\textwidth]{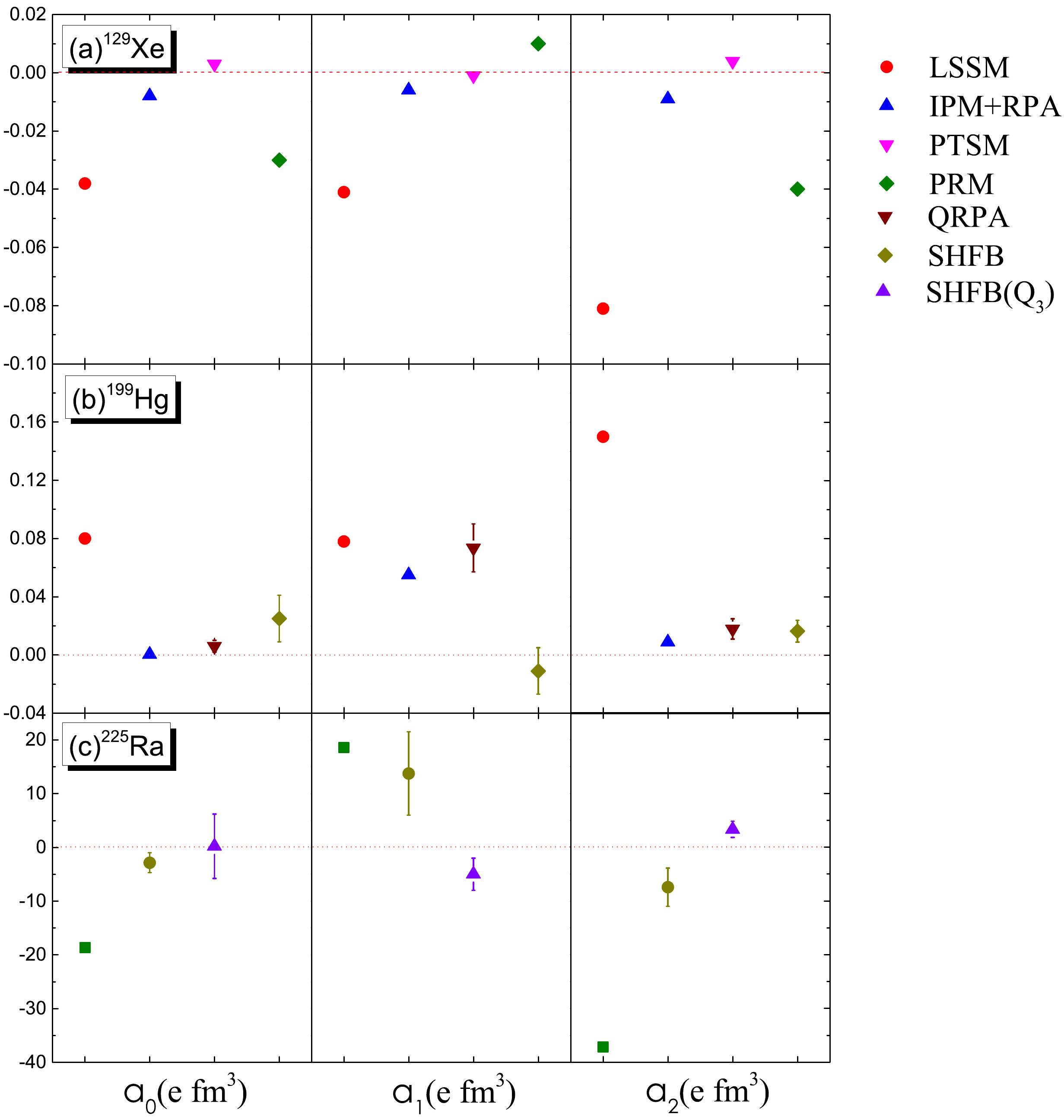}}
\caption{The coefficients ($a_0, a_1, a_2$) of nuclear Schiff moments for $^{129}$Xe, $^{199}$Hg, and $^{225}$Ra from different model calculations, including the IPM+RPA~\cite{Dmitriev:2004fk}, SHFB~\cite{Dobaczewski:2005hz,Ban:2010ea}, SHFB($Q_3$)~\cite{Dobaczewski:2018nim}, PRM~\cite{Spevak:1996tu}, LSSM\cite{Yanase:2020agg}, SHFB+QRPA\cite{deJesus:2005nb}, and \ac{PTSM}~\cite{Teruya:2017don}.}
\label{fig:NSM_comparison}
\end{figure}

\begin{itemize}
\item The collective particle-core coupling model or \ac{PRM}, in which the wave function of  ground state with a small amount of mixing of the opposite-parity component due to the PT-odd nuclear force $V_{PT}$ is approximated as~\cite{Spevak:1996tu}
\beq 
\ket{\Psi} = \frac{1}{\sqrt{2}}[(1+\alpha)\ket{JMK}+(1-\alpha)\ket{JM-K}],
\eeq
where the mixing amplitude $\alpha$ is determined by 
\beq
\alpha=\frac{\left\langle JMK(-)\left|V^{P T}\right| JMK(+)\right\rangle}{E^{+}-E^{-}},
\eeq
and $E^{+}-E^{-}$ is the energy splitting between the parity-doublet states. The  wave function of $\ket{JMK}$ is a product of collective rotation part and intrinsic part,
\beq 
\ket{JMK}
=\left(\frac{2J+1}{8 \pi^{2}}\right)^{1 / 2} D_{MK}^{J}(\varphi, \vartheta, \psi) \varphi_{K}^{(A)}\chi^{(A)}_c
\eeq
where $D_{MK}^{J}$ is the Wigner-D function,  $\chi^{(A)}_c$  is the wave function of the quadrupole and octupole deformed (reflection asymmetric) nuclear core in the intrinsic frame. The $\varphi_{ K}^{(A)}$ stands for the wave function of the unpaired nucleon in the intrinsic frame, with an angular momentum projection of $K$ along the  symmetric axis, usually determined by the Schrodinger equation with a reflection-asymmetric single-particle  plus pairing Hamiltonian in the particle-core coupling models~\cite{Auerbach:1996zd,Spevak:1996tu}. 

For simplicity, one can use the following expression to estimate the Schiff moment ~\cite{Flambaum:2020_EDM_NSM,Flambaum:2020tym},
\begin{eqnarray}
\label{eq:Schiff_moment_MF_beta23}
    S \thickapprox 1\times 10^{-4}
    \cfrac{J}{J+1}~\beta_2\beta^2_3ZA^{2/3}~\cfrac{\mathrm{keV}}{E^--E^+}~e~\eta~ \mathrm{fm}^3,
\end{eqnarray} 
 where  $\beta_2$ and $\beta_3$ are the quadrupole and octupole deformation parameters of the nucleus in intrinsic frame, $\eta$ is the dimensionless strength constant (a function of the LECs $\bar g_i$~\cite{Dmitriev:2004fk}) of the one-body mean-field potential corresponding to the  PT-violating nuclear force~\cite{Sushkov:1984,Engel:2003rz},
\beq
\bra{a}\hat U_{PT}\ket{b}
=\sum_{\epsilon_c<\epsilon_F} \bra{ac}\hat V_{PT}\ket{bc}
\simeq \frac{G}{\sqrt{2}} \frac{\eta}{2 m_N}\bra{a}(\boldsymbol{\sigma} \cdot\mathbf{\nabla}) \rho_t \ket{b}.
\eeq 
Here $G$ is the Fermi constant, $\rho_t$ is the total density of neutrons and protons.  
 
It is seen from Eq.(\ref{eq:Schiff_moment_MF_beta23}) that the Schiff moment $S$ is proportional not only  to $ZA^{2/3}$, but also to quadrupole deformation $\beta_2$ and the square of octupole deformation parameter $\beta_3$~\cite{Flambaum:2020tym}. 
This implies that the nuclear Schiff moment can be greatly enhanced by octupole deformation~\cite{Engel:2000,Flambaum:2003mv,Auerbach:2006pp}. Thus, the atoms with an odd-even heavy octupole-correlated nuclei are of particular interest for the experimental search for permanent atomic EDMs.  Indeed, it is shown in Fig.~\ref{fig:NSM_comparison} that the $a_i$ values of \nuclide[225]{Ra} are overall larger than those of \nuclide[129]{Xe} and \nuclide[199]{Hg} by two orders of magnitude.

\item The self-consistent mean-field approaches, in which all nucleons are  treated at the same foot in the mean-field approximation~\cite{Engel:2003rz,Dobaczewski:2005hz,Ban:2010ea}, various Skyrme EDFs are usually used. The Schiff moment in the mean-field approximation is given by the product of the intrinsic expectation value of $\hat S_z$ and $\hat V_{PT}$ 
\begin{eqnarray}
\label{eq:Schiff_SHFB}
    S\thickapprox -2\cfrac{J}{J+1}~\bra{\Phi(\mathbf{q})} \hat S_z
    \ket{\Phi(\mathbf{q})}~\cfrac{\bra{\Phi(\mathbf{q})} V_{PT}
    \ket{\Phi(\mathbf{q})}}{E^- - E^+},
\end{eqnarray}
where the $J/(J+1)$ comes from the transformation from intrinsic frame to laboratory frame.  $\ket{\Phi(\mathbf{q})}$ is the parity-violating wave function of mean-field energy-minimal state.
Since parity is usually violated in the mean-field states, there is no way to calculate the energy of $E_{\pm}$. Thus, the energy difference $E^- - E^+$ of two opposite parity states but with same angular momentum is taken from the data.   The mean-field potential $\hat U_{PT}$ under the zero-range approximation $(m_\pi\to \infty)$ is simplified as~\cite{Engel:2003rz}
\beqn 
\hat{U}_{P T}(\boldsymbol{r}) 
&\simeq& -\frac{g_{\pi NN}}{2 m_{\pi}^{2} m_{N}}  
\Bigg\{\sum_{i=1}^{A} \boldsymbol{\sigma}_{i} \tau_{z, i} \cdot\left[\left(\bar{g}_{0}+2 \bar{g}_{2}\right) \boldsymbol{\nabla} \rho_{1}(\boldsymbol{r})-\bar{g}_{1} \boldsymbol{\nabla} \rho_{0}(\boldsymbol{r})\right]  \nonumber\\
&&  +\frac{1}{2} \sum_{i=1}^{A} \boldsymbol{\sigma}_{i} \cdot [\left(-3 \bar{g}_{0}+\bar{g}_{1} \tau_{z, i}\right) \boldsymbol{J}_{0}(\boldsymbol{r}) \nonumber\\
&&  +\left(\bar{g}_{1}+\bar{g}_{0} \tau_{z, i}-4 \bar{g}_{2} \tau_{z, i}\right) \boldsymbol{J}_{1}(\boldsymbol{r}) ]\Bigg\}.
\eeqn
where the isoscalar and isovector densities $\rho_0=\rho_n+\rho_p$ and $\rho_1=\rho_n-\rho_p$,   and the “spin-orbit” current  $\boldsymbol{J}_{i=0, 1}(\boldsymbol{r})$ in the Skyrme EDF. The intrinsic expectation value of  $\hat V_{PT}$ is finally given by the expectation value of $\hat{U}_{P T}$.
Since the summation $\sum_i$ runs over all the nucleons, core polarization effect is automatically taken into account.

\item The \ac{BMF} approaches in which the ground-state and excited states are calculated in the laboratory frame with the configuration mixing effect.  Recently, {\em ab initio} no-core shell model (NCSM) framework has been applied to calculate EDMs of several light nuclei using chiral two- and three-body
interactions and a PT-violating Hamiltonian based on a one-meson-exchange model~\cite{Froese:2021_EDM_nuclei}. Application of this kind of studies to the Schiff moments of heavy nuclei are very challenging. So far, only the valence-space ISM~\cite{Yanase:2020agg,Yanase:2022atk} based on different approximations have been applied to nuclear Schiff moments, including the \ac{PTSM}~\cite{Yoshinaga:2013,Teruya:2017don}, where many-body configurations are composed of nucleon pairs with specific angular momenta. In these studies, the two-body matrix elements of PT-violating nucleon-nucleon interactions are evaluated explicitly. 

Given the remarkable success of the SPGCM in describing low-lying states of octupole-deformed even-even nuclei, there is considerable interest in extending this framework to investigate nuclear Schiff moments. To achieve this, it's essential to adapt the GCM for odd-mass nuclei, which has already been developed based on \ac{HFB} states utilizing Skyrme~\cite{Bally:2014} and Gogny~\cite{Borrajo:2017PLB} forces. However, these studies have not yet incorporated octupole deformation and parity projection. More recently, progress has been made in the SPGCM  based on the CDFT for quadrupole-octupole deformed odd-mass nuclei~\cite{Zhou:2023_GCM4OA}. As shown in Fig.~\ref{fig:Ra225 spectra}, this approach successfully reproduces the main features of the energy spectrum for parity doublets with $K=1/2$ in \nuclide[225]{Ra}, with the exception that the negative-parity band is  lower in energy than the positive-parity one which is different from the data. This discrepancy is expected to be resolved by incorporating the mixing of deformed states with different values of $(\beta_2, \beta_3)$.

\end{itemize}

\begin{table}[tb]
    \centering
    \tabcolsep=6pt 
    \caption{The coefficients $a_i$ ($e~\text{fm}^3$) of nuclear Schiff moments from the calculations of different nuclear models.}
 \begin{tabular}{ccccl}
   \hline
 Isotopes & $a_0$ & $a_1$ & $a_2$& Nuclear models\\
  \hline
         $^{153}$Eu  & $-9.62$& 47.3 &$-25.53$  & PRM\cite{Flambaum:2020tym}\\ 
  \hline
         $^{129}$Xe  & $-0.038$& $-0.041$ &$-0.081$  & LSSM\cite{Yanase:2020agg}\\ 
     $^{129}$Xe  & $-0.008$& $-0.006$ &$-0.009$  & IPM+RPA\cite{Dmitriev:2004fk}\\    
          $^{129}$Xe  & 0.003& $-0.001$ & 0.004  & PTSM\cite{Teruya:2017don}\\ 
    $^{129}$Xe  & $-0.03$& 0.01 & $-0.04$  & PRM\cite{Flambaum:2020Xe}\\ 
  \hline
           $^{199}$Hg  & 0.080& 0.078 &0.15  & LSSM\cite{Yanase:2020agg}\\ 
           $^{199}$Hg  & 0.0004& 0.055 &0.009  & IPM+RPA\cite{Dmitriev:2004fk,Dmitriev:2003kb}\\ 
            $^{199}$Hg  & [0.002, 0.010]& [0.057, 0.090] &[0.011, 0.025]  & SHFB+QRPA\cite{deJesus:2005nb}\\ 
        $^{199}$Hg  & [0.009, 0.041]& [$-$0.027, +0.005] &[0.009, 0.024]  & SHFB\cite{Ban:2010ea}\\
  \hline
     $^{221}$Rn   & $-0.04(10)$ & $-1.7(3)$ & 0.67(10)& SHFB($Q_3$)\cite{Dobaczewski:2018nim} \\ 
  \hline
    $^{223}$Rn   & $-62$ & 62 & $-100$ & PRM\cite{Spevak:1996tu} \\  
      $^{223}$Rn  & $-0.08(8)$& $-2.4(4)$ &0.86(10) & SHFB($Q_3$)\cite{Dobaczewski:2018nim}  \\ 
  \hline
   $^{223}$Ra    & $-25$ & 25 & $-50$& PRM\cite{Spevak:1996tu} \\ 
   \hline
    $^{223}$Fr   & $-31$ & 31 & $-62$ & PRM\cite{Spevak:1996tu} \\  
    $^{223}$Fr  & $-0.02$ & $-0.02$ & $-0.04$ & QOV \cite{Auerbach:2006pp} \\  
        $^{223}$Fr  & 0.07(20) &$-0.8(7)$ &0.05(40) & SHFB($Q_3$)\cite{Dobaczewski:2018nim} \\ 
  \hline
           $^{225}$Ra  & $-18.6$ & 18.6 & $-37.2$ & PRM \cite{Spevak:1996tu} \\ 
    $^{225}$Ra  & $[-1.0,-4.7]$ & [6.0, 21.5] &[$-3.9,-11.0$] & SHFB \cite{Dobaczewski:2005hz} \\ 
    $^{225}$Ra  & 0.2(6)& $-5(3)$ & 3.3(1.5) & SHFB($Q_3$) \cite{Dobaczewski:2018nim} \\ 
  \hline
           $^{227}$Ac  &$ -26$& 129 &$-69$ & PRM \cite{Flambaum:2020tym} \\ 
  \hline
      $^{229}$Pa   & $-1.2(3)$& $-0.9(9)$ &$-0.3(5)$ & SHFB($Q_3$) \cite{Dobaczewski:2018nim} \\ 
  \hline
             $^{235}$U   & $-7.8$& 38.7 &$-20.7$ & PRM \cite{Flambaum:2020tym} \\ 
  \hline
      $^{237}$Np    & $-15.6$& 77.4 &$-41.4$& PRM \cite{Flambaum:2020tym} \\ 
  \hline
\end{tabular}
    \label{tab:SummatyofNSM}
\end{table}

Table~\ref{tab:SummatyofNSM} summarizes the coefficients $a_{0, 1, 2}$ of nuclear Schiff moments obtained from various models for nuclei of experimental interest. Figure~\ref{fig:NSM_comparison} provides a comparison of these coefficients for three prominent isotopes \nuclide[129]{Xe}, \nuclide[199]{Hg}, and \nuclide[225]{Ra}. It is evident that the values of $a_i$ exhibit significant discrepancies, varying by up to two to three orders of magnitude and sometimes even having different signs. This substantial variation in $a_i$ introduces considerable uncertainty into the Schiff moments and, consequently, into the allowed values for the Low-Energy Constants (LECs) characterizing Parity-and-Time (PT)-violating nucleon-nucleon interactions.

In a Skyrme HFB study~\cite{Dobaczewski:2005hz}, it was observed that finite-range effects reduce the direct matrix elements of PT-odd interactions (and the corresponding $a_i$ values) from the zero-range limit by a factor of two or three. Furthermore, the different choices of Skyrme EDFs introduce an uncertainty of a similar magnitude. Minimizing the discrepancy among these models remains a primary objective in the nuclear theory community. Achieving this goal involves understanding the assumptions and approximations made by each model and quantifying the potential errors associated with them~\cite{Engel:2013PPNP}.

Recent research has unveiled strong correlations between the intrinsic Schiff moments $\bra{\Phi(\mathbf{q})} \hat S_z \ket{\Phi(\mathbf{q})}$ of odd-mass nuclei and the octupole moments $Q_{30}=\bra{\Phi(\mathbf{q})} \hat Q_{30} \ket{\Phi(\mathbf{q})}$ of neighboring even-even nuclei~\cite{Dobaczewski:2018nim}. With the correlation relation and available data on octupole moments $Q_{30}$, it has become possible to determine the intrinsic Schiff moments. This correlation significantly reduces systematic uncertainties originating from nuclear EDFs. Furthermore, linear correlation relations have been identified between the coefficients $a_i$ and $Q_{30}$. Nevertheless, despite these correlations, substantial uncertainties persist in the resulting $a_i$ values due to the significant uncertainty in the data of $Q_{30}$.

\subsection{Constraints on the LECs of PT-violating  nuclear force}

\begin{figure}[tb]
\centerline{\includegraphics[width=\textwidth]{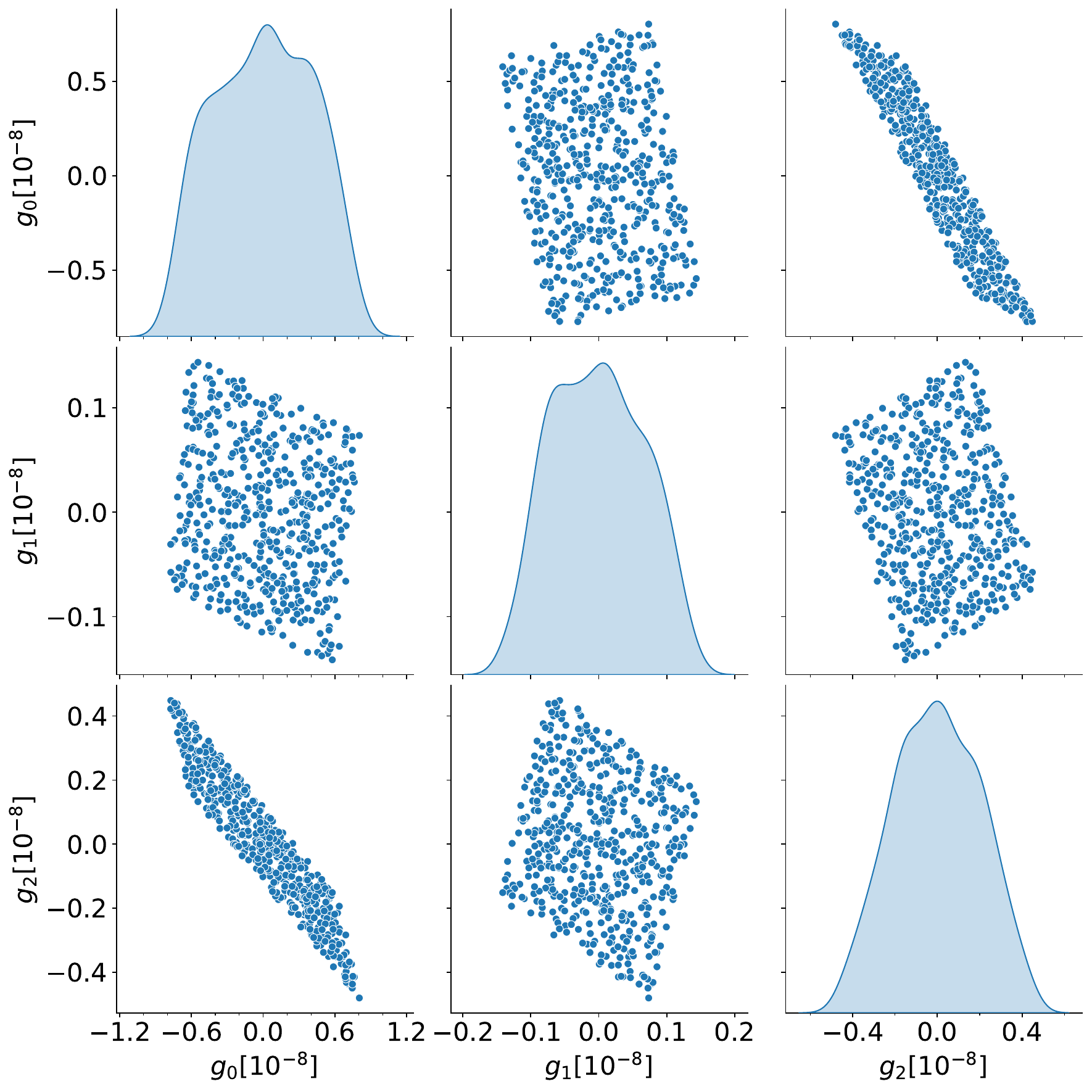}}
\caption{The distribution of the allowed values for the LECs $\bar{g}_{0,1,2}[10^{-8}]$ of the CP-violating nuclear force constrained by the latest measurements on the EDMs of $\nuclide[129]{Xe}$, $\nuclide[199]{Hg}$, and $\nuclide[225]{Ra}$ and the theoretical calculations for the atomic factors~\cite{Dzuba:2002kg,Dzuba:2009kn} and coefficients in the nuclear Schiff moments~\cite{Engel:2013PPNP}, where $a_1(\nuclide[199]{Hg})=0.02$ for $\nuclide[199]{Hg}$ is used. See text for details. }
\label{fig:EDM_LECs_constraints_p002}
\end{figure}

\begin{figure}[tb]
\centerline{\includegraphics[width=\textwidth]{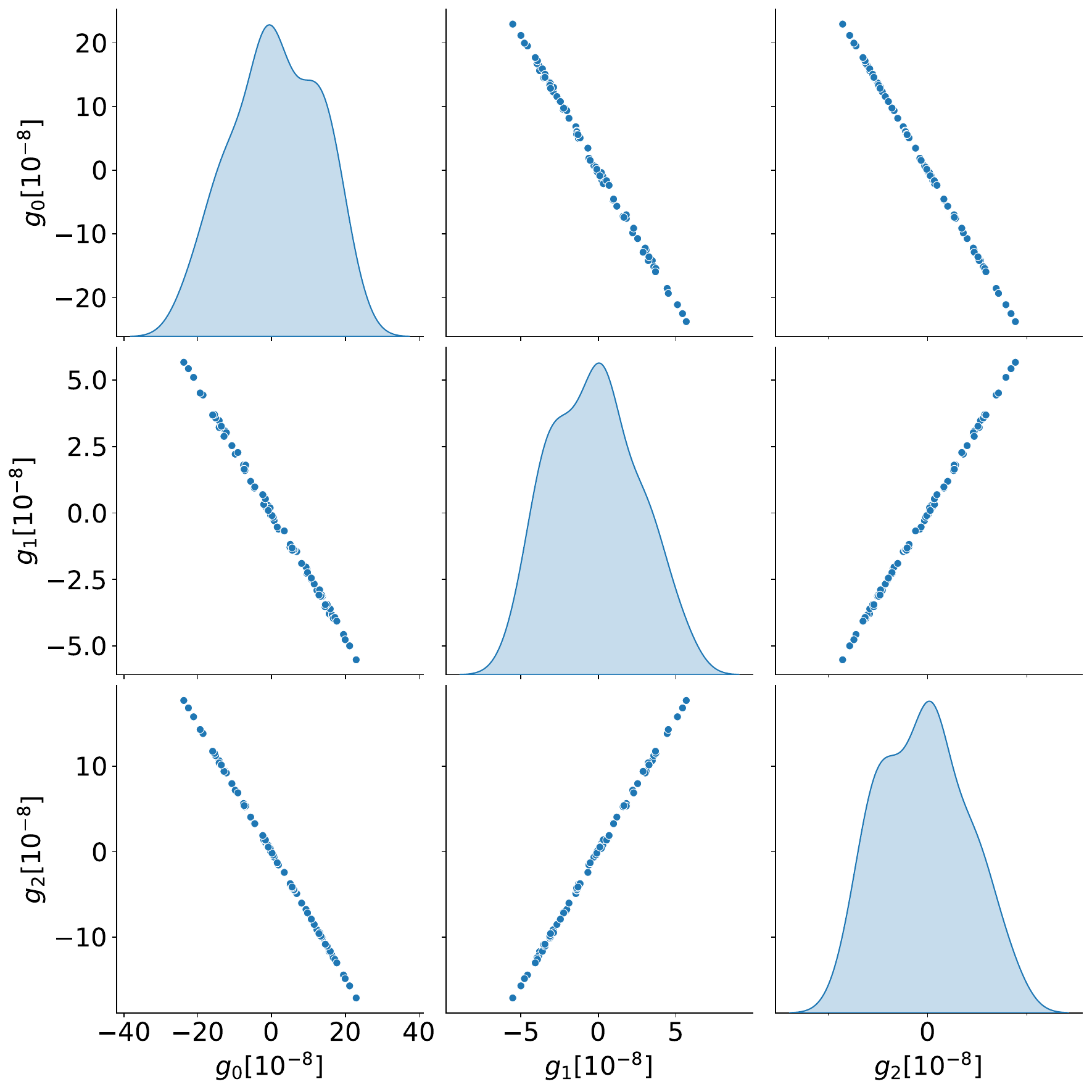}}
\caption{Same as Fig.~\ref{fig:EDM_LECs_constraints_p002}, but with $a_1=-0.02$ for $\nuclide[199]{Hg}$. }
\label{fig:EDM_LECs_constraints_n002}
\end{figure}

The best experimental upper limits (95\%C.L.) on the ground-state atomic EDMs of the three atoms $^{129}\mathrm{Xe}$~\cite{Sachdeva:2019_Xe129},  $^{199}\mathrm{Hg}$~\cite{Graner:2016_Hg199}, and $^{225}\mathrm{Ra}$~\cite{Bishof:2016_Ra225},
\begin{subequations}
\begin{align}
 |d_{A}(^{129}\mathrm{Xe})| &\le 1.4\times 10^{-27} (e~ \mathrm{cm}), \\
 |d_{A}(^{199}\mathrm{Hg})| &\le 7.4\times 10^{-30} (e~ \mathrm{cm}), \\
 |d_{A}(^{225}\mathrm{Ra})| &\le 1.4\times 10^{-23} (e~ \mathrm{cm}),   
\end{align}
\end{subequations} 
and the coefficients $a_i$ for the nuclei $^{129}$Xe, $^{199}$Hg, and $^{225}$Ra from Ref.~\cite{Engel:2013PPNP},
\begin{subequations}
\begin{align}
     S(^{129}\mathrm{Xe})&=g_{\pi NN}(-0.008 \bar{g}_0 -0.006\bar g_1-0.009\bar g_2) (e~ \mathrm{fm}^3)\\
    S(^{199}\mathrm{Hg})&=g_{\pi NN}(+0.01 \bar{g}_0 \pm 0.02\bar g_1+0.02\bar g_2)(e~ \mathrm{fm}^3)\\
     S(^{225}\mathrm{Ra})&=g_{\pi NN}(-1.5 \bar{g}_0 + 6.0\bar g_1 -4.0\bar g_2)(e~ \mathrm{fm}^3),
\end{align}
\end{subequations}  
provide constraints on the allowed values for the LECs $\bar g_i$ of PT-violating nuclear force. 

To determine the ranges of these values, we assume that the atomic EDMs $d_A$ are only contributed to by the long-range PT-violating nuclear force $V_{PT}$, c.f. Fig.~\ref{fig:TCTV}, and sample the values of $\bar{g}_{0,1,2}$ approximately $10^9$ times. The samples that can reproduce atomic EDMs below their corresponding upper limits are displayed in Figs.~\ref{fig:EDM_LECs_constraints_p002} and \ref{fig:EDM_LECs_constraints_n002}. The results for the two choices of $a_1(\nuclide[199]{Hg})=\pm0.02$ are compared. It is observed that there are correlations between the three LECs, especially in the case of $a_1(\nuclide[199]{Hg})=-0.02$. Quantitatively, the upper limits of the three atomic EDMs provide constraints on the $\bar{g}_i$, which, for the case $a_1(\nuclide[199]{Hg})=0.02$, are
\begin{subequations}
\label{eq:a1_positive}
\begin{align}
    \bar g_0 \in [-0.78, 0.80]\times 10^{-8},\\
    \bar g_1 \in [-0.14, 0.14]\times 10^{-8},\\
    \bar g_2 \in [-0.48, 0.45]\times 10^{-8}.
\end{align}
\end{subequations}
and for the case $a_1(\nuclide[199]{Hg})=-0.02$, 
\begin{subequations}
\label{eq:a1_positive}
\begin{align}
    \bar g_0 \in [-24, 23]\times 10^{-8},\\
    \bar g_1 \in [-5.5, 5.6]\times 10^{-8},\\
    \bar g_2 \in [-17, 17]\times 10^{-8}.
\end{align}
\end{subequations}
It is evident that changing the value of the coefficient $a_1$ for the nuclear Schiff moment of \nuclide[199]{Hg} from $-0.02$ to $+0.02$ reduces the ranges of $\bar{g}_i$ by approximately one order of magnitude. This underscores the importance of accurate calculations of nuclear Schiff moments for all the nuclei of interest. Furthermore, it can be observed that the LEC $\bar{g}_1$ associated with the isovector type of PT-violating $\pi N$ coupling is more tightly constrained than those of isoscalar and isotensor types.

\section{Summary and perspectives}
\label{sec:summary}

Octupole deformation enriches the structure of atomic nuclei by introducing novel collective excitation modes, the description of which has garnered significant attention in the pursuit of a profound microscopic theory. In this review, we introduce the symmetry-projected generator coordinate method (SPGCM) for studying nuclear octupole collective motions. This method is based on reference states with different reflection-asymmetric shapes obtained from self-consistent mean-field calculations. The success of SPGCM has been demonstrated through applications to phenomena such as nucleon clustering in light nuclei,  and the phase transition from octupole vibrations to rotational motions in heavy nuclei. The SPGCM has also provided important inputs to understand structure of hypernuclei and  sub-barrier fusion reactions.

The atoms with octupole-shaped odd-mass nuclei has been of great experimental interest to  search for permanent atomic electric dipole moments, as the signal is proportional to nuclear Schiff moment which could be enhanced by several orders of magnitude if the nuclei possess intrinsic octupole deformation. We have reviewed the current status of studies on nuclear Schiff moments. Significant discrepancies in nuclear Schiff moments among different phenomenological nuclear models and the impact of these uncertainties on the determination of the low energy constants of PT-violating nucleon-nucleon interactions have been discussed in detail. Given the success of the SPGCM discussed above, we have shown the great potential for extending this framework to study the low-lying states of odd-mass nuclei and nuclear Schiff moments with octupole deformation, even though numerous challenges need to be overcome. The combination of SPGCM with modern nuclear EDFs, especially the CDFT where the parameters of the time-odd currents (important for odd-mass and rotating nuclei) share the same coupling constants with those of time-even densities, or with the ab initio method of in-medium similarity renormalization group starting from nuclear interactions derived from chiral effective field theory, will offer a compelling approach for determining nuclear Schiff moments in the future.
 
\section*{Acknowledgements}
We are grateful to Y. Fu, H. Hergert,  C.F. Jiao, G. Li, Z.P. Li,  B.N. Lv, J. Meng, H. Mei, P. Ring, Y.T. Rong, L.J. Wang and X.Y. Wu for collaboration and fruitful discussions on the works reviewed in this manuscript. This work is supported in part by the National Natural Science Foundation of China (Grant Nos. 12375119 and 12141501), and the Guangdong Basic and Applied Basic Research Foundation (2023A1515010936).

 \printacronyms

 

\end{document}